\documentclass[aps,showpacs,onecolumn,floatfix,amsmath,amssymb,nofootinbib]{revtex4-1}

\usepackage{amssymb}
\usepackage{amsmath}
\usepackage{epsfig}
\usepackage{graphicx}
\usepackage{color}
\usepackage{latexsym}
\usepackage{bm,latexsym}
\usepackage{mathrsfs}
\usepackage{float}
\usepackage{pbox}

\begin{document}


\title{ Semiclassical  Self Consistent  Treatment of  the Emergence of  Seeds  of  Cosmic  Structure. The second order construction.}

\author{Pedro Ca\~nate$^1$ $^2$}
\email{pcanate@utb.edu.co}

\author{Erandy Ramirez$^1$}
\email{ena.ramirez@correo.nucleares.unam.mx}

\author{Daniel Sudarsky$^3$ $^1$}
\email{sudarsky@nucleares.unam.mx}

\affiliation{$^1$Instituto de Ciencias Nucleares, Universidad Nacional
Aut\'onoma de M\'exico, A.P. 70-543, M\'exico D.F. 04510, M\'exico \\
$^2$Facultad de Ciencias B\'asicas, Universidad Tecnol\'ogica de Bol\'ivar, Campus Tecnol\'ogico Km 1. V\'ia Turbaco, Cartagena, Colombia\\ 
$3$ Department of Philosophy, New York University,  New York, NY 10003, United States of America \\
}


\date{\today}

    
\begin{abstract}

In this work we extend the results of  \cite{DiezTaSudarD}  where,
  Semiclassical Selfconsistent Configurations  (SSC)   formalism was  introduced.  The scheme  combines  quantum field theory on a background space-time, semiclassical  treatment of    gravitation and   spontaneous collapse  theories. The approach is applied to the context of early universe cosmology using a formal description of the transition from an initial inflationary stage characterized by a spatially homogeneous and isotropic (H\&I) universe, to another where inhomogeneities are present  in association  with  quantum fluctuations of the field driving inflation.
In  that work two constructions  are produced.   One   of them  describes  a    universe  that is completely   spatially homogeneous and isotropic, and the other  is  characterized by a slight excitation of the particular inhomogeneous and anisotropic perturbation. Finally, a  characterization of   their gluing  to each  other  is  provided   as  representing  the transition as  a result from  a spontaneous  collapse of the  state  of the  quantum  field, following the  hypothesis originally introduced in \cite{PeSahSu}.


Specifically, in \cite{DiezTaSudarD} this construction is carried out by using cosmological perturbation theory and working up to linear order in the perturbation.  However, given the nonlinear  nature of   gravitation,  we  should  in principle  explore the  application of the formalism  in a nonlinear regime.  To this end and as a first step, we study in this  work the transition from a spatially homogeneous and isotropic $(H\&I)$ Semiclassical Self-Consistent Configuration (SSC-I) to one SSC-II that is not spatially $(H\&I)$,  working this time up to second order in perturbation theory.  We find that the self consistent construction now requires consideration of the so called tensor modes, as well as a nontrivial mixing of modes   that  made the  analysis  much more difficult and which   could not a priori  be  warranted to work out in detail. The present  work   shows that this  is  indeed the case.  

\end{abstract}

\pacs{
04.62.+v, 
98.80.Qc, 
04.30.-w 
}

\maketitle



\section{Introduction}

The interplay between quantum theory and  gravitation continues to  be one of the  most intriguing open issues in   our   present   understanding  of the  fundamental  laws of   physics.    The fact   that, despite   impressive   developments  in the  various approaches \cite{String,LQG,Carlip,Page}, we  still do not have  at hand a  fully satisfactory and   workable   theory of  quantum gravity \cite{GambiniPullin, KozamehParisi, AlfMor-TecUrrut,CollPerSudUrrutVuc},   forces   us to  adopt    various  kinds of  half baked   approaches  to specific  questions  in the hope that they  will offer  adequate approximated treatments of the   relevant  situations.  One  of the  best known,   among   such    approaches, is the so called  semiclassical gravity  where,    matter is treated  in the  language  of quantum field  theory in curved  spacetime   while gravitation itself is   described in terms of a classical spacetime metric.   There is however, a widespread view  that such an approach is  essentially  untenable,   as   a result, in part of the   work \cite{Page},   which is   often quoted   as  conclusively indicating that   semiclassical gravity is  either,  in glaring  conflict  with  experiment, or internally inconsistent. In fact, the   paper shows  that,  unless one considers  that the quantum state of matter,  presumably prepared to yield    a superposition of  two   space-time metrics,  undergoes   a collapse   upon observation,  the theory  leads to conflict with  experimental  results. On the other hand,   if the     quantum state of matter   does   undergo   such  a collapse,  the semi-classical   equations  could not hold   while  it is  occurring,   as  one  side  is    divergence free  by construction and  the other  will generically  not be  so. 

On the other hand,  considerations  about  foundational  difficulties on quantum theory   have  led to the development of proposals   incorporating spontaneous   collapse of the wave function \cite{Pearle, Pearle06, GhiPERi} as a mechanism to address the so called  measurement problem (or  as it is  more accurately called, the "macro-objectification problem") .  There are  good,  but not yet conclusive,  arguments   suggesting that  gravity might  have a fundamental  role in such  process \cite{RPenrose, penrose}.  Moreover, consideration of the  "macro-objectification problem"  in the context of inflationary cosmology  has  led   to   works  \cite{DiezTaSudarD, PeSahSu, CaPeSu, UnaSu, LESU, sud2, LandaSU} based on the  posture that  the  incorporation  of     modified  versions of  quantum   theory,  in  particular    theories   involving  spontaneous collapse, are  essential  for   the adequate  account for  the emergence of the seeds of cosmic  structure   out of  fundamental quantum uncertainties ( sometimes called    "quantum fluctuations" ) in  the    inflationary   era  of the universe (see \cite{KieferPolarski} for  dissenting opinion, and   \cite{shortcomings} for  a  detailed  response ).  The problem in fact,   afflicts  approaches  such as the  most common one \cite{AlbFerr, PolStaro},  that do not relay on   strict semiclassical gravity  (and is  acknowledged  explicitly by  some   well-known authors \cite{weinberg,mukbook}), and  even  models that  try  to  account for the  emergence of  the so called  "primordial fluctuations",  without   relying on inflation \cite{HollWald},  as  discussed  in \cite{GaPeSUd}. 
 
 We  should stress that  in the present   manuscript we  will be working   with a fixed gauge  rather than relying  on  the  so  called   gauge invariant   cosmological   perturbation  approach  of \cite{jamesBardeen}  ( an  approach implementing   gauge  invariance at the perturbative level).    Because,   on the one  hand,  the  latter is not  very well suited  to   implement a semiclassical   type of treatment  ( in which the  metric  is not to be quantized)   that we advocate, and   on the other hand,    we want to   work with a formalism  which, despite having to rely on perturbation theory in practical applications,   is,  at least,    in principle,  suitable  for full fledged non-perturbative treatment.   Works  deviating  from strict  semiclassical approach, while   still  incorporating  spontaneous  collapse  in  inflationary   context  can be found  in  \cite{AlberGAbriDANIEL, MarVenn, DasLoch}.

Regarding the   viability of  semiclassical gravity,  in  various  previous  works  we  have     been advocating    a posture  that,   we  believe,   is much less  radical than the usual one, and  which is  based on viewing   semiclassical gravity not as  a fundamental theory,  but   only  as an effective and approximated  description, in  analogy with say, the    characterization of  a fluid in terms of   standard hydrodynamical concepts and  take  to  satisfy the  Navier-Stokes (NS) equations. It is clear that, although such characterization  and the results that emerge from  the use of the NS  equations,  are often  very  reliable and  provide  a sufficiently  good  approximated  description  of  a fluid  under  a large set of circumstances,  such a description is neither fundamental, nor  can it be expected to  provide a good characterization of the physics  in  every situation. When drastic  turbulence   ensues,  or  when     waves  break in the ocean, the fluid  description brakes down  and   the treatment using NS  is  simply   unviable. We take the view  that,  at  the  fundamental level, the   complete  and  precise characterization of the  situations  at the interface  between quantum theory and  gravitation,   requires   a quantum  theory of gravitation, which  quite   likely will be expressed in terms  that  are thus distinct from   the  metric  characterization of spacetime, as the  ultimate description in terms of quarks leptons  and   gauge bosons  are from the   notions of volume elements of  fluids  with  a given   density  and   moving with a certain  velocity. In fact,  it  is  likely that spacetime   notions of  general relativity  would only    have  an effective validity   and  appear  only  as   emergent properties. In that case,  it might be   that when  we arrive  at  the level of approximation in the description   where   spacetime notions  become   at all relevant, we  would    be already deep in the semiclassical regime.   This of course  does not   remove the need to    have a  solid    framework for the usage of  semiclassical gravity in the various contexts  we intend to.   Just as  it is desirable  to have a solid conceptual and   mathematical  formulation of   hydrodynamics  and the NS equations, even if they do  not represent  a    fundamental characterization of  the nature of fluids, a  similar  characterization of semiclassical gravity  seems  highly  desirable.    Moreover,  as we have seen,    it is  describable that such  formalism  be  able to  incorporate the  collapse  of the quantum  state of matter,    which we  will  be   considering in the context of the so-called  dynamical  or spontaneous  collapse  theories that  have been designed  to deal with the  measurement problem  in quantum theory, and in which    the   reduction of the quantum state is not  fundamentally tied to  an  external observer (see \cite{Bassi}  for extensive  discussion of such theories,  the motivation  development and  experimental  tests). 
   
Such   basic  formalism  has been  schematically proposed  in  \cite{DiezTaSudarD}  and   employed in the context of   cosmic  inflation where  the  spontaneous   collapse of the quantum state of the  inflaton field   is taken to play a fundamental role in generating the primeval inhomogeneities and anisotropies that   must  act as seeds of all the  structure in the universe \cite{PeSahSu}\footnote{We  should  point out that,  although    further discussing  the following issue  is not the  objective  of the present manuscript, there  are   some  conceptual  difficulties    in the   standard  treatments   of the  inflationary  account of emergence of the seeds of  cosmic  structure. These  difficulties  are tied to the so-called  "quantum  classical transition"  a question that continues to generate  spirited  debates. For   discussion of  our perspective  on the  subject, we point the reader to \cite{shortcomings, benefits, CastFortLauSud}. }.

The formalism is based on the concept of  the  semi-classical  self-consistent configuration (SSC),  defined  as follows:

{\bf DEFINITION: }The set 
$\{ \mathcal{M},  g_{\mu\nu}(x),\hat{\varphi}(x), \hat{\pi}(x), \mathscr{H}, \vert \xi \rangle \in\mathscr{H} \}$
represents a 
SSC if and only if $\{\mathcal{M}, g_{\mu\nu}(x)\}$ is  a  globally  hyperbolic  spacetime,  {$\hat{\varphi}(x)$} and {$\hat{\pi}(x)$}  correspond  to quantum field operators  over the  Hilbert space and {$\mathscr{H}$} 
constructed  acceding to the   standard   QFT over  the  curved  space-time  with  metric {$g_{\mu\nu}(x)$} (as described in, say  in \cite{Wald:1994}), and  the  state {$\vert\xi\rangle$} in {$\mathscr{H}$},  is  such that:

\begin{equation}\label{EinsU1}
G_{\mu\nu}[g(x)]=8\pi G\langle\xi\vert \hat{T}_{\mu\nu}[g(x),\hat{\varphi}(x),\hat{\pi}(x)]\vert\xi\rangle 
\end{equation}

On the one  hand, this   can be thought of as  simply   describing  semiclassical gravity, where all the matter  is   quantum mechanical (i.e. where  no  classical source of the energy momentum is  considered), but at the same time,   it    might  be  considered as   the  GR version of Schr\"odinger-Newton equation \cite{Diosi:1984}. In that  scheme   where   one considers the  Schr\"odingier equation  for  the  state  $\psi$ of a particle  subject to the gravitational  interaction,   described in terms of a Newtonian  potential $\Phi_N$.  That is 
$ i\frac{\partial \psi }{\partial t}= -\frac{1}{2M} \nabla^2\psi + M\Phi_N\psi $
  where   at the same time the wave function of the particle  is  considered as  characterizing    the distribution of   gravitating mass,  so that  its  square   acts as the  source of  the  Newtonian potential, according to 
 $ \nabla^2\Phi_N =  4\pi G M |\psi|^2$.
 It is  worth  mentioning  that the  non linearity  implied  by these  equations leads  to  some interesting   behavior \cite{van-Metter:2011}, and    therefore, it is natural to expect  that the  general  relativistic  version   will  share    some  of those  features.

The point  however,  is that the formalism  described up   to this point,  is not  flexible enough  for the    incorporation  of spontaneous    collapses  of the quantum state  of the matter fields. Recall   however, that  we    should   not expect   the theory to   be  valid during the collapses, just as   the NS  equations  are not expected to hold  when an  ocean wave breaks at the beach. Nevertheless, and  just as the hydrodynamical   description  should hold,  to  a good  approximation,  just   before  and well after the   breaking of the ocean wave,  so  should  the  SSC  formalism  hold    before and  after the spontaneous collapse. When we consider   the limit in which the collapse can be thought of as  occurring instantaneously, we  are faced  with  the  task of producing a new  SSC  out of the previous one.  And the   collapse dynamics (which  we  here take as given  by some  kind of generalization of the GRW model of spontaneous  collapses  adapted to quantum field theory  as in \cite{QFT-Collapse, Pearle07, Myrvold, Tumulka, Pearle14, BediModSud})  and   describe the gluing of the   two SSC's,   that can be   taken as  an effective,  and  rough description of the   collapse process (just as gluing the  patterns  describing the   body of water in the ocean  as the    wave  breaks  on the beach).

In particular,   we   must  note    that  any sudden  change  in the quantum state  will lead,  generically, to a  sudden change in  the expectation value of the energy  momentum tensor, and  thus to  a change in the  space-time  metric.  However,  such a   change  would   in   general,  require  also a   change in the quantum field theory construction,  and thus   a  change of  Hilbert space  to which the state  can belong. This  fact  prevents  us  from taking the  state  $\vert\zeta^{(I)}_{g}\rangle$ 
  that results from the collapse  dynamics  acting on the  original quantum sate  of the matter fields $\vert\xi^{(I)}\rangle$,  of the first SSC (SSC-I),  as a viable state  for the second SSC (SSCs), simply  because the former will be  a state in  $\mathscr{H}^{(I)}$,  while the latter  needs to be a state in  $\mathscr{H}^{(II)}$.  We deal with this  issue  by  regarding the state $\vert\zeta^{(I)}_{g}\rangle\in\mathscr{H}^{(I)}$   
  as  encoding  the  energy momentum tensor   that  characterizes the  state  $\vert\xi^{(II)} \rangle\in \mathscr{H}^{(II)}$  of the  SSC-II.  We  thus   refer  to  $\vert\zeta^{(I)}_{g}\rangle $ as the target state (the state  the collapse theory would have produced  if allowed to   provide a state in the same Hilbert space), and  use  it to  specify  the   state  of the new SSC.
 
 That is, in order to adhere to the  formalism as  closely as possible, we are forced to contemplate a transition  from  one  complete  SSC  to  another  one,  rather than  simple ``jumps" in states in one  Hilbert space.  That is,  the   jumps  should  be characterized as  taking the system  from  one  complete  SSC  construction  to another,  SSC-I $\to$ SSC-II. The two configurations should then be glued according to the following prescription guidelines \cite{DiezTaSudarD}:

The  gluing process  was taken in \cite{DiezTaSudarD}    to be described   as follows:
 
We assume  that the   specific collapse theory   provides  a   choice  of the  collapse  hypersurface $\Sigma_c$ in the manifold $\mathcal{M}^{(I)}$ of the SSC-I.  We  consider  now   gluing the spacetimes  of  the SSC-I  and  that of the  SSC-II  along  $\Sigma_c$  and  following  the thin shell  formalism  of \cite{Israel} we demand  continuity of the  induced  metric on the gluing $\Sigma_c$   and  allow the  extrinsic  curvature to undergo a sudden jump  across  it.   

That is,  $\Sigma_c$ has   an induced  metric, a   three-metric $\hskip.1cm ^{(\Sigma)}\!g^{(I)}_{ab}$    that results from  its  embedding on the    pseudo-Riemmanian manifold $(\mathcal{M}^{(I)},   g_{\mu \nu}^{(I)})$,  and  we  require that the  space-time  $ (\mathcal{M}^{(II)},   g_{\mu \nu}^{(II)})$    to contain an   embedded  hypersurface  $\Sigma_c'$  whose induced  metric $\hskip.1cm ^{(\Sigma)}\!g^{(II)}_{ab}$  be  isometric to $\hskip.1cm ^{(\Sigma)}\!g^{(I)}_{ab}$.

One the collapse hypersurface, the state of the system   is   supposed to change,  abruptly, as   dictated  by the theory  from $\vert\xi^{(I)}\rangle$ to $\vert\zeta^{(I)}_{g}\rangle$.  We then demand that  the    quantum  state $\vert\xi^{(II)}\rangle$ of the SSC-II be such that   
   
\begin{equation}
\langle \zeta^{(I)}_{g}  |\hat{T}^{(I)}_{\mu\nu}[g^{(I)},\hat{\varphi}^{(I)},\hat{\pi}^{(I)}] | \zeta^{(I)}_{g} \rangle \Big|_{\Sigma_{c}} = \langle \xi^{(II)} | \hat{T}^{(II)}_{\mu\nu}[g^{(II)},\hat{\varphi}^{(II)},\hat{\pi}^{(II)}] | \xi^{(II)} \rangle\Big|_{\Sigma_{c}}. \label{EM identification}
\end{equation}

 
We note that, as expected,  Einstein's  semi-classical  equations will  in general   be  violated at $\Sigma_c$.
In following  with the hydrodynamical analogy,  we take this to mean that the  description  we are working   with,  is  starting to break down and   must be   viewed as a step gap  procedure characterizing the need  for a  deeper level   description,  which at this  time  is  unavailable.

We   must  acknowledge that  the characterization  of the gluing process    we have  provided  up to  this point, is not  precise enough  for   general  situations.  In fact, there is an  ongoing  research  project  dealing with this problem, part of it   represented in the   work \cite{Juarez-Kay-Tonatiuh} with   the  rest   of the   necessary developments  expected to  be   forthcoming soon. However, as it was  shown in  \cite{DiezTaSudarD},  and we  will see    in the reminder  of the paper, the   simplicity and the  high  degree of  symmetry   of the cosmological situation   of interest here, will permit  the  completion of the  construction in all detail. 
 
 In fact,  \cite{DiezTaSudarD}  corresponds to a  simple  application   to inflationary cosmology,  and represents the single  instance  where the program has   been   carried  out   in detail,   which moreover  was  based on   a perturbative scheme,   developed  only to first order in perturbation theory.  Although  encouraging,  that    work    raises the  natural  question   about the viability of  extending such  perturbative treatment to higher orders, not to mention the question of  convergence of the   whole   perturbative expansion.  In this paper, we take a step forward  by showing explicitly and in detail, that the program can be extended  to second  order in  perturbation theory. As  a by-product,  this  work ended up  uncovering    some   initially unexpected   features that  were absent in the first order  treatment \cite{tesisPedro}, \cite{DiezTaSudarD}.
 
 There is one more issue that needs  discussing. The fact that, in order for the whole  formalism to make  any sense,   we  must  use in the RHS  of  Einstein's semiclassical equations (\ref{EinsU1}), the    renormalized   energy momentum tensor.  As it  is  well known,  the    general   procedure   for the renormalization of the energy momentum tensor in  arbitrary spacetimes is a rather  intricate  one, and involves    both,   correction  terms  that    ensure conservation  as well as  essential   ambiguities. However, as it was shown  in  \cite{
 WALD-T-Anomaly, WALD-QFTCS, Decanini}
 the difference   in the value of the renormalized   energy momentum tensor between two states is  ambiguity-free.  We take advantage of this fact and  the circumstance that the  unperturbed  FRW space-time   possess  a high degree  of symmetry in order to take   the renormalized  energy-momentum tensor  for the vacuum state  of the  usual Bunch-Davies construction to   vanish. This  can  be  thought of as  simply fixing the   cosmological constant  to  zero,  which    for the   context at hand  can  be  considered as  a very  good  approximation (in fact  one can  absorb the cosmological constant  in the  scalar field  potential,   which is  taken to  have a minimum  at  zero in the present context). Moreover, our gluing recipe   regarding the  construction of
 the  post-collapse  SSC  out of the  pre-collapse   SSC,   can be   thought of as  specifying the ambiguities  in the renormalized  energy-momentum tensor   for the new construction.   
 As  discussed  in \cite{Benito-Bernard},  if the states involved  are Hadamard,  one can   obtain  the expectation value of  the  renormalized  energy  momentum tensor in that  state,  from the   corresponding  value on the vacuum state of the  QFT construction and  a  normal  ordering prescription based on such vacuum.   This  together with the fact  that we will be  dealing with  coherent states \cite{PC-Benito},    simplifies  the  whole   renormalization  related issues in the treatment at hand.     
 
 
The  manuscript is  organized  as follows: In section II we review the ideas regarding the setting of cosmological perturbation theory and we also explain how to construct the semi-classical self-consistent configurations (SSCs). In section IIA we carry out the construction of a homogeneous and isotropic SSC-I, whereas in section IIB an inhomogeneous and anisotropic SSC-II will be constructed. The matching at the collapse hypersurface of the two SSCs constructed in sections IIA and IIB will be analyzed in section IIC. Finally, we discuss our results in section III. 
Our conventions are: (-,+,+,+) signature for the space-time metric and Wald's convention for the Riemann tensor. We will use natural units with $c=1$, but will keep the gravitational constant $G$ explicit throughout the paper.

\section{SSC's  in inflationary cosmology}

We will be working in the setting of  cosmological perturbation theory. Therefore, we assume that the  space-time metric $g_{\mu\nu}(\boldsymbol{x})$ can be taken as a small deviation from   that of a FRW spacetime. This metric is characterized by scalar, vector and tensor perturbations. In so far, we are considering that only the scalar perturbations $\Phi(\boldsymbol{x})$ and  $\Psi(\boldsymbol{x})$ are relevant for the cosmological problem at hand.   We  will work  with a  specific  gauge,    rather than relying on  the  so called "gauge invariant formalism"  both,  for simplicity and   because the  extension of the later to arbitrarily higher orders has not been worked out  in detail \cite{SteWalker, Bert, MukFeldBran, MalikWands} (although  important work   exists for  the extension to  the second order perturbation theory  \cite{Nakam03} ), and  our  ultimate goal is to   establish the  viability  of the SSC  formalism to all orders in perturbation theory.    The  aim here is  to  explore the  issue  up to  second-order perturbation theory,  and   thus,  not unexpectedly, we will  show that, a \textbf{SSC} can only be constructed  if, metric tensor perturbations $h_{ij}(\boldsymbol{x})$ are included in our treatment \cite{tesisPedro}. It is convenient to  work in the generalized longitudinal gauge, (~\cite{Acquaviva}, \cite{Kouji} and \cite{Kouji2}).
Thus we write, 

\begin{equation}\label{71old}
ds^{2} = a^{2}(\eta)\Big( 1 - 2\varepsilon ^{2}a_{(2)}(\eta) \Big)\Big\{-\Big(1+ 2\Phi(\boldsymbol{x})\Big)d\eta^{2} + \Big(1 - 2\Psi(\boldsymbol{x})\Big)\delta_{ij}dx^{i}dx^{j} + h_{ij}(\boldsymbol{x})dx^{i}dx^{j}\Big\},
\end{equation}
where $a_{(2)}(\eta)$ is a correction of the scale factor, the metric potential $\Phi(\boldsymbol{x})$, $\Psi(\boldsymbol{x})$ and $h_{ij}(\boldsymbol{x})$  will be written as a power series expansion of the parameter $\varepsilon$;

\begin{eqnarray}\label{ordpert}
\Phi(\boldsymbol{x}) &=& \varepsilon\Phi_{(1)}(\boldsymbol{x}) + \frac{\varepsilon^{2}}{2}\Phi_{(2)}(\boldsymbol{x}) + \mathcal{O}(\varepsilon^{3}), \\ 
\Psi(\boldsymbol{x}) &=& \varepsilon\Psi_{(1)}(\boldsymbol{x}) + \frac{\varepsilon^{2}}{2}\Psi_{(2)}(\boldsymbol{x}) + \mathcal{O}(\varepsilon^{3}), \\
h_{ij}(\boldsymbol{x}) &=& \varepsilon h_{ij}^{(1)}(\boldsymbol{x}) + \frac{\varepsilon^{2}}{2}h_{ij}^{(2)}(\boldsymbol{x}) + \mathcal{O}(\varepsilon^{3}).
\end{eqnarray}
In this way $\varepsilon$ can be used to control the order of the power series. 
Working up to zeroth-order in $\varepsilon$, the metric (\ref{71}) is reduced to FRW, this means that $\Phi(\boldsymbol{x}) = \Psi(\boldsymbol{x}) = h_{ij}(\boldsymbol{x}) = 0,$. Meanwhile,  up to first order in  $\varepsilon$ the semiclassical Einstein equations lead to  $\Phi_{(1)}(\boldsymbol{x})\neq0$, $\Psi_{(1)}(\boldsymbol{x})\neq0$ and $h_{ij}^{(1)}(\boldsymbol{x}) = 0.$ Nevertheless, working up to second order in $\varepsilon$  will make it necessary to consider the functions $a_{(2)}(\eta)$, $\Phi_{(2)}(\boldsymbol{x})$, $\Psi_{(2)}(\boldsymbol{x})$ and $h_{ij}^{(2)}(\boldsymbol{x})$ as non trivial. \\
 
 For further simplification, we define

\begin{eqnarray}
\tilde{\Phi}(\boldsymbol{x}) &=& \varepsilon\Phi_{(1)}(\boldsymbol{x}) + \frac{\varepsilon^{2}}{2}\Phi_{(2)}(\boldsymbol{x}) - \varepsilon^{2}a_{(2)}(\eta), \\ 
\tilde{\Psi}(\boldsymbol{x}) &=& \varepsilon\Psi_{(1)}(\boldsymbol{x}) + \frac{\varepsilon^{2}}{2}\Psi_{(2)}(\boldsymbol{x}) + \varepsilon^{2}a_{(2)}(\eta), 
\end{eqnarray}

then, up to second order in $\varepsilon$, we have the metric (\ref{71old}) written as:

\begin{equation}\label{71} 
ds^{2} = a^{2}(\eta)\Big\{-\Big(1+ 2\tilde{\Phi}(\boldsymbol{x})\Big)d\eta^{2} + \Big(1 - 2\tilde{\Psi}(\boldsymbol{x})\Big)\delta_{ij}dx^{i}dx^{j} + h_{ij}(\boldsymbol{x})dx^{i}dx^{j}\Big\}.
\end{equation}

On the other side, 
the inflaton field satisfies the Klein-Gordon equation,  

\begin{equation}\label{KGEcua}
g^{\mu\nu}\nabla_{\mu}\nabla_{\nu}\phi - m^{2}\phi = 0,
\end{equation}
where $\nabla_{\mu}$ is the covariant derivative compatible with the metric (\ref{71}). \\

Starting from the solutions $\phi_{1}(\boldsymbol{x})$ and $\phi_{2}(\boldsymbol{x})$ of the above equation, the symplectic product is defined by 

\begin{equation}\label{simplec}
(\phi_{1},\phi_{2})_{simpl} \equiv -i\int_{\Sigma} [\phi_{1}\partial_{\mu}\phi_{2}^{\ast} - \phi_{2}^{\ast}\partial_{\mu}\phi_{1}]d\Sigma^{\mu}, 
\end{equation}
where $\Sigma$ is a 3-dimensional Cauchy hypersurface, $d\Sigma^{\mu} = n^{\mu}d\Sigma$ with $d\Sigma = \sqrt{g_{(\Sigma)}}d^{3}x$ being the volume element over $\Sigma$. While  $n^{\mu}$ is a time-like, future-directed, normalized 4-vector orthogonal to $\Sigma$. In terms of the conjugate momentum $\pi$ associated to the inflaton field $\phi$, this is; $\pi = \sqrt{g_{(\Sigma)}}n^{\mu}\partial_{\mu}\phi$ the symplectic product can be written as:

\begin{equation}\label{simplecPI}
((\phi_{1},\pi_{1}),(\phi_{2},\pi_{2}))_{simpl} \equiv -i\int_{\Sigma} [\phi_{1}\pi_{2}^{\ast} - \phi_{2}^{\ast}\pi_{1}]d^{3}x.
\end{equation}

Concerning quantization, the classical fields $\phi$ and $\pi$ are promoted
to quantum operators $\hat{\phi}$ and $\hat{\pi}$ on which we impose the canonical commutation relations

\begin{equation}\label{conmutacionU7}
[\hat{\phi}(\eta,\vec{x}),\hat{\phi}(\eta,\vec{y})] = [\hat{\pi}(\eta,\vec{x}),\hat{\pi}(\eta,\vec{y})] = 0,\hspace{.2cm}[\hat{\phi}(\eta,\vec{x}),\hat{\pi}(\eta,\vec{y})] = i\delta(\vec{x}-\vec{y}).
\end{equation}
The operators $\hat{\phi}(\boldsymbol{x})$ and $\hat{\phi}(\boldsymbol{x})$ will be written in terms of time independent operators $\hat{a}_{\vec{k}}$ and $\hat{a}_{\vec{k}}^{\dag}$ defined as annihilation and creation operators respectively;

\begin{eqnarray}
&& \hat{\phi}(\boldsymbol{x}) = \sum_{\vec{k}} [u_{\vec{k}}(\boldsymbol{x})\hat{a}_{\vec{k}} + u_{\vec{k}}^{\ast}(\boldsymbol{x})\hat{a}_{\vec{k}}^{\dag}], \label{ucuanticoU7ante} \\
&&
\hat{\pi}(\boldsymbol{x}) = \sum_{\vec{k}} [q_{\vec{k}}(\boldsymbol{x})\hat{a}_{\vec{k}} + q_{\vec{k}}^{\ast}(\boldsymbol{x})\hat{a}_{\vec{k}}^{\dag}]. 
\label{ucuanticoU7}
\end{eqnarray}
Sums run over the vectors $\vec{k}$ with rectangular components given by $k_{n} = 2\pi j_{n}/L,$ where $j_{n} = 0, \pm1, \pm2, \pm3,.. \in\mathbb{Z}$ and $n$ stands for $n=x,y,z$. This means that the quantization is taken over all the wave modes in a cubic cavity of space with side $L$.
\\

However, for practical reasons, Eqns.~(\ref{ucuanticoU7ante}) and (\ref{ucuanticoU7}) will be written as:

\begin{eqnarray}\label{pol}
\hat{\phi}(\boldsymbol{x}) &=& \sum_{ \vec{k}\in\mathbb{R}^{3} } [u_{\vec{k}}(\boldsymbol{x})\hat{a}_{\vec{k}} + u_{\vec{k}}^{\ast}(\boldsymbol{x})\hat{a}_{\vec{k}}^{\dag}] = \sum_{ \vec{k}\in \mathbb{R}^{+^{3}} }[ u_{\vec{k}}(\boldsymbol{x})\hat{a}_{\vec{k}} + u_{-\vec{k}}^{\ast}(\boldsymbol{x})\hat{a}_{-\vec{k}}^{\dag} ] + \sum_{ \vec{k}\in \mathbb{R}^{+^{3}} }
[ u_{-\vec{k}}(\boldsymbol{x})\hat{a}_{-\vec{k}} + u_{\vec{k}}^{\ast}(\boldsymbol{x})\hat{a}_{\vec{k}}^{\dag} ],    
\end{eqnarray}

then, defining for all 3-vector $\vec{k}$ such that $\vec{k}\in \mathbb{R}^{+^{3}}$, where the region $\mathbb{R}^{+^{3}}$ is the positive half-plane and has been defined as $ \mathbb{R}^{+^{3}} = \{\vec{q} = (q,\theta,\phi) / q \in [0,\infty), \theta\in [0,\pi/2), \phi\in [0,2\pi) \}$.

On the other hand, defining the operators $\hat{\phi}_{\vec{k}}(\boldsymbol{x})$ and $\hat{\phi}^{\dag}_{\vec{k}}(\boldsymbol{x})$, as $\hat{\phi}_{\vec{k}}(\boldsymbol{x}) = u_{\vec{k}}(\boldsymbol{x})\hat{a}_{\vec{k}} + u_{-\vec{k}}^{\ast}(\boldsymbol{x})\hat{a}_{-\vec{k}}^{\dag}$ and $\hat{\phi}^{\dag}_{\vec{k}}(\boldsymbol{x}) = u_{-\vec{k}}(\boldsymbol{x})\hat{a}_{-\vec{k}} + u_{\vec{k}}^{\ast}(\boldsymbol{x})\hat{a}_{\vec{k}}^{\dag}$, equation~(\ref{pol})  can be written as :

\begin{equation}\label{largos}
\hat{\phi}(\boldsymbol{x}) = \sum_{ \vec{k}\in \mathbb{R}^{+^{3}} 
}[ \hat{\phi}_{\vec{k}}(\boldsymbol{x}) + \hat{\phi}^{\dag}_{\vec{k}}(\boldsymbol{x}) ].
\end{equation}

Proceeding in a similar manner with $\hat{\pi}(\boldsymbol{x})$, we find :

\begin{eqnarray}
&&\hat{\pi}(\boldsymbol{x}) = \sum_{\vec{k}\in \mathbb{R}^{+^{3}}
} [ \hat{\pi}_{\vec{k}}(\boldsymbol{x}) + \hat{\pi}_{\vec{k}}^{\dag}(\boldsymbol{x})], \hskip.1cm \textup{ where we define} \hskip.2cm \hat{\pi}_{\vec{k}}(\boldsymbol{x}) = q_{\vec{k}}(\boldsymbol{x})\hat{a}_{\vec{k}} + q_{-\vec{k}}^{\ast}(\boldsymbol{x})\hat{a}_{-\vec{k}}^{\dag}, \nonumber\\
&&\hskip1cm \textup{along with} \hskip.2cm \hat{\pi}^{\dag}_{\vec{k}}(\boldsymbol{x}) = q_{-\vec{k}}(\boldsymbol{x})\hat{a}_{-\vec{k}} + q_{\vec{k}}^{\ast}(\boldsymbol{x})\hat{a}_{\vec{k}}^{\dag},\hskip.1cm \textup{ for all vector $\vec{k}$ such that} \hskip.2cm \vec{k}\in \mathbb{R}^{+^{3}}. 
\end{eqnarray}

The  mode functions $u_{\vec{k}}(\boldsymbol{x})$ satisfy the classical equations of motion (\ref{KGEcua}),

\begin{equation}\label{75exp}
\Big( g^{\mu\nu}\nabla_{\mu}\nabla_{\nu} - m^{2}\Big)u_{\vec{k}}(\boldsymbol{x})  = 0,    
\end{equation}
and are normalized according to the symplectic product;

\begin{equation}\label{76}
 (u_{\vec{k}},u_{\vec{k}^{\prime}})_{simpl} = \delta_{\vec{k},\vec{k}^{\prime}} 
\end{equation}
with $\delta_{\vec{k},\vec{k}^{\prime}}$ defined as $\delta_{\vec{k},\vec{k}^{\prime}} = 0$ if $\vec{k} \neq \vec{k}^{\prime}$, while that $\delta_{\vec{k},\vec{k}^{\prime}} = 1$ if $\vec{k} = \vec{k}^{\prime}$,
hence, the functions $u_{\vec{k}}(\boldsymbol{x})$  are  a complete set of normal modes required for the construction of the quantum field theory. The functions $q_{\vec{k}}(\boldsymbol{x})$ are related with the function $u_{\vec{k}}(\boldsymbol{x})$ through $q_{\vec{k}}(\boldsymbol{x}) = \sqrt{g_{(\Sigma)}}n^{\mu}\partial_{\mu}u_{\vec{k}}(\boldsymbol{x})$.
\\
Instead of using the canonical commutation relations (\ref{conmutacionU7}) in terms of the operators $\hat{\phi}(\boldsymbol{x})$ and $\hat{\pi}(\boldsymbol{x})$, we use the canonical commutation relations in terms of $\hat{a}_{\vec{k}}$ and $\hat{a}_{\vec{k}'}$;

\begin{equation}\label{conmutacionU7a}
[\hat{a}_{\vec{k}},\hat{a}_{\vec{k}'}] = [\hat{a}^{\dag}_{\vec{k}},\hat{a}^{\dag}_{\vec{k}'}] = 0, \hskip.4cm [\hat{a}_{\vec{k}},\hat{a}^{\dag}_{\vec{k}'}] = \delta_{\vec{k},\vec{k}'}. 
\end{equation}
We see that the Klein-Gordon field may be viewed as an infinite collection of decoupled harmonic oscillators with time-dependent frequency  $\omega_{\vec{k}}(\eta)$. The ground state of the oscillator with $\vec{k}$ mode will be denoted by $|0_{\vec{k}}\rangle$, while the vacuum state of the system, conformed by the set of all the modes $\vec{k}$ will be denoted by; 

\begin{eqnarray}\label{estadovaciodoSSC}
|0\rangle &=& ..|0_{-2\vec{k}_{1}}\rangle\otimes|0_{-\vec{k}_{1}}\rangle\otimes..|0_{-2\vec{k}_{0}}\rangle\otimes|0_{-\vec{k}_{0}}\rangle\otimes|0_{0}\rangle
\otimes|0_{\vec{k}_{0}}\rangle\otimes|0_{2\vec{k}_{0}}\rangle.. \otimes|0_{\vec{k}_{1}}\rangle\otimes|0_{2\vec{k}_{1}}\rangle..
\end{eqnarray}
and is the state defined as $\hat{a}_{\vec{k}}|0\rangle = 0$ \hskip.05cm $\forall\vec{k}$.
This vacuum state $|0\rangle$ is not unique, whatever the complete  set of solutions $u_{\alpha}$, (\ref{75exp}) and (\ref{76}) are, we have to determine a choice of the state $|0\rangle$. 

Consequently, the Hilbert space $\mathscr{H}$ can be built a la Fock, through successive applications of the creation operator $\hat{a}^{\dag}_{\vec{k}}$ over the state $|0\rangle$.\\

For simplicity, and without loss of generality, we consider that tensor perturbations describe a gravitational wave with longitudinal modes $h_{L}(\boldsymbol{x})$ and transverse modes $h_{T}(\boldsymbol{x})$ propagating along the $z$ axis with wavelength vector given by $2\vec{k}_{0}$. That is, we assume that:

\begin{equation}\label{vectordeOndaK1}
\vec{k}_{0} = k_0 \hat{z},
\end{equation}
meanwhile, the non-zero components are $h_{ij}(\boldsymbol{x})$ with: $h_{xy}(\boldsymbol{x}) = h_{yx}(\boldsymbol{x}) = h_{T}(\boldsymbol{x}),$ $h_{xx}(\boldsymbol{x})=h_{yy}(\boldsymbol{x})=-\frac{1}{2}h_{zz}(\boldsymbol{x})=h_{L}(\boldsymbol{x})$. Hence,  components $g_{\mu\nu}$ of (\ref{71}) given as a matrix array, will take the following form:


\begin{equation*}\label{gabnHyI}
{ g_{\mu\nu} } = a^{2}(\eta)\left(\begin{array}{cccc}
-(1+2\tilde{\Phi}(\boldsymbol{x}))&0&0&0\\
0&(1-2\tilde{\Psi}(\boldsymbol{x}) + h_{L}(\boldsymbol{x}))&h_{T}(\boldsymbol{x})&0\\
0&h_{T}(\boldsymbol{x})&(1-2\tilde{\Psi}(\boldsymbol{x}) + h_{L}(\boldsymbol{x}))&0\\
0&0&0&(1-2\tilde{\Psi}(\boldsymbol{x}) -2h_{L}(\boldsymbol{x}))\\
\end{array} 
\right)  
\end{equation*}

On the other hand, the inverse matrix components are,


\begin{equation*}\label{IgabnHyI}
{ g^{\mu\nu} } = a^{-2}(\eta)\left(\begin{array}{cccc}
-\frac{1}{1+2\tilde{\Phi}(\boldsymbol{x}) }&0&0&0\\
0&\frac{-(1-2\tilde{\Psi}(\boldsymbol{x}) + h_{L}(\boldsymbol{x}))}{-(1-2\tilde{\Psi}(\boldsymbol{x}) + h_{L}(\boldsymbol{x}))^{2} + h_{T}^{2}(\boldsymbol{x})}&\frac{h_{T}(\boldsymbol{x})}{-(1-2\tilde{\Psi}(\boldsymbol{x}) + h_{L}(\boldsymbol{x}))^{2} + h_{T}^{2}(\boldsymbol{x})}&0\\
0&\frac{h_{T}(\boldsymbol{x})}{-(1-2\tilde{\Psi}(\boldsymbol{x}) + h_{L}(\boldsymbol{x}))^{2} + h_{T}^{2}(\boldsymbol{x})}&\frac{-(1-2\tilde{\Psi}(\boldsymbol{x}) + h_{L}(\boldsymbol{x}))}{-(1-2\tilde{\Psi}(\boldsymbol{x}) + h_{L}(\boldsymbol{x}))^{2} + h_{T}^{2}(\boldsymbol{x})}&0\\
0&0&0&\frac{1}{1-2\tilde{\Psi}(\boldsymbol{x}) -2h_{L}(\boldsymbol{x})}\\
\end{array} 
\right)  
\end{equation*}

Now, we proceed to write the equations as a power series in $\varepsilon$. In this paper we will work up to second order in this parameter. 

We first note that the $g^{\mu\nu}$ components are not linear in the $\tilde{\Phi}$, $\tilde{\Psi}$, $h_{T}$ and $h_{L}$ perturbations. Also, (\ref{ordpert}) implies that the lowest order at which the coefficients $\Phi^{(n)}$ and $\Psi^{(n)}$ are non zero is $n=1$, whilst for $h^{(n)}_{L}$ and $h^{(n)}_{T}$ is $n=2$. This means that, in working  to second order in perturbation theory, we are allowed to neglect terms of the form  $\tilde{\Psi} h_{T}$, $\tilde{\Psi} h_{L}$, $\tilde{\Phi} h_{T}$, $\tilde{\Phi} h_{L}$, $\tilde{\Psi}\tilde{\Psi}^{2}$, $\tilde{\Psi}\tilde{\Phi}^{2}$, $\tilde{\Phi}\tilde{\Psi}^{2}$, $\tilde{\Phi}\tilde{\Phi}^{2}$, $h_{T}h_{L}$, $h^{2}_{T}$ and $h^{2}_{L}$ onwards.

\underline{Notation:} we use $^{[n]}\!\!A(\boldsymbol{x})$ to indicate that we are working with a power series expansion of the function $A(\boldsymbol{x})$ in the parameter $\varepsilon$ and that we are truncating such series at the n-th order in the expansion. Whilst with $^{(n)}\!\!A(\boldsymbol{x})$ we denote the n-th order of the expansion.  In this way, $^{[2]}\!g^{\mu\nu}$ denotes $g^{\mu\nu}$ to second order in $\varepsilon$ and is given by:


\begin{equation*}\label{RIgabnHyI}
\Big\{\hskip.1cm ^{[2]}\!g^{\mu\nu} \hskip.1cm\Big\} = a^{-2}(\eta)\left(\begin{array}{cccc}
-\frac{1}{1+2\tilde{\Phi}(\boldsymbol{x})}&0&0&0\\
0&\frac{1}{1-2\tilde{\Psi}(\boldsymbol{x}) + h_{L}(\boldsymbol{x})}&-h_{T}(\boldsymbol{x})&0\\
0&-h_{T}(\boldsymbol{x})&\frac{1}{1-2\tilde{\Psi}(\boldsymbol{x}) + h_{L}(\boldsymbol{x})}&0\\
0&0&0&\frac{1}{1-2\tilde{\Psi}(\boldsymbol{x}) -2h_{L}(\boldsymbol{x})}\\
\end{array} 
\right)  
\end{equation*}

In this way, we are interested in expanding (\ref{75exp}) up to $\varepsilon^{2}$. Moreover (by the symmetries of the problem), considering that the metric potential only depends on coordinates $\eta$ and $z$, we find that the differential equation which the mode functions $u_{\vec{k}}(\boldsymbol{x})$ satisfy is:

\begin{eqnarray}\label{KleinGordonP}
&&(1 + 2\tilde{\Phi} -4\tilde{\Psi} - 8\tilde{\Phi}\tilde{\Psi} + 4\tilde{\Psi}^{2})\Big\{ u''_{\vec{k}} + 2\frac{a'}{a}u'_{\vec{k}} \Big\} - ( 1 + 4\tilde{\Phi} - 2\tilde{\Psi} + 4\tilde{\Phi}^{2} - 8\tilde{\Phi}\tilde{\Psi} - h_{L})\nabla^{2}u_{\vec{k}}  \nonumber \\ 
&& - 3h_{L}\partial^{2}_{z}u_{\vec{k}} - \{ (1 + 2\tilde{\Phi} - 2\tilde{\Psi})\partial_{z}\tilde{\Phi} - (1 + 4\tilde{\Phi})\partial_{z}\tilde{\Psi} + 2\partial_{z}h_{L} \}\partial_{z}u_{\vec{k}} - \{3(1-2\tilde{\Psi} + 2\tilde{\Phi})\tilde{\Psi}' \nonumber\\
&&+ ( 1-4\tilde{\Psi})\tilde{\Phi}' \}u'_{\vec{k}} + 2h_{T}\partial^{2}_{xy}u_{\vec{k}} + a^{2}m^{2}( 1 + 4\tilde{\Phi} - 4\tilde{\Psi} + 4\tilde{\Phi}^{2} + 4\tilde{\Psi}^{2} - 16\tilde{\Phi}\tilde{\Psi})u_{\vec{k}} = 0. 
\end{eqnarray}

Now, we proceed to find the expression for the orthonormalization condition induced by the symplectic product 
(\ref{76}),

\begin{equation}
 (u_{\vec{k}},u_{\vec{k}^{\prime}})_{simpl} = \delta_{\vec{k},\vec{k}^{\prime}}.
\end{equation}
Then, using the definition for the symplectic product (\ref{simplec}),

\begin{equation}\label{simplecPIuk}
(u_{\vec{k}}, u_{\vec{k}})_{simpl} \equiv -i\int_{\Sigma} [u_{\vec{k}}\partial_{\mu}u_{\vec{k}'}^{\ast} - u_{\vec{k}'}^{\ast}\partial_{\mu}u_{\vec{k}}] n^{\mu} d\Sigma, 
\end{equation}

where the Cauchy hypersurface $\Sigma$ is defined as $\Sigma = \{\boldsymbol{x}\in\mathcal{M}\hskip.2cm / \hskip.2cm\eta=\eta_{c}\}$. Whereas the induced metric on $\Sigma$ is:

\begin{eqnarray}\label{Sigmametric} 
ds^{2}_{(\Sigma)} = \hskip.2cm ^{(\Sigma)}\!g_{ij}dx^{i}dx^{j}
=a^{2}(\eta)\Big\{\Big(1 - 2\tilde{\Psi}(\boldsymbol{x})\Big)\delta_{ij}dx^{i}dx^{j} + h_{ij}(\boldsymbol{x})dx^{i}dx^{j}\Big\},
\end{eqnarray}

whereby, the volume element  $d\Sigma = \sqrt{g_{(\Sigma)}}d^{3}x$, with $g_{(\Sigma)} = det\{\hskip.1cm ^{(\Sigma)}\!g_{ij}\hskip.1cm\}$, and considering contributions only to second order in $\varepsilon$, 
meaning $\sqrt{^{[2]}\!g_{(\Sigma)}} = a^{3}(\eta)(1-2\tilde{\Psi}(\boldsymbol{x}))^{3/2}$. Therefore, $^{[2]}\!d\Sigma$ will be written as $^{[2]}\!d\Sigma = a^{3}(\eta)(1-2\tilde{\Psi}(\boldsymbol{x}))^{3/2} d^{3}x$. \\

We are going to write the unitary four-vector $\boldsymbol{n}$ perpendicular to the hypersurface $\Sigma$ 
as $\boldsymbol{n} = n^{\mu}\partial_{\mu}$.  This quantity can also be expressed as a linear combination of a part $b^{i}_{\parallel}$ parallel to $\Sigma$ and a perpendicular one $b_{\perp}$. That is, $\boldsymbol{n} = b_{\perp}(\boldsymbol{x})\partial_{\eta} + b^{i}_{\parallel}(\boldsymbol{x})\partial_{i}$, but by the requirement that  $\boldsymbol{n}$ be normal to $\Sigma$, we conclude that $b^{i}_{\parallel}(\boldsymbol{x})=0$. Therefore, $\boldsymbol{n} = b_{\perp}(\boldsymbol{x})\partial_{\eta}$.  Given that $\boldsymbol{n}$ is a time, unitary vector, means that, $ds^{2}(\boldsymbol{n},\boldsymbol{n}) =-1$, from which we obtain:

\begin{equation}
b_{\perp} = \frac{1}{a(\eta)\sqrt{1 + 2\tilde{\Phi}(\boldsymbol{x})}} \Rightarrow \boldsymbol{n} =  \frac{1}{a(\eta)\sqrt{1 + 2\tilde{\Phi}(\boldsymbol{x})}}\partial_{\eta},  
\end{equation}

therefore, the normalization condition for the mode functions is:

\begin{eqnarray}
\hskip.1cm^{[2]}\!(u_{\vec{k}}, u_{\vec{k}})_{simpl} = -i a^{2}(\eta)\int_{\Sigma} (u_{\vec{k}}\partial_{\eta}u_{\vec{k}'}^{\ast} - u_{\vec{k}'}^{\ast}\partial_{\eta}u_{\vec{k}}) \frac{ (1-2\tilde{\Psi})^{\frac{3}{2}} }{ (1 + 2\tilde{\Phi})^{\frac{1}{2}} } d^{3}x \hskip.1cm\Bigg|_{\eta=\eta_{c}}
= \delta_{\vec{k},\vec{k}'}.
\end{eqnarray}

Expanding up to second order in $\varepsilon$ on the right-hand side of the first equality above, we find:

\begin{equation}\label{NorsimplecPI}
\int_{\Sigma} \Big(u_{\vec{k}}\partial_{\eta}u_{\vec{k}'}^{\ast} - u_{\vec{k}'}^{\ast}\partial_{\eta}u_{\vec{k}} \Big) \Big(1-3\tilde{\Psi}-\tilde{\Phi} + \frac{3}{2}\tilde{\Psi}^{2} + 3\tilde{\Psi}\tilde{\Phi} + \frac{3}{2}\tilde{\Phi}^{2}\Big) d^{3}x  
= \frac{i}{a^{2}(\eta_{c})}\delta_{\vec{k},\vec{k}'}.
\end{equation}

The quantity $\hat{T}_{\mu\nu}[g(\boldsymbol{x}),\hat{\phi}(\boldsymbol{x}),\hat{\pi}(\boldsymbol{x})]$ and the state $|\xi\rangle\in\mathscr{H}$, are expected to satisfy equation~(\ref{EinsU1}) given by:

\begin{equation}\label{EinsUu2}
G_{\mu}\hskip.02cm^{\nu}[g(\boldsymbol{x})] = 8\pi G \langle\xi|\hat{T}_{\mu}\hskip.02cm^{\nu}[g(\boldsymbol{x}),\hat{\phi}(\boldsymbol{x}),\hat{\pi}(\boldsymbol{x})]|\xi\rangle,
\end{equation}
just as in the classical context, we are dealing with a scalar field $\phi$ minimally coupled to gravity. Then the energy-momentum tensor $T_{\mu}\hskip.02cm^{\nu}$ is given by:

\begin{equation}
T_{\mu}\hskip.02cm^{\nu} = g\hskip.02cm^{\alpha\nu} \partial_{\mu}\phi \hskip.1cm \partial_{\alpha}\phi - \Big( \frac{1}{2}g\hskip.02cm^{ab}\partial_{a}\phi \hskip.1cm \partial_{b}\phi + V[\phi] \Big) g_{\mu}\hskip.02cm^{\nu}. 
\end{equation}

Once the classical fields $\phi$ and $\pi$ are promoted to operators, in a similar direct manner $T_{\mu}\hskip.02cm^{\nu}$ will be promoted to the operator $\hat{T}_{\mu}\hskip.02cm^{\nu}$, which can be expressed as:

\begin{equation}\label{TunfullQm}
\hat{T}_{\mu}\hskip.02cm^{\nu} = g\hskip.02cm^{\alpha\nu} \partial_{\mu}\hat{\phi} \hskip.1cm \partial_{\alpha}\hat{\phi} - \Big( \frac{1}{2}g\hskip.02cm^{ab}\partial_{a}\hat{\phi} \hskip.1cm \partial_{b}\hat{\phi} + V[\hat{\phi}] \Big) g_{\mu}\hskip.02cm^{\nu}, 
\end{equation}
particularly, we use $V[\hat{\phi}] = \frac{1}{2} m^{2} \hat{\phi}^{2}$. 
\subsection{Construction of a Homogeneous and Isotropic SSC-I}
Following Refs \cite{DiezTaSudarD}, in the same manner, we will start by assuming that at a time corresponding to a few e-foldings after inflation achieves the slow-roll phase, the relevant region of the universe for observations can be described by a SSC homogeneous and isotropic, which we call now SSC-I. For this configuration, all metric potentials are equal to zero,

\begin{equation}\label{SSC1frw}
\tilde{\Psi}(\boldsymbol{x})=\tilde{\Phi}(\boldsymbol{x})=h_{ij}(\boldsymbol{x})=0.
\end{equation}

In order to carry out the construction, we will take the space-time metric to be that of a Robertson-Walker universe, with a pre-established (nearly) de Sitter scale factor. The small deviation from the exact de Sitter expansion will be parametrized by  $\epsilon^{(I)} = 1 - \mathcal{H}^{'(I)}/\mathcal{H}^{2(I)}$ where $\mathcal{H} = a'/a$ and $\epsilon^{(I)} $ is known as a slow-roll parameter.
During slow-roll  $0<\epsilon^{(I)}\ll 1$. For all the situations analyzed in this paper,
it will be sufficient to consider the scale factor; $a^{(I)}(\eta) = \Big(-\frac{1}{ H_{0}^{(I)} \eta } \Big)^{ 1 + \epsilon^{(I)}}$. To first order in the slow-roll
parameter, it is found that the Hubble rate is given by; $\mathcal{H} = -(1+\epsilon^{(I)})/\eta$, remember that during inflation $\eta$ takes on the values  $-\infty<\eta<0$.

The quantum field theory construction requires a complete set of modes $u^{(I)}_{\vec{k}}(\boldsymbol{x})$, which given the symmetries of the spatial background, we take to be of the form; 

\begin{equation}\label{modoshynh}
u^{(I)}_{\vec{k}}(\boldsymbol{x}) = v^{(I)}_{\vec{k}}\!(\eta) e^{i\vec{k}\cdot\vec{x}}/L^{3/2},
\end{equation}
then, by substitution of (\ref{SSC1frw}) and the above ansatz in the motion equation (\ref{KleinGordonP}), we find;

\begin{equation}\label{ceropertur}
v^{''(I)}_{\vec{k}}\!(\eta) + 2\mathcal{H}^{(I)} v^{'(I)}_{\vec{k}}\!(\eta) + (k^{2} + a^{2(I)}m^{2})v^{(I)}_{\vec{k}}\!(\eta) = 0,
\end{equation}
while,  by substitution of (\ref{SSC1frw}) and the ansatz (\ref{modoshynh}) in the equation (\ref{NorsimplecPI}), we obtain:

\begin{equation}\label{NorsimplecPIcero}
\Big(v_{\vec{k}}^{(I)}\!(\eta)\partial_{\eta}v_{\vec{k}'}^{(I)\ast}\!(\eta) - v_{\vec{k}'}^{(I)\ast}\!(\eta)\partial_{\eta}v^{(I)}_{\vec{k}}\!(\eta) \Big)\int_{\Sigma} \frac{e^{i(\vec{k}-\vec{k}')\cdot\vec{x}}}{L^{3}} d^{3}x \hskip.1cm\Bigg|_{\eta=\eta_{c}} = \frac{i}{a^{2}(\eta_{c})}\delta_{\vec{k},\vec{k}'}.
\end{equation}
Now, given that $\int_{\Sigma} \frac{e^{i(\vec{k}-\vec{k}')\cdot\vec{x}}}{L^{3}} d^{3}x = \delta_{\vec{k},\vec{k}'}$, the equation (\ref{NorsimplecPIcero}) becomes, 

\begin{equation}\label{NormasimplecPI} 
\Big(v_{\vec{k}}^{(I)}\!(\eta)\partial_{\eta}v_{\vec{k}}^{(I)\ast}\!(\eta) - v_{\vec{k}}^{(I)\ast}\!(\eta)\partial_{\eta}v^{(I)}_{\vec{k}}\!(\eta) \Big) \hskip.1cm\Big|_{\eta=\eta_{c}} = \frac{i}{a^{2(I)}(\eta_{c})}.
\end{equation}
For the modes $\vec{k}\neq0$ the general solution of equation~(\ref{ceropertur}) up to first order in the slow-roll parameter, yields;

\begin{equation}\label{ceroperturSol}
v^{(I)}_{\vec{k}}\!(\eta) \approx \eta^{ 3/2  + \epsilon^{(I)} }\{ c_{1} H^{(1)}_{\nu}(-k\eta) +  c_{2} H^{(2)}_{\nu}(-k\eta) \},
\end{equation}
where $c_{1}$ and $c_{2}$ are integration constants, while $H^{(1)}_{\nu}$ and $H^{(2)}_{\nu}$ are the Hankel functions of first and  second kind, with $\nu = 3/2 + \epsilon^{(I)} - m^{2}/(3H^{2(I)}_{0})$.\\
A choice of $c_{1}$ and $c_{2}$ corresponds to an election of the vacuum.
But following the standard literature on the subject we will take the Bunch-Davis
vacuum, which working to the lowest non-vanishing order in the slow-roll parameters $\epsilon^{(I)}$, implies $\nu = 3/2$, and the mode function $v^{(I)}_{\vec{k}}\!(\eta)$ can be written as;

\begin{equation}\label{ceroperturBD}
v^{(I)}_{\vec{k}}\!(\eta) \approx \sqrt{ \frac{1}{2k} }\Big(-H_{0}^{(I)}\eta\Big) \Big( 1 - \frac{i}{k\eta} \Big) e^{-ik\eta}, \hskip.2cm \forall \vec{k}\neq0. 
\end{equation}
Whereas  for $\vec{k}=0$, the general solution of the equation (\ref{ceropertur}) is given by;

\begin{equation}\label{ceroperturSol0}
v^{(I)}_{0}\!(\eta) = c_{3}\eta^{ 3/2  + \epsilon^{(I)} - \nu }  +  c_{4}\eta^{  3/2  + \epsilon^{(I)} + \nu }. 
\end{equation}
The choice of the pair $c_{3}$ and $c_{4}$ is an arbitrary one, provided $v^{(I)}_{0}$ has positive symplectic norm. Here we take;
 
\begin{equation}\label{ceroperturSolPar0}
v^{(I)}_{0}\!(\eta) = \sqrt{ \frac{1}{ H^{(I)}_{0} } }\Big[ 1 - \frac{i}{6}\Big(-H_{0}^{(I)}\eta\Big)^{3} \Big]\Big( -H_{0}^{(I)}\eta \Big)^{ m^{2}/(3H_{0}^{2(I)}) },  
\end{equation}
where $v^{(I)}_{0}$ has been normalized by using the relation (\ref{NormasimplecPI}).\\

However, we still need to find a state $|\xi^{(I)}\rangle\in\mathrm{H}^{(I)}$, with expectation value for the energy-momentum tensor that leads to the desired nearly de
Sitter, homogeneous and isotropic cosmological expansion.

Due to the homogeneity and isotropy symmetries, the state  $|\xi^{(I)}\rangle$ should be such that the expectation values of the operators $\hat{\phi}^{(I)}$ and $\hat{\pi}^{(I)}$ only depend on the temporal coordinate 
$\eta$ and not on the spatial coordinates $(x,y,z)$. Therefore, from Eq.~(\ref{modoshynh}) along with (\ref{ucuanticoU7ante}) and (\ref{ucuanticoU7}), we conclude that the only mode vector $\vec{k}$ contributing  
in this way is $\vec{k}=0.$ Hence, the considered state can only have the  $\vec{k}=0$ mode excited, while the other mode vectors $\vec{k}\neq0$  should be in their background state $|0^{(I)}_{\vec{k}}\rangle$. It is worth noticing that the vacuum state corresponding to the system formed by all the modes $\vec{k}$, including $\vec{k}=0$ in the SSC-I, according to (\ref{estadovaciodoSSC}) is written as :

\begin{eqnarray}\label{estadovaciodoSSCI}
|0^{(I)}\rangle &=& ..|0^{(I)}_{-2\vec{k}_{1}}\rangle\otimes|0^{(I)}_{-\vec{k}_{1}}\rangle\otimes..|0^{(I)}_{-2\vec{k}_{0}}\rangle\otimes|0^{(I)}_{-\vec{k}_{0}}\rangle\otimes|0^{(I)}_{0}\rangle
\otimes|0^{(I)}_{\vec{k}_{0}}\rangle\otimes|0^{(I)}_{2\vec{k}_{0}}\rangle.. \otimes|0^{(I)}_{\vec{k}_{1}}\rangle\otimes|0^{(I)}_{2\vec{k}_{1}}\rangle..\otimes|0^{(I)}_{\vec{k}_{2}}\rangle\otimes..     
\end{eqnarray}
Therefore, the state $|\xi^{(I)}\rangle$ can be expressed as: 

\begin{equation}\label{xi000estado}
|\xi^{(I)}\rangle = \mathcal{F}(\xi^{(I)}_{0}\hat{a}^{(I)\dag}_{0})|0^{(I)}\rangle,
\end{equation}

where $\mathcal{F}(\hat{X})$ represents in principle, a generic function of the operators $\hat{X}$.  For this task, we will use the function $\mathcal{F}(\hat{X})$ associated with the coherent states 
$\mathcal{F}(\hat{X}) \propto e^{\hat{X}}$.
\\

We should  clarify  here  that   although  Eq.~(\ref{xi000estado}) is a  specific choice   we are making  for definiteness  in  the present work,  our results  at the desired order, depend only on  a few  expectation values, and there are many more quantum states for which those expectations agree we the one we have chosen. 
\\

Finally, making use of the above equation 
together with (\ref{ucuanticoU7ante}) and (\ref{modoshynh}), we can 
calculate the expectation value of the field $\hat{\phi}^{(I)}(\boldsymbol{x})$ as :

\begin{equation}\label{expHI} 
\phi^{(I)}_{\xi}(\boldsymbol{x}) = \langle \xi^{(I)}| \hat{\phi}^{(I)}(\boldsymbol{x}) |\xi^{(I)}\rangle = \xi^{(I)}_{0} v_{0}^{(I)}(\eta)/L^{3/2} + c.c, 
\end{equation} 
which can be written as, 
 
\begin{equation}
\phi^{(I)}_{\xi}(\boldsymbol{x})=\phi^{(I)}_{\xi,0}(\eta),   \textup{  with  }  \phi^{(I)}_{\xi,0}(\eta) = \xi^{(I)}_{0} v_{0}^{(I)}(\eta)/L^{3/2} + c.c. 
\end{equation} 
For the state $|\xi^{(I)}\rangle$ in which the only excited mode $\vec{k}$ is $\vec{k}=0$, the only 
non-trivial  part of  semi-classical Einstein's equations are the components $\eta\eta$, $i=j$, which are simplified to:

\begin{eqnarray}
3\mathcal{H}^{2(I)} &=& 4\pi G \Big( (\phi^{'2(I)})_{\xi,0} + a^{2(I)}m^{2}(\phi^{2(I)})_{\xi,0} \Big),\label{asimi} \\
\mathcal{H}^{2(I)} + 2\mathcal{H}^{'(I)}&=& -4\pi G \Big( (\phi^{'2(I)})_{\xi,0} - a^{2(I)}m^{2}(\phi^{2(I)})_{\xi,0} \Big).\label{similares}
\end{eqnarray}
We remind the reader that in the above,  we  need to use  a renormalized expectation  value of  the energy momentum tensor. The renormalization   procedure   here   is   simplified  by the fact   that the correction terms  and the  ambiguities that are  usually  present in  the  vacuum  expectation value,   are limited by  the  symmetry of the spacetime,  while the  difference  between    that  and  the   expectation value  in other states is   obtained  by  a normal ordering  procedure. Finally, the  fact that we are using   coherent states  reduces the   problem to that of  evaluating the corresponding  classical energy momentum tensor of the corresponding "classical  configuration"\cite{PC-Benito}.

Then, in  the above expression, $(\phi^{'2(I)})_{\xi,0} = \langle \xi^{(I)}|(\partial_{\eta}\hat{\phi}^{(I)})^{2}|\xi^{(I)}\rangle,$ and $(\phi^{2(I)})_{\xi,0} = \langle \xi^{(I)}| (\hat{\phi}^{(I)})^{2}|\xi^{(I)}\rangle,$ only depend on the variable $\eta$. Equations~(\ref{asimi}) and (\ref{similares}), are analogue, (but not exactly the same) to the Friedmann equations obtained in the context of classical field theory. The reason being, that despite the Ehrenfest relations provide equations of motion for expectation values of the inflaton field, the relations 

\begin{equation}\label{relclasicas}
(\phi^{'2(I)})_{\xi,0} = (\phi^{'(I)}_{\xi,0})^{2},\hskip.3cm
 \textup{ and } \hskip.3cm (\phi^{2(I)})_{\xi,0} = (\phi^{(I)}_{\xi,0})^{2}, 
\end{equation}
satisfied at a classical level, in general, at a quantum level  are not satisfied. This happens since $\langle\xi^{(I)}|\hat{a}^{(I)}_{0}|\xi^{(I)}\rangle = \xi^{(I)}_{0}$, does not necessarily imply that $\langle \xi^{(I)}|\hat{a}^{2(I)}_{0}|\xi^{(I)}\rangle = \xi^{2(I)}_{0}$. However, for the state $|\xi^{(I)}\rangle$ defined in  Eq.~(\ref{xi000estado}) which is a coherent state, strongly localized around the classical configuration $\phi^{(I)}_{\xi,0}$ and $\pi^{(I)}_{\xi,0}$, with only the $\vec{k}=0$ mode excited (that is, $\hat{a}^{(I)}_{0}|\xi^{(I)}\rangle = \xi^{(I)}_{0}|\xi^{(I)}\rangle$, being $\xi^{(I)}_{0}\in\mathbb{C}$ ),  the classical relations (\ref{relclasicas}) are satisfied. For this case, Equations~(\ref{similares}) are reduced to the standard Friedmann equations. Then by using  (\ref{similares}), we observe that during slow-roll inflation, the equation for  $\epsilon^{(I)} = 1 - \mathcal{H}^{'(I)}/\mathcal{H}^{2(I)}$ can be written as:

\begin{equation}\label{ecuparconst}
\epsilon^{(I)}\mathcal{H}^{2(I)} = 4\pi G\hskip.2cm (\phi^{'(I)}_{\xi,0})^{2},
\end{equation}
solving for $\mathcal{H}^{2(I)}$ in the above equation and substituting it into Eq.~ (\ref{asimi}), we find:

\begin{equation}
(3- \epsilon^{(I)})(\phi^{'(I)}_{\xi,0})^{2} = \epsilon^{(I)}a^{2(I)}m^{2}(\phi^{(I)}_{\xi,0})^{2},
\end{equation}
given that, during slow-roll inflation $0<\epsilon\ll1$, the previous equation takes the following approximated form :

\begin{equation}\label{ecuparconst2}
3(\phi^{'(I)}_{\xi,0})^{2} \approx \epsilon^{(I)}a^{2(I)}m^{2}(\phi^{(I)}_{\xi,0})^{2},
\end{equation}
now substituting $a^{(I)} = \Big(-\frac{1}{H_{0}^{(I)}\eta}\Big)^{1 + \epsilon^{(I)}}\approx -\frac{1}{H_{0}^{(I)}\eta}$ in the previous equation, where the scale factor is written at the lowest-order in slow-roll $\epsilon^{(I)}$ given that the right-hand side of Eq.~(\ref{ecuparconst2}) is already first order in $\epsilon^{(I)}$.  We find :

\begin{equation}
3(\phi^{'(I)}_{\xi,0})^{2} \approx \frac{\epsilon^{(I)} m^{2} }{H_{0}^{2(I)}\eta^{2}} (\phi^{(I)}_{\xi,0})^{2} \Rightarrow  \frac{\phi^{'(I)}_{\xi,0}}{\phi^{(I)}_{\xi,0}} \approx \sqrt{ \frac{\epsilon^{(I)}}{3} }\frac{ m }{H^{(I)}_{0}}\frac{1}{|\eta|},
\end{equation}
its first integral with respect to conformal time indicates:

\begin{equation}\label{resul71}
\phi^{(I)}_{\xi,0}(\eta) \propto |\eta|^{ \sqrt{ \frac{\epsilon^{(I)}}{3} }\frac{m}{H_{0}^{(I)} }  }.
\end{equation}
On the other hand, assuming that the parameter $\xi^{(I)}_{0}\in\mathbb{R}$, and using Eq.~(\ref{ceroperturSolPar0}) in Eq.~ (\ref{expHI}), we obtain:

\begin{equation}\label{expHIcomparar} 
\phi^{(I)}_{\xi,0}(\eta) = \frac{2\xi^{(I)}_{0}}{L^{3/2}}\sqrt{ \frac{1}{ H^{(I)}_{0} } }\Big( -H_{0}^{(I)}\eta \Big)^{ m^{2}/(3H_{0}^{2(I)}) }.
\end{equation}
Now, demanding consistency between equations~(\ref{resul71}) and (\ref{expHIcomparar}), and recalling that during inflation $-\infty<\eta<0$ meaning that $-\eta = |\eta|$,  we obtain:

\begin{equation}\label{fsrp}
\epsilon^{(I)} = \frac{m^{2}}{ 3H^{2(I)}_{0} }.  
\end{equation}
We note   that the standard treatment's   expression for the first slow-roll parameter  in terms of the potential is :

\begin{equation}\label{fpsrp}
\epsilon^{(I)}  = \frac{1}{16\pi G    }\left(\frac{V'}{V}\right)^2
\end{equation}

 which  we specialized  to  a quadratic potential, gives:

\begin{equation}\label{fpsrppq}
\epsilon^{(I)}  =  \frac{1}{4\pi G}\frac{1}{\phi_{\xi,0}^{2(I)}}
\end{equation}

which is consistent with Eq.~(\ref{fsrp})  because as we   will see below  $\phi_{\xi,0}^{2(I)} \sim \frac{3}{4\pi G} \frac{H_0^{2(I)}}{m^2}  $ as is expected in the slow-roll approximation to the Friedman equation.

For other type of potentials, Eq.~(\ref{fpsrp}) might be applicable  as   an approximation,  as long as one is working in the standard cosmology and a minimally coupled scalar field in the slow-roll regime and  in the regime of interest  the potential is  susceptible to a power series expansion.  The point is  however, that   deviating from  quadratic potentials  invalidates the explicit   construction of the Hilbert space which,    as  is  well known  can not   in general  be carried out  for   interacting field theories  which  relay on  a perturbative  treatment.  In principle  we   do not see a clear  obstacle to  combining  such perturbation with the perturbative nature of our SSC constructions  but it is clear that   the completion of such  exercise  would be  a much more difficult task  which certainly falls  beyond the scope of the present work.
\\

Lastly, in order to determine the proportionality constant missing in Eq.~ (\ref{resul71}), we proceed in the following manner. We solve for $\epsilon^{(I)}$ in Eq.~(\ref{ecuparconst}) and substitute it into Eq.~(\ref{ecuparconst2}),

\begin{equation}
\epsilon^{(I)} = 4\pi G\hskip.2cm \bigg( \frac{\phi^{'(I)}_{\xi,0}}{\mathcal{H}^{(I)}}\bigg)^{2} \approx  \frac{3}{a^{2(I)}m^{2}} \bigg( \frac{\phi^{'(I)}_{\xi,0}}{\phi^{(I)}_{\xi,0}}\bigg)^{2} \Rightarrow  
\frac{4\pi G\hskip.01cm m^{2}}{3} \bigg( \frac{a^{(I)}}{ \mathcal{H}^{(I)} }\bigg)^{2} \approx \bigg( \frac{1}{\phi^{(I)}_{\xi,0}}\bigg)^{2},
\end{equation}
therefore, we obtain $\phi^{(I)}_{\xi,0}\approx \sqrt{\frac{ 3 }{ 4\pi G } }\frac{H^{(I)}}{m}$. Now substituting $a^{(I)} = \Big(-\frac{1}{H^{(I)}_{0}\eta}\Big)^{1 + \epsilon^{(I)}}$ to calculate $H^{(I)} = \frac{ \mathcal{H}^{(I)} }{ a^{(I)} },$ obtaining $a^{'(I)} = \frac{1}{H^{(I)}_{0}\eta^{2} }\Big(-\frac{1}{H^{(I)}_{0}\eta}\Big)^{\epsilon^{(I)}}$, then $H^{(I)} = (-H^{(I)}_{0}\eta)^{\epsilon^{(I)}} H^{(I)}_{0}$, hence $\phi^{(I)}_{\xi,0}$ can be written as: 

\begin{equation}\label{expHIcomparar3}
\phi^{(I)}_{\xi,0}\approx \sqrt{\frac{ 3 }{ 4\pi G } }\frac{H^{(I)}_{0}}{m}(-H^{(I)}_{0}\eta)^{ \epsilon^{(I)} },
\end{equation}
which  at  the desired order  implies $\phi^{(I)}_{\xi,0}\approx \sqrt{\frac{ 3 }{ 4\pi G } }\frac{H^{(I)}_{0}}{m}$. Once more again,  consistency between equations~(\ref{expHIcomparar}) and (\ref{expHIcomparar3}), implies: 

\begin{equation}\label{expHIcomparar4}
\frac{2\xi^{(I)}_{0}}{L^{3/2}}\sqrt{ \frac{1}{ H^{(I)}_{0} } } \approx \sqrt{\frac{ 3 }{ 4\pi G } }\frac{H^{(I)}_{0}}{m} =  \sqrt{\frac{ 1 }{ 4\pi G \epsilon^{(I)} } } 
\end{equation}
which in  turn leads to:

\begin{equation}\label{expHIcomparar5}
\xi^{(I)}_{0} \approx\sqrt{\frac{  L^{3}H^{(I)}_{0} }{ 16\pi G \epsilon^{(I)}} } = \sqrt{\frac{  L^{3}H^{(I)}_{0} }{ 2 \epsilon^{(I)} t_{p}^{2} } },
\end{equation}
with $t_{p} = \sqrt{8\pi G}$ being the Planck time.

In this way, in having found $\xi^{(I)}_{0}$, and then substituting it into (\ref{xi000estado}) the state $|\xi^{(I)}\rangle$ will be completely determined.

\subsection{Construction of a SSC-II in an  ``almost FRW"  space time (``slightly inhomogeneous and anisotropic")}

We are now making the construction of a SSC-II, corresponding to a perturbed FRW space time, where we will truncate all power-series expansions of perturbations up to second order in  $\varepsilon$. This is in contrast with work presented in  \cite{DiezTaSudarD} which was performed at linear-order in perturbations.  In  \cite{DiezTaSudarD} the only  modes  that  are excited modes  are  those with $\vec{k} = 0$ and $\vec{k} = \pm\vec{k}_0$, whereas here, we assume that   excitations of the metric potentials characterized by the modes $\vec{k}=0$, $\pm\vec{k}_{0}$ and $\pm2\vec{k}_{0}$ are potentially  relevant when  working  up to second order in $\varepsilon$ and thus that  the  metric  perturbations,  can  be written  in  the form:

\begin{eqnarray}
\tilde{\Phi}(\boldsymbol{x}) &=& -\varepsilon^{2}a_{(2)}(\eta) + [( \varepsilon P_{(1)}(\eta)e^{i\vec{k}_{0}\cdot\vec{x}} + \varepsilon^{2}P_{(2)}(\eta)e^{2i\vec{k}_{0}\cdot\vec{x}} ) + c.c ] ,\label{ordpertHas2Phi} \\ 
\tilde{\Psi}(\boldsymbol{x}) &=& \hskip.35cm\varepsilon^{2}a_{(2)}(\eta) + [( \varepsilon P_{(1)}(\eta)e^{i\vec{k}_{0}\cdot\vec{x}} + \varepsilon^{2}P_{(2)}(\eta)e^{2i\vec{k}_{0}\cdot\vec{x}} ) + c.c ],\label{ordpertHas2Psi} \\
h_{ij}(\boldsymbol{x}) &=& \hskip.35cm\varepsilon^{2}h_{ij}^{(2)}(\eta) + (\varepsilon^{2} H_{ij}^{(2)}(\eta)e^{2i\vec{k}_{0}\cdot\vec{x}} + c.c). \label{ordpertHas2Ht}
\end{eqnarray}
 We impose the  condition that  the quantities  $a_{(2)}(\eta)$, $h_{ij}^{(2)}(\eta)$, $P_{(1)}(\eta)$ are real valued. We, will see that  in order to construct the SSC-II up to second order in perturbation theory, it will be necessary to 
allow  the functions $P_{(2)}(\eta)$, $F_{(2)}(\eta)$, $H_{ij}^{(2)}(\eta)$ to be complex valued  indicating the presence  of  physically meaningful relative  phase. 

As a first step for the construction of the SSC-II, we will perform the quantum field theory over the new configuration of the metric spacetime (\ref{71}), characterized by the metric potentials (\ref{ordpertHas2Phi}), (\ref{ordpertHas2Psi}) and (\ref{ordpertHas2Ht}). \\ 
Considering that (\ref{ordpertHas2Phi}), (\ref{ordpertHas2Psi}) and (\ref{ordpertHas2Ht}) we propose the following ansatz for the mode functions:

\begin{eqnarray}
&&u^{(II)}_{\vec{k}}(\boldsymbol{x})L^{\frac{3}{2}} = \delta^{(0)}v^{(II)}_{\vec{k}}(\eta)e^{i\vec{k}\cdot\vec{x}} + \varepsilon^{2}\Sigma_{ij}\delta^{(2)}w^{(II)}_{ij}(\eta,\vec{k})e^{i\vec{k}\cdot\vec{x}} + \varepsilon \delta^{(1)}v^{(II)\pm}_{\vec{k}}(\eta)e^{i( \vec{k} \pm \vec{k}_{0})\cdot\vec{x}} \nonumber\\ 
&& \hskip2cm + \hskip.08cm \varepsilon^{2}\delta^{(2)}v^{(II)\pm}_{\vec{k}}(\eta)e^{i( \vec{k} \pm 2\vec{k}_{0})\cdot\vec{x}} + \varepsilon^{2}\Sigma_{ij}\delta^{(2)}W^{(II)\pm}_{ij}(\eta,\vec{k})e^{i(\vec{k} \pm 2\vec{k}_{0})\cdot\vec{x}} + \mathcal{O}(\varepsilon^{3}), 
\end{eqnarray}
for simplicity, we rename $\theta^{(II)}_{\vec{k}} = \Sigma_{ij}\delta^{(2)}w^{(II)}_{ij}(\eta,\vec{k})$ and $\delta^{(2)}\tilde{v}^{(II)\pm}_{\vec{k}}(\eta) = \delta^{(2)}v^{(II)\pm}_{\vec{k}}(\eta) +  \Sigma_{ij}\delta^{(2)}W_{ij}^{(II)\pm}(\eta,\vec{k})$ then $u^{(II)}_{\vec{k}}(\eta,\vec{x})$ can be written as;

\begin{eqnarray}\label{Uk2eps}
u^{(II)}_{\vec{k}}(\boldsymbol{x}) L^{3/2}&=& (\delta^{(0)}v^{(II)}_{\vec{k}}(\eta) + \varepsilon^{2}\theta^{(II)}_{\vec{k}})e^{i\vec{k}\cdot\vec{x}} + \varepsilon \delta^{(1)}v^{(II)\pm}_{\vec{k}}(\eta)e^{i( \vec{k} \pm \vec{k}_{0})\cdot\vec{x}} + \varepsilon^{2}\delta^{(2)}\tilde{v}^{(II)\pm}_{\vec{k}}(\eta)e^{i( \vec{k} \pm 2\vec{k}_{0})\cdot\vec{x}} + \mathcal{O}(\varepsilon^{3}).
\end{eqnarray}
Substituting now (\ref{ordpertHas2Phi}), (\ref{ordpertHas2Psi}), (\ref{ordpertHas2Ht}) and (\ref{Uk2eps}) into (\ref{KleinGordonP}), and working order by order in the $\varepsilon$ parameter, we find that at zeroth order in  $\varepsilon$ the equation of motion for the mode functions $\delta^{(0)}v^{(II)}_{\vec{k}}$ is;

\begin{equation}\label{ceroperturII}
\delta^{(0)}v^{''(II)}_{\vec{k}}(\eta) + 2\mathcal{H}^{(II)} \delta^{(0)}v^{'(II)}_{\vec{k}}(\eta) + (k^{2} + a^{2(II)}m^{2})\delta^{(0)}v^{(II)}_{\vec{k}}(\eta) = 0.
\end{equation}
As we are showing zeroth-order results in the expansion, perturbations do not contribute for the moment and
the previous equation has the same form as Eq.~(\ref{ceropertur}).

Whereas performing the same substitutions into (\ref{NorsimplecPI}), leads to obtaining the normalization condition for the functions $\delta^{(0)}v^{(II)}_{\vec{k}}(\eta)$. Following the same steps as in (\ref{NorsimplecPIcero}) and (\ref{NormasimplecPI}) results in;

\begin{equation}\label{NormasimplecPI0eps} 
\Big( \delta^{(0)}v_{\vec{k}}^{(II)}(\eta)\delta^{(0)}v_{\vec{k}}^{'(II)\ast}(\eta) - \delta^{(0)}v^{'(II)}_{\vec{k}}(\eta)\delta^{(0)}v_{\vec{k}}^{(II)\ast}(\eta) \Big) \hskip.1cm\Big|_{\eta=\eta_{c}} = \frac{i}{a^{2(I)}(\eta_{c})}.
\end{equation}
A particular solution for the equations Eq.~(\ref{ceroperturII}) normalized according to Eq.~(\ref{NormasimplecPI0eps}), for modes such that $\vec{k}\neq0$, is; 

\begin{equation}\label{ceroperturBDII}
\delta^{(0)}v^{(II)}_{\vec{k}}(\eta) \approx \sqrt{ \frac{1}{2k} }\Big(-H_{0}^{(II)}\eta\Big) \Big( 1 - \frac{i}{k\eta} \Big) e^{-ik\eta}. 
\end{equation}
Whereas for the mode $\vec{k}=0$, is;  

\begin{equation}\label{ceroperturSolPar0II}
\delta^{(0)}v^{(II)}_{0}(\eta) = \sqrt{ \frac{1}{ H^{(II)}_{0} } }\Big[ 1 - \frac{i}{6}\Big(-H_{0}^{(II)}\eta\Big)^{3} \Big]\Big( -H_{0}^{(II)}\eta \Big)^{ m^{2}/(3H_{0}^{2(II)}) }. 
\end{equation}
We now study the contributions given at first order in perturbations. That is, we consider contributions up to first order in $\varepsilon$.

Substituting the ansatz (\ref{ordpertHas2Phi}), (\ref{ordpertHas2Psi}), (\ref{ordpertHas2Ht}) and (\ref{Uk2eps}) into (\ref{KleinGordonP}), after manipulations, keeping terms at first order in $\varepsilon$ in Eq.~(\ref{KleinGordonP}), we obtain; 

\begin{eqnarray}\label{primperturII}
&&\Big(\delta^{(1)}v^{''(II)\pm}_{\vec{k}}(\eta) + 2\mathcal{H}^{(II)} \delta^{(1)}v^{'(II)\pm}_{\vec{k}}(\eta) \Big)e^{ i(\vec{k}\pm\vec{k}_{0})\cdot\vec{x}} - \Big( \delta^{(0)}v^{''(II)}_{\vec{k}} + 2\mathcal{H}^{(II)}\delta^{(0)}v^{'(II)}_{\vec{k}} \Big)\Big(2P_{(1)}(\eta)e^{ i\vec{k}_{0}\cdot\vec{x}} \nonumber\\ 
&& + c.c\Big)e^{ i\vec{k}\cdot\vec{x}} + ||\vec{k} \pm \vec{k}_{0}||^{2}\delta^{(1)}v^{(II)\pm}_{\vec{k}}(\eta)e^{ i(\vec{k}\pm\vec{k}_{0})\cdot\vec{x}} + \Big(2P_{(1)}(\eta)e^{ i\vec{k}_{0}\cdot\vec{x}} + c.c\Big)k_{0}^{2}\delta^{(0)}v^{(II)}_{\vec{k}} e^{ i\vec{k}\cdot\vec{x}} \nonumber \\
&& - 4\Big(2P'_{(1)}(\eta)e^{ i\vec{k}_{0}\cdot\vec{x}} + c.c\Big)\delta^{(0)}v^{'(II)}_{\vec{k}} e^{i\vec{k}\cdot\vec{x}} + a^{2}m^{2}\delta^{(1)}v^{(II)\pm}_{\vec{k}}(\eta)e^{ i(\vec{k}\pm\vec{k}_{0})\cdot\vec{x} } = 0, 
\end{eqnarray}
the previous equation corresponds to a linear combination of the form:

\begin{equation}
\mathcal{C}_{+}(\eta,\vec{k},\vec{k}_{0})e^{ i(\vec{k}+\vec{k}_{0})\cdot\vec{x} } + \mathcal{C}_{-}(\eta,\vec{k},\vec{k}_{0})e^{ i(\vec{k} - \vec{k}_{0})\cdot\vec{x} }  = 0.
\end{equation}
Given that the functions $e^{ i(\vec{k} + \vec{k}_{0})\cdot\vec{x} }\hskip.03cm$ and $\hskip.03cm e^{ i(\vec{k} - \vec{k}_{0})\cdot\vec{x} }$ are linearly independent,  meaning  $\mathcal{C}_{+}(\eta,\vec{k},\vec{k}_{0}) = \mathcal{C}_{-}(\eta,\vec{k},\vec{k}_{0}) = 0$. In this way, the equations of motion for
 $\delta^{(1)}v^{(II)+}_{\vec{k}}(\eta)$ are found to be;

\begin{eqnarray}\label{Klv1masII}
&&\delta^{(1)}v^{''(II)+}_{\vec{k}}(\eta) + 2\mathcal{H}^{(II)} \delta^{(1)}v^{'(II)+}_{\vec{k}}(\eta) + \Big\{ a^{2}m^{2} + ||\vec{k} + \vec{k}_{0}||^{2} \Big\}\delta^{(1)}v^{(II)+}_{\vec{k}}(\eta) 
\nonumber \\ 
&& - 2\Big( \delta^{(0)}v^{''(II)}_{\vec{k}} + 2\mathcal{H}^{(II)}\delta^{(0)}v^{'(II)}_{\vec{k}} - k_{0}^{2}\delta^{(0)}v^{(II)}_{\vec{k}} \Big)P_{(1)}(\eta) - 4\delta^{(0)}v^{'(II)}_{\vec{k}}P'_{(1)}(\eta)  = 0,
\end{eqnarray}
whereas the equation of motion for $\delta^{(1)}v^{(II)-}_{\vec{k}}$ is;

\begin{eqnarray}\label{Klv1menII}
&&\delta^{(1)}v^{''(II)-}_{\vec{k}}(\eta) + 2\mathcal{H}^{(II)} \delta^{(1)}v^{'(II)-}_{\vec{k}}(\eta) + \Big\{ a^{2}m^{2} + ||\vec{k} - \vec{k}_{0}||^{2} \Big\}\delta^{(1)}v^{(II)-}_{\vec{k}}(\eta) \nonumber \\ 
&&- 2\Big( \delta^{(0)}v^{''(II)}_{\vec{k}} + 2\mathcal{H}^{(II)}\delta^{(0)}v^{'(II)}_{\vec{k}} - k_{0}^{2}\delta^{(0)}v^{(II)}_{\vec{k}} \Big)P_{(1)}^{\ast}(\eta)  
- 4\delta^{(0)}v^{'(II)}_{\vec{k}}P^{'\ast}_{(1)}(\eta)  = 0. 
\end{eqnarray}
We observe that in general $\delta^{(1)}v^{(II)+}_{\vec{k}}$ and $\delta^{(1)}v^{(II)-}_{\vec{k}}$ satisfy different equations of motion. However, given  our requirement that $P_{(1)}(\eta)\in\mathbb{R}$ then $P'_{(1)}(\eta)\in\mathbb{R}$,   indicates that this is  necessary to  permit  the construction of the SSC-II up to second order in perturbation theory. Therefore,  it  is only for the mode $\vec{k} = 0$, that  the equations of motion  will be  simplified  to;

\begin{eqnarray}
&&\delta^{(1)}v^{''(II)+}_{0}(\eta) + 2\mathcal{H}^{(II)} \delta^{(1)}v^{'(II)+}_{0}(\eta) + \Big\{ a^{2}m^{2} + k_{0}^{2} \Big\}\delta^{(1)}v^{(II)+}_{0}(\eta) \nonumber \\ &&- 2\Big( \delta^{(0)}v^{''(II)}_{0} + 2\mathcal{H}^{(II)}\delta^{(0)}v^{'(II)}_{0} - k_{0}^{2}\delta^{(0)}v^{(II)}_{\vec{k}}\Big)P_{(1)}(\eta) 
- 4\delta^{(0)}v^{'(II)}_{0}P'_{(1)}(\eta) = 0,\label{primperturIIceromodo} \\
&&\delta^{(1)}v^{''(II)-}_{0}(\eta) + 2\mathcal{H}^{(II)} \delta^{(1)}v^{'(II)-}_{0}(\eta) + \Big\{ a^{2}m^{2} + k_{0}^{2} \Big\}\delta^{(1)}v^{(II)-}_{0}(\eta) \nonumber \\ &&- 2\Big( \delta^{(0)}v^{''(II)}_{0} + 2\mathcal{H}^{(II)}\delta^{(0)}v^{'(II)}_{0} - k_{0}^{2}\delta^{(0)}v^{(II)}_{\vec{k}}\Big)P_{(1)}(\eta) 
- 4\delta^{(0)}v^{'(II)}_{0}P'_{(1)}(\eta)  = 0. \label{primperturIIceromodo2} 
\end{eqnarray}
In particular,  for the mode $\vec{k} =0$, one finds  that $\delta^{(1)}v^{(II)+}_{0}$ and $\delta^{(1)}v^{(II)-}_{0}$ satisfy the same equation of motion.\\

On the other hand, working with (\ref{NorsimplecPI}) to first order in $\varepsilon$, the normalization condition can be obtained from the functions $\delta^{(1)}v^{(II)\pm}_{\vec{k}}(\eta)$,

\begin{eqnarray}\label{privarnorma}
\int_{\Sigma}d^{3}x\Big\{ \delta^{(0)}v^{(II)}_{\vec{k}}(\delta^{(1)}v^{'(II)\pm}_{\vec{k}'})^{\ast} e^{ i\{ \vec{k} - ( \vec{k}'\pm\vec{k}_{0} ) \} \cdot\vec{x} }  +  \delta^{(1)}v^{(II)\pm}_{\vec{k}}(\delta^{(0)}v^{'(II)}_{\vec{k}'})^{\ast} e^{ i\{ \vec{k} \pm \vec{k}_{0} - \vec{k}' \} \cdot\vec{x} } + \nonumber\\  
-  \delta^{(0)}v^{'(II)}_{\vec{k}}(\delta^{(1)}v^{(II)\pm}_{\vec{k}'})^{\ast} e^{ i\{ \vec{k} - ( \vec{k}'\pm\vec{k}_{0} ) \} \cdot\vec{x} } - \delta^{(1)}v^{'(II)\pm}_{\vec{k}}(\delta^{(0)}v^{(II)}_{\vec{k}'})^{\ast} e^{ i\{ \vec{k} \pm \vec{k}_{0} - \vec{k}' \} \cdot\vec{x} } + \nonumber\\
- 4 ( \delta^{(0)}v_{\vec{k}}^{(II)}\delta^{(0)}v_{\vec{k}'}^{'(II)\ast} - \delta^{(0)}v^{'(II)}_{\vec{k}}\delta^{(0)}v_{\vec{k}'}^{(II)\ast} )(P_{(1)}e^{ i\vec{k}_{0}\cdot\vec{x}} + c.c )e^{ i(\vec{k}-\vec{k}')\cdot\vec{x}} \Big\} = 0.
\end{eqnarray}
Now we use once more the result for the vectors $\vec{A}$ and $\vec{B}$ of the form $\vec{V} = V_{x}\hat{e}_{x} + V_{y}\hat{e}_{y} + V_{z}\hat{e}_{z}$, with $V_{u} = \frac{2\pi n_{u}}{L}$, $u: x, y, z$ and $n_{u} \in\mathbb{Z}$;

\begin{equation}
\frac{1}{L^{3}}\int_{\Sigma}d^{3}x e^{i(\vec{A} - \vec{B})\cdot\vec{x}} = \frac{1}{L^{3}}\int_{0}^{L}\int_{0}^{L}\int_{0}^{L}d^{3}x e^{i(\vec{A} - \vec{B})\cdot\vec{x}} = \delta_{\vec{A},\vec{B}} 
\end{equation}
Making use of this result, we find that the only non-trivial integrals appearing in (\ref{privarnorma}), are present when the vector $\vec{k}'$ takes on the values $\vec{k}' = \vec{k} \pm \vec{k}_{0}$. \\

\begin{center}
\textbf{$\vec{k}' = \vec{k} + \vec{k}_{0}$.}
\end{center}

Evaluating $\vec{k}' = \vec{k} + \vec{k}_{0}$ on the equation~(\ref{privarnorma}) we find;

\begin{eqnarray}\label{1privarnorma}
&&\int_{\Sigma}d^{3}x\Big\{ \delta^{(0)}v^{(II)}_{\vec{k}}(\delta^{(1)}v^{'(II)\pm}_{ \vec{k} + \vec{k}_{0} })^{\ast} e^{-i( \vec{k}_{0}\pm\vec{k}_{0})\cdot\vec{x} }  +  \delta^{(1)}v^{(II)\pm}_{\vec{k}}(\delta^{(0)}v^{'(II)}_{\vec{k} + \vec{k}_{0}})^{\ast} e^{ i(\pm\vec{k}_{0} - \vec{k}_{0}) \cdot\vec{x} }  \nonumber\\  
&&-  \delta^{(0)}v^{'(II)}_{\vec{k}}(\delta^{(1)}v^{(II)\pm}_{\vec{k} + \vec{k}_{0}})^{\ast} e^{-i(\vec{k}_{0}\pm\vec{k}_{0})\cdot\vec{x} } - \delta^{(1)}v^{'(II)\pm}_{\vec{k}}(\delta^{(0)}v^{(II)}_{\vec{k} + \vec{k}_{0}})^{\ast} e^{ i(  \pm \vec{k}_{0} - \vec{k}_{0} ) \cdot\vec{x} }  \nonumber\\
&&- 4 ( \delta^{(0)}v_{\vec{k}}^{(II)}\delta^{(0)}v_{\vec{k} + \vec{k}_{0}}^{'(II)\ast} - \delta^{(0)}v^{'(II)}_{\vec{k}}\delta^{(0)}v_{\vec{k} + \vec{k}_{0}}^{(II)\ast} )(P_{(1)}e^{ i(\vec{k}_{0} - \vec{k}_{0})\cdot\vec{x}}  + P_{(1)}^{\ast}e^{-i(\vec{k}_{0} + \vec{k}_{0})\cdot\vec{x}} ) 
\Big\} = 0.
\end{eqnarray}
Now, after calculating the corresponding integrals, we find

\begin{eqnarray}\label{1privarnormaRes}
&&\Big\{ \delta^{(0)}v^{(II)}_{\vec{k}}(\delta^{(1)}v^{'(II)-}_{ \vec{k} + \vec{k}_{0} })^{\ast}  +  \delta^{(1)}v^{(II)+}_{\vec{k}}(\delta^{(0)}v^{'(II)}_{\vec{k} + \vec{k}_{0}})^{\ast} 
- \delta^{(0)}v^{'(II)}_{\vec{k}}(\delta^{(1)}v^{(II)-}_{\vec{k} + \vec{k}_{0}})^{\ast} + \nonumber\\ 
&&- \delta^{(1)}v^{'(II)+}_{\vec{k}}(\delta^{(0)}v^{(II)}_{\vec{k} + \vec{k}_{0}})^{\ast} 
- 4 \Big( \delta^{(0)}v_{\vec{k}}^{(II)} \delta^{(0)}v_{\vec{k} + \vec{k}_{0}}^{'(II)\ast} - \delta^{(0)}v^{'(II)}_{\vec{k}}\delta^{(0)}v_{\vec{k} + \vec{k}_{0}}^{(II)\ast} \Big)P_{(1)} \Big\}  
\Bigg|_{\eta=\eta_{c}} = 0.
\end{eqnarray}
We observe that the simple choice $(\delta^{(1)}v^{'(II)-}_{\vec{k}}(\eta_{c}))^{\ast}$, $(\delta^{(1)}v^{(II)-}_{\vec{k}}(\eta_{c}))^{\ast}$, $\delta^{(1)}v^{'(II)+}_{\vec{k}}(\eta_{c})$,  and $\delta^{(1)}v^{(II)+}_{\vec{k}}(\eta_{c})$, that satisfies the previous equations, corresponds to:

\begin{eqnarray}
&&(\delta^{(1)}v^{'(II)-}_{\vec{k}}(\eta_{c}))^{\ast} = \delta^{(1)}v^{'(II)+}_{\vec{k}}(\eta_{c}) = 0, \hskip.3cm  
\delta^{(1)}v^{(II)+}_{\vec{k}}(\eta_{c}) = 4\delta^{(0)}v^{(II)}_{\vec{k}}(\eta_{c})P_{(1)}(\eta_{c}),\label{AcondicionInP}\\
&&(\delta^{(1)}v^{(II)-}_{\vec{k}}(\eta_{c}))^{\ast} = 4\delta^{(0)}v^{(II)\ast}_{\vec{k}}(\eta_{c})P_{(1)}(\eta_{c}).\label{condicionInP}
\end{eqnarray}

On the other hand, we need to verify that the previous initial conditions are also compatible with the normalization condition  (\ref{privarnorma}) for $\vec{k}' = \vec{k} - \vec{k}_{0}$. Then evaluating $\vec{k}' = \vec{k} - \vec{k}_{0}$ in the equation (\ref{privarnorma}) we find:

\begin{eqnarray}\label{2privarnorma}
&&\int_{\Sigma}d^{3}x\Big\{ \delta^{(0)}v^{(II)}_{\vec{k}}(\delta^{(1)}v^{'(II)\pm}_{\vec{k} - \vec{k}_{0}})^{\ast} e^{ -i( - \vec{k}_{0}\pm\vec{k}_{0})\cdot\vec{x} }  +  \delta^{(1)}v^{(II)\pm}_{\vec{k}}(\delta^{(0)}v^{'(II)}_{\vec{k} - \vec{k}_{0}})^{\ast} e^{ i( \pm \vec{k}_{0} + \vec{k}_{0} ) \cdot\vec{x} }  \nonumber\\  
&& -  \delta^{(0)}v^{'(II)}_{\vec{k}}(\delta^{(1)}v^{(II)\pm}_{\vec{k} - \vec{k}_{0}})^{\ast} e^{ -i(- \vec{k}_{0}\pm\vec{k}_{0} )\cdot\vec{x} } - \delta^{(1)}v^{'(II)\pm}_{\vec{k}}(\delta^{(0)}v^{(II)}_{\vec{k} - \vec{k}_{0}})^{\ast} e^{ i(\pm \vec{k}_{0} + \vec{k}_{0})\cdot\vec{x} }  \nonumber\\
&& - 4 ( \delta^{(0)}v_{\vec{k}}^{(II)}\delta^{(0)}v_{\vec{k} - \vec{k}_{0}}^{'(II)\ast} - \delta^{(0)}v^{'(II)}_{\vec{k}}\delta^{(0)}v_{\vec{k} - \vec{k}_{0}}^{(II)\ast} )(P_{(1)}e^{ i(\vec{k}_{0}+\vec{k}_{0}) \cdot\vec{x}} + P_{(1)}^{\ast} e^{ i(\vec{k}_{0}-\vec{k}_{0}) \cdot\vec{x}} ) \Big\} = 0,
\end{eqnarray}
after calculating the corresponding integrals, we obtain:

\begin{eqnarray}\label{2privarnormaRes}
&&\Big\{ \delta^{(0)}v^{(II)}_{\vec{k}}(\delta^{(1)}v^{'(II)+}_{\vec{k} - \vec{k}_{0}})^{\ast} +  \delta^{(1)}v^{(II)-}_{\vec{k}}\delta^{(0)}v^{'(II)\ast}_{\vec{k} - \vec{k}_{0}} - \delta^{(0)}v^{'(II)}_{\vec{k}}(\delta^{(1)}v^{(II)+}_{\vec{k} - \vec{k}_{0}})^{\ast} + \nonumber\\
&&- \delta^{(1)}v^{'(II)-}_{\vec{k}}\delta^{(0)}v^{(II)\ast}_{\vec{k} - \vec{k}_{0}} - 4 \Big( \delta^{(0)}v_{\vec{k}}^{(II)}\delta^{(0)}v_{\vec{k} - \vec{k}_{0}}^{'(II)\ast} - \delta^{(0)}v^{'(II)}_{\vec{k}}\delta^{(0)}v_{\vec{k} - \vec{k}_{0}}^{(II)\ast} \Big)P_{(1)}^{\ast} \Big\}\Bigg|_{\eta=\eta_{c}} = 0.
\end{eqnarray}
Straightforward choices  for $(\delta^{(1)}v^{'(II)+}_{\vec{k}}(\eta_{c}))^{\ast}$, $(\delta^{(1)}v^{(II)+}_{\vec{k}}(\eta_{c}))^{\ast}$, $\delta^{(1)}v^{'(II)-}_{\vec{k}}(\eta_{c})$,  and $\delta^{(1)}v^{(II)-}_{\vec{k}}(\eta_{c})$, that satisfy the previous equation are;

\begin{eqnarray}
&&(\delta^{(1)}v^{'(II)+}_{\vec{k}}(\eta_{c}))^{\ast} = \delta^{(1)}v^{'(II)-}_{\vec{k}}(\eta_{c}) = 0, \hskip.3cm  
\delta^{(1)}v^{(II)-}_{\vec{k}}(\eta_{c}) = 4\delta^{(0)}v^{(II)}_{\vec{k}}(\eta_{c})P_{(1)}^{\ast}(\eta_{c}),\label{AcondicionInS}\\
&&(\delta^{(1)}v^{(II)+}_{\vec{k}}(\eta_{c}))^{\ast} = 4\delta^{(0)}v^{(II)\ast}_{\vec{k}}(\eta_{c})P_{(1)}^{\ast}(\eta_{c}).\label{condicionInS}
\end{eqnarray}
We find that (\ref{AcondicionInP}) and (\ref{condicionInP}) correspond to the complex conjugate of (\ref{AcondicionInS}) and (\ref{condicionInS}), therefore, they are the same initial conditions. Now, given that $P_{(1)}(\eta)$ is a function defined over the set of real numbers, for all modes $\vec{k}$ we can write in a compact manner:

\begin{eqnarray}\label{condicionIn} 
\delta^{(1)}v^{'(II)\pm}_{\vec{k}}(\eta_{c}) = 0, \hskip.7cm  
\delta^{(1)}v^{(II)\pm}_{\vec{k}}(\eta_{c}) = 4\delta^{(0)}v^{(II)}_{\vec{k}}(\eta_{c})P_{(1)}(\eta_{c}). 
\end{eqnarray}
We had previously found in (\ref{primperturIIceromodo}) and (\ref{primperturIIceromodo2}) that the equation of motion for the variable $\delta^{(1)}v^{(II)+}_{0}(\eta)$ is identical to that of the variable $\delta^{(1)}v^{'(II)-}_{0}(\eta)$. While (\ref{condicionIn}) represents an initial condition compatible with the normalization condition   (\ref{privarnorma}), such that; $\delta^{(1)}v^{(II)+}_{0}(\eta_{c}) = \delta^{(1)}v^{(II)-}_{0}(\eta_{c})$, and $\delta^{(1)}v^{'(II)+}_{0}(\eta_{c}) = \delta^{(1)}v^{'(II)-}_{0}(\eta_{c})$. Therefore, by the unitary and existence 
 theorem, we conclude that the solution of the system  (\ref{primperturIIceromodo}) and (\ref{primperturIIceromodo2}), that satisfy the conditions (\ref{condicionIn}), is such that; $\delta^{(1)}v^{(II)+}_{0}(\eta) = \delta^{(1)}v^{(II)-}_{0}(\eta)$.
 
Nevertheless, for the remaining $\delta^{(1)}v^{(II)\pm}_{\vec{k}}$ functions, once $P_{(1)}(\eta)$ is found, it is substituted into  (\ref{Klv1masII}) and (\ref{Klv1menII}), constructing so the equations of motion for $\delta^{(1)}v^{(II)\pm}_{\vec{k}}$. The initial conditions for these equations are obtained by evaluating $P_{(1)}(\eta)$ in $\eta_{c}$ to be then substituted in (\ref{condicionIn}).

We now proceed to examine the second order in perturbations. This means, working up to second order in 
$\varepsilon$. Therefore, the $\varepsilon^{2}$ order in the equation  (\ref{KleinGordonP}), corresponds to:

\begin{eqnarray}\label{segperturII}
&&\Big( \delta^{(2)}\tilde{v}^{''(II)\pm}_{\vec{k}}(\eta) + 2\mathcal{H}^{(II)} \delta^{(2)}\tilde{v}^{'(II)\pm}_{\vec{k}}(\eta) \Big)e^{ i(\vec{k} \pm 2\vec{k}_{0})\cdot\vec{x} } +   
\Big(\theta^{''(II)\pm}_{\vec{k}}(\eta) + 2\mathcal{H}^{(II)} \theta^{'(II)\pm}_{\vec{k}}(\eta) \Big)e^{ i\vec{k}\cdot\vec{x} } - \Big( \delta^{(1)}v^{''(II)\pm}_{\vec{k}}(\eta) \nonumber \\ 
&& + 2\mathcal{H}^{(II)}\delta^{(1)}v^{'(II)\pm}_{\vec{k}}(\eta) \Big)\Big(2P_{(1)}(\eta)e^{ i\vec{k}_{0}\cdot\vec{x}} + c.c\Big)e^{ i(\vec{k} \pm \vec{k}_{0})\cdot\vec{x}} - \Big\{ [P_{(2)}(\eta) e^{2i\vec{k}_{0}\cdot\vec{x}} + c.c] + 4(P_{(1)}(\eta)e^{ i\vec{k}_{0}\cdot\vec{x}} + c.c)^{2} \nonumber\\ 
&& + 3a_{(2)}(\eta) \Big\} \Big( \delta^{(0)}v^{''(II)}_{\vec{k}} + 2\mathcal{H}^{(II)} \delta^{(0)}v^{'(II)}_{\vec{k}}  \Big)e^{ i\vec{k}\cdot\vec{x} } + ||\vec{k} \pm 2\vec{k}_{0}||^{2}\delta^{(2)}\tilde{v}^{(II)\pm}_{\vec{k}}(\eta)e^{ i(\vec{k}\pm2\vec{k}_{0})\cdot\vec{x}} + 2||\vec{k} \pm \vec{k}_{0}||^{2}( \nonumber\\ 
&& P_{(1)}(\eta)e^{ i\vec{k}_{0}\cdot\vec{x} }   
+ c.c)\delta^{(1)}v^{(II)\pm}_{\vec{k}}(\eta)e^{ i(\vec{k}\pm\vec{k}_{0})\cdot\vec{x}} + 
k^{2}\theta^{(II)}_{\vec{k}}(\eta)e^{ i\vec{k}\cdot\vec{x} } + \Big\{[P_{(2)}(\eta)e^{2i\vec{k}_{0}\cdot\vec{x}}  + c.c] - 4(P_{(1)}(\eta)e^{ i\vec{k}_{0}\cdot\vec{x}} + c.c)^{2} \nonumber\\ 
&& - 3a_{(2)}(\eta) - 0.5h_{L}^{(2)}(\eta) - 0.5( H_{L}^{(2)}(\eta)e^{ 2i\vec{k}_{0}\cdot\vec{x}} + c.c ) \Big\}k^{2}\delta^{(0)}v^{(II)}_{\vec{k}}e^{i\vec{k}\cdot\vec{x}} + 1.5\Big[h_{L}^{(2)}(\eta) + (H_{L}^{(2)}(\eta)e^{ 2i\vec{k}_{0}\cdot\vec{x}} + c.c ) \Big] \times \nonumber\\ 
&&
k_{z}^{2}\delta^{(0)}v^{(II)}_{\vec{k}}e^{i\vec{k}\cdot\vec{x}} 
+ \Big\{   
4k_{0z}(P_{(1)}(\eta)e^{i\vec{k}_{0}\cdot\vec{x}} + c.c)(iP_{(1)}(\eta)e^{i\vec{k}_{0}\cdot\vec{x}} + c.c) - 2k_{0z} (iH_{L}^{(2)}(\eta)e^{2i\vec{k}_{0}\cdot\vec{x}} + c.c) \Big\}ik_{z}  \delta^{(0)}v^{(II)}_{\vec{k}}e^{i\vec{k}\cdot\vec{x}} \nonumber\\ 
&&
-4(P'_{(1)}(\eta)e^{i\vec{k}_{0}\cdot\vec{x}} + c.c ) \delta^{(1)}v^{'(II)\pm}_{\vec{k}}(\eta)e^{ i(\vec{k}\pm\vec{k}_{0})\cdot\vec{x} } 
- \Big\{ a'_{(2)}(\eta) + 2[P'_{(2)}(\eta)e^{2i\vec{k}_{0}\cdot\vec{x}} + c.c] - 4(P_{(1)}(\eta)e^{i\vec{k}_{0}\cdot\vec{x}}  \nonumber\\
&& 
+ c.c)(P'_{(1)}(\eta)e^{i\vec{k}_{0}\cdot\vec{x}} + c.c) \Big\}\delta^{(0)}v^{'(II)}_{\vec{k}}e^{i\vec{k}\cdot\vec{x}} - k_{x}k_{y}\Big\{ h_{T}^{(2)}(\eta) + (H_{T}^{(2)}(\eta)e^{2i\vec{k}_{0}\cdot\vec{x}} + c.c) \Big\}\delta^{(0)}v^{(II)}_{\vec{k}}e^{i\vec{k}\cdot\vec{x}} + a^{2(II)} \times \nonumber\\
&& 
m^{2} \Big\{ \delta^{(2)}\tilde{v}^{(II)\pm}_{\vec{k}}(\eta)e^{ i(\vec{k}\pm2\vec{k}_{0})\cdot\vec{x} } + \theta_{\vec{k}}^{(II)}(\eta)e^{ i\vec{k}\cdot\vec{x} } \Big\} - a^{2(II)} m^{2}\Big\{ 8(P_{(1)}(\eta)e^{i\vec{k}_{0}\cdot\vec{x}} + c.c)^{2} 
+ 4a_{(2)}(\eta) \Big\} \delta^{(0)}v^{(II)}_{\vec{k}}e^{i\vec{k}\cdot\vec{x}} = 0, 
\nonumber\\
\end{eqnarray} 
which can be written as a linear combination of the form:

\begin{eqnarray}\label{combinacionlineal2movi}
&&\mathcal{C}_{2+}(\eta,\vec{k},\vec{k}_{0},2\vec{k}_{0})e^{ i(\vec{k}+2\vec{k}_{0})\cdot\vec{x} } + \mathcal{C}_{0}(\eta,\vec{k},\vec{k}_{0})e^{ i\vec{k}\cdot\vec{x} } + \mathcal{C}_{2-}(\eta,\vec{k},\vec{k}_{0},2\vec{k}_{0})e^{ i(\vec{k} - 2\vec{k}_{0})\cdot\vec{x} } = 0.
\end{eqnarray}
Therefore, by linear independence we obtain the equation $\mathcal{C}_{2+}(\eta,\vec{k},\vec{k}_{0},2\vec{k}_{0}) = 0$:

\begin{eqnarray}\label{segperturIIC2masV}
&&\delta^{(2)}\tilde{v}^{''(II)+}_{\vec{k}}(\eta) + 2\mathcal{H}^{(II)} \delta^{(2)}\tilde{v}^{'(II)+}_{\vec{k}}(\eta) + \Big(||\vec{k} + 2\vec{k}_{0}||^{2} + a^{2(II)}m^{2}  \Big)\delta^{(2)}\tilde{v}^{(II)+}_{\vec{k}}(\eta) 
- 2 P_{(1)}(\eta) \Big( \delta^{(1)}v^{''(II)+}_{\vec{k}}(\eta) \nonumber \\
&& + 2\mathcal{H}^{(II)}\delta^{(1)}v^{'(II)+}_{\vec{k}}(\eta) \Big) - \Big( P_{(2)}(\eta) + 4P_{(1)}^{2}(\eta) \Big)\Big( \delta^{(0)}v^{''(II)+}_{\vec{k}} 
+ 2\mathcal{H}^{(II)} \delta^{(0)}v^{'(II)+}_{\vec{k}} \Big) 
+ 2||\vec{k} + \vec{k}_{0}||^{2}\times\nonumber \\
&&
P_{(1)}(\eta)\delta^{(1)}v^{(II)+}_{\vec{k}}(\eta) + \Big( P_{(2)}(\eta) - 4P_{(1)}^{2}(\eta) \Big)k^{2}\delta^{(0)}v^{(II)}_{\vec{k}} 
- 4(\vec{k}_{0}\cdot\vec{k})P_{(1)}^{2}(\eta)\delta^{(0)}v^{(II)}_{\vec{k}} - 4P_{(1)}'(\eta) \delta^{(1)}v^{'(II)+}_{\vec{k}}(\eta) - \nonumber\\
&& \Big( 2P'_{(2)}(\eta) 
- 4P_{(1)}(\eta)P'_{(1)}(\eta)\Big)\delta^{(0)}v^{'(II)}_{\vec{k}} - 8 a^{2(II)}m^{2} 
 P_{(1)}^{2}(\eta)\delta^{(0)}v^{(II)}_{\vec{k}} 
+  \Big[ \Big(\hskip.002cm 2\vec{k}_{0}\cdot\vec{k} - 0.5k^{2} 
+ 1.5k_{z}^{2} \hskip.01cm \Big) H_{L}^{(2)}(\eta) \nonumber\\
&& 
- k_{x}k_{y} H_{T}^{(2)}(\eta) \Big]\delta^{(0)}v^{(II)}_{\vec{k}} = 0. 
\end{eqnarray}

While equation $\mathcal{C}_{2-}(\eta,\vec{k},\vec{k}_{0},2\vec{k}_{0}) = 0,$ corresponds to;

\begin{eqnarray}\label{segperturIIC2menV}
&&\delta^{(2)}\tilde{v}^{''(II)-}_{\vec{k}}(\eta) + 2\mathcal{H}^{(II)} \delta^{(2)}\tilde{v}^{'(II)-}_{\vec{k}}(\eta) + \Big(||\vec{k} - 2\vec{k}_{0}||^{2} + a^{2(II)}m^{2}  \Big)\delta^{(2)}\tilde{v}^{(II)-}_{\vec{k}}(\eta)  
- 2 P^{\ast}_{(1)}(\eta) \Big( \delta^{(1)}v^{''(II)-}_{\vec{k}}(\eta) \nonumber \\
&& 
+ 2\mathcal{H}^{(II)}\delta^{(1)}v^{'(II)-}_{\vec{k}}(\eta) \Big) - \Big( P^{\ast}_{(2)}(\eta) + 4P^{\ast 2}_{(1)}(\eta) \Big)(\delta^{(0)}v^{ ''(II) }_{ \vec{k} } 
+ 2\mathcal{H}^{(II)} \delta^{(0)}v^{ '(II) }_{ \vec{k} })  
+ 2||\vec{k} - \vec{k}_{0}||^{2} \times \nonumber \\
&&
P^{ \ast }_{(1)}(\eta)\delta^{(1)}v^{(II)-}_{\vec{k}}(\eta) + \Big( P^{\ast}_{(2)}(\eta)   
- 4P_{(1)}^{\ast 2}(\eta) \Big)k^{2}\delta^{(0)}v^{(II)}_{\vec{k}}  
-4( \vec{k}_{0}\cdot\vec{k} )P_{(1)}^{\ast 2}(\eta)\delta^{(0)}v^{(II)}_{ \vec{k} }  
- 4P^{ '\ast }_{(1)}(\eta) \delta^{ (1) }v^{ '(II)- }_{\vec{k}}(\eta) - \nonumber\\
&&\Big( 2P^{'\ast}_{(2)}(\eta) 
- 4P^{\ast}_{(1)}(\eta)P^{'\ast}_{(1)}(\eta) \Big)\delta^{(0)}v^{'(II)}_{\vec{k}} -8a^{2(II)}m^{2}  
P_{(1)}^{\ast 2}(\eta)\delta^{(0)}v^{(II)}_{\vec{k}} 
+  \Big[ \Big(\hskip.002cm 2\vec{k}_{0}\cdot\vec{k} - 0.5k^{2} + 1.5k_{z}^{2} \hskip.01cm \Big) H_{L}^{(2)\ast}(\eta) \nonumber\\
&& - k_{x}k_{y} H_{T}^{(2)\ast}(\eta)  \Big]\delta^{(0)}v^{(II)}_{\vec{k}} = 0. 
\end{eqnarray}

Lastly, concerning the coefficient $\mathcal{C}_{0}(\eta,\vec{k},\vec{k}_{0})$, the relation (\ref{combinacionlineal2movi}), leads us to the equation:

\begin{eqnarray}\label{segperturIIC0}
&&\theta^{''(II)}_{\vec{k}}(\eta) + 2\mathcal{H}^{(II)} \theta^{'(II)}_{\vec{k}}(\eta)  + \Big(k^{2} + a^{2(II)}m^{2} \Big)\theta^{(II)}_{\vec{k}}(\eta) - 2\Big( \delta^{(1)}v^{''(II)-}_{ \vec{k} }(\eta) + 2\mathcal{H}^{ (II) }\delta^{(1)}v^{ '(II)- }_{ \vec{k} }(\eta) \Big)P_{(1)}(\eta) \nonumber \\
&& - 2\Big( \delta^{(1)}v^{''(II)+}_{ \vec{k} }(\eta) + 2\mathcal{H}^{ (II) }\delta^{(1)}v^{ '(II)+ }_{ \vec{k} }(\eta) \Big)P^{\ast}_{(1)}(\eta) - 8 (P_{(1)}(\eta))^{2}\Big( \delta^{(0)}v^{''(II)}_{ \vec{k} } + 2\mathcal{H}^{ (II) }\delta^{(0)}v^{ '(II) }_{ \vec{k} } \Big) + \nonumber \\
&& 2||\vec{k} - \vec{k}_{0}||^{2} P_{(1)}(\eta)\delta^{(1)}v^{(II)-}_{ \vec{k} }(\eta) + 2||\vec{k} + \vec{k}_{0}||^{2} P^{\ast}_{(1)}(\eta)\delta^{(1)}v^{(II)+}_{ \vec{k} }(\eta) - \Big\{\Big( 8(P_{(1)}(\eta))^{2} + h^{(2)}_{L}(\eta)  \Big)k^{2} - 3k_{z}^{2}h^{(2)}_{L}(\eta) \nonumber \\
&& + 2 k_{x}k_{y} h^{(2)}_{T}(\eta) + a^{2(II)}m^{2} \Big( 4a_{(2)}(\eta) + 16(P_{(1)}(\eta))^{2} \Big)\Big\}\delta^{(0)}v^{(II)}_{ \vec{k} } - 4P^{'}_{(1)}(\eta)\delta^{(1)}v^{'(II)-}_{\vec{k} }(\eta) - 4P^{'\ast}_{(1)}(\eta)\delta^{(1)}v^{'(II)+}_{ \vec{k} }(\eta) \nonumber \\
&& = 0. 
\end{eqnarray}

The next step consists in expanding the normalization condition (\ref{NorsimplecPI}), up to second order in $\varepsilon$ to be able to work with perturbations at second order. Performing this, we obtain the normalization condition of the functions $\delta^{(2)}v^{(II)\pm}_{\vec{k}}$ and $\theta^{(II)}_{\vec{k}}$, that is;  

\begin{eqnarray}\label{normalizacionsegorden}
&&\int_{\Sigma}\Big\{ \delta^{(0)}v^{(II)}_{\vec{k}} \Big[ \theta^{'(II)\ast}_{\vec{k}'}e^{ i(\vec{k} - \vec{k}')\cdot\vec{x} } + (\delta^{(2)}v^{'(II)\pm}_{\vec{k}'})^{\ast} e^{ i(\vec{k} -( \vec{k}' \pm 2\vec{k}_{0} ))\cdot\vec{x} } \Big] 
+ \Big( \theta^{(II)}_{\vec{k}}e^{ i(\vec{k} - \vec{k}')\cdot\vec{x} } + \nonumber\\  
&& \delta^{(2)}v^{(II)\pm}_{\vec{k}} e^{ i(\vec{k} \pm 2\vec{k}_{0} - \vec{k}' )\cdot\vec{x} } \Big) (\delta^{(0)}v^{'(II)}_{\vec{k}'})^{\ast}
- \Big( \theta^{'(II)}_{\vec{k}}e^{ i(\vec{k} - \vec{k}')\cdot\vec{x} } + \delta^{(2)}v^{'(II)\pm}_{\vec{k}'} e^{ i(\vec{k} \pm 2\vec{k}_{0} - \vec{k}')\cdot\vec{x} }  
\nonumber\\ 
&& 
\Big) (\delta^{(0)}v^{(II)}_{\vec{k}'})^{\ast} 
- (\delta^{(0)}v^{'(II)}_{\vec{k}}) \Big[ \theta^{(II)\ast}_{\vec{k}'}e^{ i(\vec{k} - \vec{k}')\cdot\vec{x} } + (\delta^{(2)}v^{(II)\pm}_{\vec{k}'})^{\ast} e^{ i(\vec{k} - ( \vec{k}' \pm 2\vec{k}_{0} ) )\cdot\vec{x} }  
\Big] + 
\nonumber\\ 
&& \Big[ \delta^{(1)}v^{(II)\pm}_{\vec{k}}(\delta^{(1)}v^{'(II)\pm}_{\vec{k}'})^{\ast} - \delta^{(1)}v^{'(II)\pm}_{\vec{k}}(\delta^{(1)}v^{(II)\pm}_{\vec{k}'})^{\ast} \Big] e^{ i(\vec{k} \pm \vec{k}_{0} - (\vec{k}' \pm \vec{k}_{0} ))\cdot\vec{x} }  \nonumber\\
&&
- \Big[ \delta^{(0)}v^{(II)}_{\vec{k}}(\delta^{(1)}v^{'(II)\pm}_{\vec{k}'})^{\ast} - \delta^{(0)}v^{'(II)}_{\vec{k}}(\delta^{(1)}v^{(II)\pm}_{\vec{k}'})^{\ast} \Big]\Big(4P_{(1)}e^{ i\vec{k}_{0}\cdot\vec{x} } + c.c\Big)e^{ i(\vec{k} - (\vec{k}' \pm \vec{k}_{0} ))\cdot\vec{x} } \nonumber\\
&&
- \Big[ \delta^{(1)}v^{(II)\pm}_{\vec{k}}(\delta^{(0)}v^{'(II)}_{\vec{k}'})^{\ast} - \delta^{(1)}v^{'(II)\pm}_{\vec{k}}(\delta^{(0)}v^{(II)}_{\vec{k}'})^{\ast} \Big]\Big( 4P_{(1)}e^{ i\vec{k}_{0}\cdot\vec{x} } + c.c\Big)e^{ i(\vec{k} - \vec{k}' \pm \vec{k}_{0} )\cdot\vec{x} } \nonumber\\
&&
+ \Big[ \delta^{(0)}v^{(II)}_{\vec{k}}(\delta^{(0)}v^{'(II)}_{\vec{k}'})^{\ast} - \delta^{(0)}v^{'(II)}_{\vec{k}}(\delta^{(0)}v^{(II)}_{\vec{k}'})^{\ast} \Big]\Big( 6[ P_{(1)}e^{ i\vec{k}_{0}\cdot\vec{x} }  + c.c]^{2} - 2a_{(2)}  \nonumber\\
&& - 2[P_{(2)} e^{ 2i\vec{k}_{0}\cdot\vec{x} } + c.c ] \Big)e^{ i(\vec{k}-\vec{k}')\cdot\vec{x} } 
\Big\} d^{3}x = 0.
\end{eqnarray} 

The integrals involved in the previous equation will be non trivial  only for the following values of the vector 
$\vec{k}'$, which are $\vec{k}' = \vec{k}$ and $\vec{k}' = \vec{k} \pm 2\vec{k}_{0}$.

For $\vec{k}' = \vec{k}$ the integral (\ref{normalizacionsegorden}) gives as a result;

\begin{eqnarray}\label{normlizasgorden1}
&&\bigg\{ \delta^{(0)}v^{(II)}_{\vec{k}}\theta^{'(II)\ast}_{\vec{k}} + \theta^{(II)}_{\vec{k}}  (\delta^{(0)}v^{'(II)}_{\vec{k}})^{\ast} - \theta^{'(II)}_{\vec{k}}  (\delta^{(0)}v^{(II)}_{\vec{k}})^{\ast} - (\delta^{(0)}v^{'(II)}_{\vec{k}}) \theta^{(II)\ast}_{\vec{k}} 
+ \nonumber\\ 
&& \delta^{(1)}v^{(II)+}_{\vec{k}}(\delta^{(1)}v^{'(II)+}_{\vec{k}})^{\ast} - \delta^{(1)}v^{'(II)+}_{\vec{k}}(\delta^{(1)}v^{(II)+}_{\vec{k}})^{\ast} + \delta^{(1)}v^{(II)-}_{\vec{k}}(\delta^{(1)}v^{'(II)-}_{\vec{k}})^{\ast} - \nonumber\\
&& \delta^{(1)}v^{'(II)-}_{\vec{k}}(\delta^{(1)}v^{(II)-}_{\vec{k}})^{\ast}  
-4 \Big[ \delta^{(0)}v^{(II)}_{\vec{k}}(\delta^{(1)}v^{'(II)+}_{\vec{k}})^{\ast} - \delta^{(0)}v^{'(II)}_{\vec{k}}(\delta^{(1)}v^{(II)+}_{\vec{k}})^{\ast} \Big] P_{(1)} \nonumber\\
&&
-4 \Big[ \delta^{(0)}v^{(II)}_{\vec{k}}(\delta^{(1)}v^{'(II)-}_{\vec{k}})^{\ast} - \delta^{(0)}v^{'(II)}_{\vec{k}}(\delta^{(1)}v^{(II)-}_{\vec{k}})^{\ast} \Big] P^{\ast}_{(1)} - 4 \Big[ \delta^{(1)}v^{(II)+}_{\vec{k}}(\delta^{(0)}v^{'(II)}_{\vec{k}})^{\ast}  \nonumber\\
&& - \delta^{(1)}v^{'(II)+}_{\vec{k}}(\delta^{(0)}v^{(II)}_{\vec{k}})^{\ast} \Big] P^{\ast}_{(1)} -4 \Big[ \delta^{(1)}v^{(II)-}_{\vec{k}}(\delta^{(0)}v^{'(II)}_{\vec{k}})^{\ast} - \delta^{(1)}v^{'(II)-}_{\vec{k}}(\delta^{(0)}v^{(II)}_{\vec{k}})^{\ast} \Big]P_{(1)} \nonumber\\
&&
+ \Big[ \delta^{(0)}v^{(II)}_{\vec{k}}(\delta^{(0)}v^{'(II)}_{\vec{k}})^{\ast} - \delta^{(0)}v^{'(II)}_{\vec{k}}(\delta^{(0)}v^{(II)}_{\vec{k}})^{\ast} \Big]\Big( 12|P_{(1)}|^{2} - 2a_{(2)} \Big) \bigg\} \Bigg|_{\eta=\eta_{c}}= 0. 
\end{eqnarray}
Using (\ref{condicionIn}) on the previous equation, allows us to see that initial conditions $\theta^{(II)}_{\vec{k}}(\eta_{c})$ and $\theta^{'(II)}_{\vec{k}}(\eta_{c})$ compatible with the condition (\ref{normlizasgorden1}) are;

\begin{eqnarray}
&&\theta^{(II)}_{\vec{k}}(\eta_{c}) = 4 \delta^{(1)}v^{(II)+}_{\vec{k}}(\eta_{c})P^{\ast}_{(1)}(\eta_{c}) + 4 \delta^{(1)}v^{(II)-}_{\vec{k}}(\eta_{c})P_{(1)}(\eta_{c}) + 
\delta^{(0)}v^{(II)}_{\vec{k}}(\eta_{c}) [ 2a_{(2)}(\eta_{c}) - 12|P_{(1)}(\eta_{c})|^{2} ],\label{Athetanormaliza}\\
 &&\theta^{'(II)}_{\vec{k}}(\eta_{c}) = 0 \label{thetanormaliza}.
\end{eqnarray}

On the other hand, the integral (\ref{normalizacionsegorden}) for $\vec{k}' = \vec{k} + 2\vec{k}_{0}$, reduces to;

\begin{eqnarray}\label{normlizasgorden2V}
&&\bigg\{ \delta^{(0)}v^{(II)}_{\vec{k}}(\delta^{(2)}\tilde{v}^{'(II)-}_{\vec{k} + 2\vec{k}_{0}})^{\ast} + \delta^{(2)}\tilde{v}^{(II)+}_{ \vec{k} } (\delta^{(0)}v^{'(II)}_{\vec{k} + 2\vec{k}_{0} })^{\ast} + \delta^{(1)}v^{(II)+}_{\vec{k}}(\delta^{(1)}v^{'(II)-}_{\vec{k} + 2\vec{k}_{0}})^{\ast}  \nonumber\\ 
&&
-\delta^{(0)}v^{'(II)}_{\vec{k}}(\delta^{(2)}\tilde{v}^{(II)-}_{\vec{k} + 2\vec{k}_{0}})^{\ast} - \delta^{(2)}\tilde{v}^{'(II)+}_{ \vec{k} } (\delta^{(0)}v^{(II)}_{\vec{k} + 2\vec{k}_{0} })^{\ast} - \delta^{(1)}v^{'(II)+}_{\vec{k}}(\delta^{(1)}v^{(II)-}_{\vec{k} + 2\vec{k}_{0}})^{\ast}  \nonumber\\ 
&&
-4 \Big[ \delta^{(0)}v^{(II)}_{\vec{k}}(\delta^{(1)}v^{'(II)-}_{\vec{k} + 2\vec{k}_{0} })^{\ast} - \delta^{(0)}v^{'(II)}_{\vec{k}}(\delta^{(1)}v^{(II)-}_{\vec{k} + 2\vec{k}_{0} })^{\ast} \Big] P_{(1)} - 4 \Big[ \delta^{(1)}v^{(II)+}_{\vec{k}}(\delta^{(0)}v^{'(II)}_{\vec{k} + 2\vec{k}_{0}})^{\ast}  \nonumber\\
&& - \delta^{(1)}v^{'(II)+}_{\vec{k}}(\delta^{(0)}v^{(II)}_{\vec{k} + 2\vec{k}_{0}})^{\ast} \Big]P_{(1)} + \Big[ \delta^{(0)}v^{(II)}_{\vec{k}}(\delta^{(0)}v^{'(II)}_{\vec{k} + 2\vec{k}_{0} })^{\ast} - \delta^{(0)}v^{'(II)}_{\vec{k}}(\delta^{(0)}v^{(II)}_{\vec{k} + 2\vec{k}_{0}})^{\ast} \Big]\Big[ 6P_{(1)}^{2} \nonumber\\
&&- 2P_{(2)} \Big] \bigg\} \Bigg|_{\eta=\eta_{c}}= 0. 
\end{eqnarray}

By using (\ref{condicionIn}) on the previous equation, we see that the initial conditions $\delta^{(2)}\tilde{v}^{(II)+}_{\vec{k}}(\eta_{c})$, $(\delta^{(2)}\tilde{v}^{(II)-}_{\vec{k}}(\eta_{c}))^{\ast}$, $\delta^{(2)}\tilde{v}^{'(II)+}_{\vec{k}}(\eta_{c})$, and $(\delta^{(2)}\tilde{v}^{'(II)-}_{\vec{k}}(\eta_{c}))^{\ast}$ compatible with the normalization condition (\ref{normlizasgorden2V}) are;

\begin{eqnarray}
&&\delta^{(2)}\tilde{v}^{'(II)+}_{\vec{k}}(\eta_{c}) = (\delta^{(2)}\tilde{v}^{'(II)-}_{\vec{k}}(\eta_{c}))^{\ast} = 0 \label{AVanormaliza22V} \\
&&\delta^{(2)}\tilde{v}^{(II)+}_{\vec{k}}(\eta_{c}) = 4 \delta^{(1)}v^{(II)+}_{\vec{k}}(\eta_{c})P_{(1)}(\eta_{c}) + \delta^{(0)}v^{(II)}_{\vec{k}}(\eta_{c}) [2P_{(2)}(\eta_{c}) - 6P^{2}_{(1)}(\eta_{c})],\label{BVanormaliza22V} \\
&&(\delta^{(2)}\tilde{v}^{(II)-}_{\vec{k}}(\eta_{c}))^{\ast} = 4 (\delta^{(1)}v^{(II)-}_{\vec{k}}(\eta_{c}))^{\ast}P_{(1)}(\eta_{c}) + \delta^{(0)}v^{(II)\ast}_{\vec{k}}(\eta_{c}) [2P_{(2)}(\eta_{c})  
- 6P^{2}_{(1)}(\eta_{c})]\label{Vanormaliza22V}.
\end{eqnarray}

Whereas for $\vec{k}' = \vec{k} - 2\vec{k}_{0}$ from (\ref{normalizacionsegorden}) we find;

\begin{eqnarray}\label{normlizasgorden3V}
&&\bigg\{ \delta^{(0)}v^{(II)}_{\vec{k}}(\delta^{(2)}\tilde{v}^{'(II)+}_{\vec{k} - 2\vec{k}_{0}})^{\ast} + \delta^{(2)}\tilde{v}^{(II)-}_{ \vec{k} } (\delta^{(0)}v^{'(II)}_{\vec{k} - 2\vec{k}_{0} })^{\ast} + \delta^{(1)}v^{(II)-}_{\vec{k}}(\delta^{(1)}v^{'(II)+}_{\vec{k} - 2\vec{k}_{0}})^{\ast}  \nonumber\\ 
&&
-\delta^{(0)}v^{'(II)}_{\vec{k}}(\delta^{(2)}\tilde{v}^{(II)+}_{\vec{k} - 2\vec{k}_{0}})^{\ast} - \delta^{(2)}\tilde{v}^{'(II)-}_{ \vec{k} } (\delta^{(0)}v^{(II)}_{\vec{k} - 2\vec{k}_{0} })^{\ast} - \delta^{(1)}v^{'(II)-}_{\vec{k}}(\delta^{(1)}v^{(II)+}_{\vec{k} - 2\vec{k}_{0}})^{\ast}  \nonumber\\ 
&&
-4 \Big[ \delta^{(0)}v^{(II)}_{\vec{k}}(\delta^{(1)}v^{'(II)+}_{\vec{k} - 2\vec{k}_{0} })^{\ast} - \delta^{(0)}v^{'(II)}_{\vec{k}}(\delta^{(1)}v^{(II)+}_{\vec{k} - 2\vec{k}_{0} })^{\ast} \Big] P^{\ast}_{(1)} - 4 \Big[ \delta^{(1)}v^{(II)-}_{\vec{k}}(\delta^{(0)}v^{'(II)}_{\vec{k} - 2\vec{k}_{0}})^{\ast}  \nonumber\\
&& - \delta^{(1)}v^{'(II)-}_{\vec{k}}(\delta^{(0)}v^{(II)}_{\vec{k} - 2\vec{k}_{0}})^{\ast} \Big]P^{\ast}_{(1)} + \Big[ \delta^{(0)}v^{(II)}_{\vec{k}}(\delta^{(0)}v^{'(II)}_{\vec{k} - 2\vec{k}_{0} })^{\ast} - \delta^{(0)}v^{'(II)}_{\vec{k}}(\delta^{(0)}v^{(II)}_{\vec{k} - 2\vec{k}_{0}})^{\ast} \Big]\Big[ 6P_{(1)}^{\ast2} \nonumber\\
&& - 2P^{\ast}_{(2)} \Big] \bigg\} \Bigg|_{\eta=\eta_{c}}= 0. 
\end{eqnarray}

Consequently, by using (\ref{condicionIn}) on the previous equation, we find that the initial condition $(\delta^{(2)}\tilde{v}^{(II)+}_{\vec{k}}(\eta_{c}))^{\ast}$, $\delta^{(2)}\tilde{v}^{(II)-}_{\vec{k}}(\eta_{c})$, $(\delta^{(2)}\tilde{v}^{'(II)+}_{\vec{k}}(\eta_{c}))^{\ast}$, and $\delta^{(2)}\tilde{v}^{'(II)-}_{\vec{k}}(\eta_{c})$ compatible with the normalization condition (\ref{normlizasgorden3V}) are;

\begin{eqnarray}
&&(\delta^{(2)}\tilde{v}^{'(II)+}_{\vec{k}}(\eta_{c}))^{\ast} = \delta^{(2)}\tilde{v}^{'(II)-}_{\vec{k}}(\eta_{c}) = 0\label{AVanormaliza22aV}
\\
&&(\delta^{(2)}\tilde{v}^{(II)+}_{\vec{k}}(\eta_{c}))^{\ast} = 4 (\delta^{(1)}v^{(II)+}_{\vec{k}}(\eta_{c}))^{\ast}P^{\ast}_{(1)}(\eta_{c}) + \delta^{(0)}v^{(II) \ast}_{\vec{k}}(\eta_{c}) [2P^{\ast}_{(2)}(\eta_{c}) 
- 6P^{\ast2}_{(1)}(\eta_{c})],\label{BVanormaliza22aV} \\
&&\delta^{(2)}\tilde{v}^{(II)-}_{\vec{k}}(\eta_{c}) = 4 \delta^{(1)}v^{(II)-}_{\vec{k}}(\eta_{c})P^{\ast}_{(1)}(\eta_{c}) + \delta^{(0)}v^{(II)}_{\vec{k}}(\eta_{c}) [2P^{\ast}_{(2)}(\eta_{c})  
- 6P^{\ast2}_{(1)}(\eta_{c})]\label{Vanormaliza22aV}
\end{eqnarray}
Which are completely equivalent to those found in (\ref{AVanormaliza22V}), (\ref{BVanormaliza22V}) and (\ref{Vanormaliza22V}).
 
\begin{center}
\textbf{Determining the state $|\xi^{(II)}\rangle\in\mathscr{H}^{(II)}$ }
\end{center}

To conclude the construction of the SSC-II, we will find the state  $|\xi^{(II)}\rangle\in\mathscr{H}^{(II)}$ such that; 

\begin{equation}\label{EinsU2ord}
G_{\mu}\hskip.02cm^{\nu}[g^{(II)}(\boldsymbol{x})] = 8\pi G \langle\xi^{(II)}|\hat{T}_{\mu}\hskip.02cm^{\nu}[g^{(II)}(\boldsymbol{x}),\hat{\phi}^{(II)}(\boldsymbol{x}),\hat{\pi}^{(II)}(\boldsymbol{x})]|\xi^{(II)}\rangle,
\end{equation}
where the metric $g^{(II)}_{\mu\nu}(\boldsymbol{x})$ is given by (\ref{71}) and is characterized by the metric potentials $\tilde{\Phi}^{(II)}(\boldsymbol{x})$, $\tilde{\Psi}^{(II)}(\boldsymbol{x})$ and $h^{(II)}_{ij}(\boldsymbol{x})$  which have been written in (\ref{ordpertHas2Phi}), (\ref{ordpertHas2Psi}) and (\ref{ordpertHas2Ht}) as power series expansions to second order in $\varepsilon$. In other words, we aim to find the state $|\xi^{(II)}\rangle\in\mathscr{H}^{(II)}$ such that, 

\begin{eqnarray}\label{EinsU2ordclaro}
\delta^{(0)}G_{\mu}\hskip.01cm^{\nu}[\hskip.1cm^{[0]}g^{(II)}(\boldsymbol{x})] &=& 8\pi G \langle\xi^{(II)}|\delta^{(0)}\hat{T}_{\mu}\hskip.01cm^{\nu}[\hskip.1cm^{[0]}g^{(II)}(\boldsymbol{x}),\hskip.1cm^{[0]}\hat{\phi}^{(II)}(\boldsymbol{x}),\hskip.1cm^{[0]}\hat{\pi}^{(II)}(\boldsymbol{x})]|\xi^{(II)}\rangle, \nonumber\\
\delta^{(1)}G_{\mu}\hskip.01cm^{\nu}[\hskip.1cm^{[1]}g^{(II)}(\boldsymbol{x})] &=& 8\pi G \langle\xi^{(II)}|\delta^{(1)}\hat{T}_{\mu}\hskip.01cm^{\nu}[\hskip.1cm^{[1]}g^{(II)}(\boldsymbol{x}),\hskip.1cm^{[1]}\hat{\phi}^{(II)}(\boldsymbol{x}),\hskip.1cm^{[1]}\hat{\pi}^{(II)}(\boldsymbol{x})]|\xi^{(II)}\rangle, \nonumber\\
\delta^{(2)}G_{\mu}\hskip.01cm^{\nu}[\hskip.1cm^{[2]}g^{(II)}(\boldsymbol{x})] &=& 8\pi G \langle\xi^{(II)}|\delta^{(2)}\hat{T}_{\mu}\hskip.02cm^{\nu}[\hskip.1cm^{[2]}g^{(II)}(\boldsymbol{x}),\hskip.1cm^{[2]}\hat{\phi}^{(II)}(\boldsymbol{x}),\hskip.1cm^{[2]}\hat{\pi}^{(II)}(\boldsymbol{x})]|\xi^{(II)}\rangle. 
\end{eqnarray}
Meaning that we will separate and solve the problem at different orders in $\varepsilon$.

Once more,  we  should  recall  that in the above expressions,  we  need to use the  renormalized   expressions for the expectation value of the energy momentum tensor.  In  the  present case,    we  can not  expect  the   renormalization procedure   to be   simplified  by  exact  symmetries and thus in principle,  one  should have to deal   with  the   usual correction terms and  the  ambiguities  that arise in the  renormalization  procedure.
 However,  once again, these  ambiguities   would  already  be   present in  the  vacuum  expectation value.  We  know  these would  appear  as higher order  corrections  constructed  out of curvature terms, and  would  thus   be  negligibly small,  given that the  spacetime,     although not exactly symmetric   is  perturbatively  close to a highly symmetric one.  Thus,   once  more,   we can  expect  these corrections   and   ambiguities to  be  of   higher order,  except those   which  are  compatible with the  symmetries of  H  \& I  (and those  are limited to ones that can be reabsorbed in  renormalized  values of  constants  such as $ G_N$   and $\Lambda$  and dealt  with as discussed in the introduction).    Once again, the  difference  between   the  renormalized     expectation value  of the energy momentum   in the vacuum state  of the  construction, and   the   renormalized  expectation value,  in other states, is   obtained  directly  using the  normal ordering  procedure.  Once  more, this  together with  the   fact that we are using   coherent states,  reduces the    problem to that of  evaluating the corresponding  classical energy momentum tensor of the corresponding "classical  configuration" \cite{PC-Benito}.  Thus,  we might proceed  as   we did  in the construction of the H\& I  SSC of section IIA, which we do in the following.

According to the ansatz (\ref{ordpertHas2Phi}), (\ref{ordpertHas2Psi}) and (\ref{ordpertHas2Ht}), to zeroth order in $\varepsilon$ only the $\vec{k}=0$ mode contributes to the construction. At first order in $\varepsilon$, the modes 
$\vec{k}=0$ and $\pm\vec{k}_{0}$ contribute. Whereas at second order in $\varepsilon$, the modes $\vec{k}=0$, $\pm\vec{k}_{0}$ and $\pm2\vec{k}_{0}$ must  be taken to contribute. That  of course  is  a necessary but not a sufficient  condition for  the construction of the SSC-II up to second order in perturbation theory. Now, given the non linearity of subsequent orders in perturbation theory (the higher the order in $\varepsilon$ we take into account, the more cross products between perturbations we encounter. Products such as scalars$\times$scalars, scalars$\times$tensors and tensors$\times$tensors) contributions from new modes $\vec{k}$ will emerge which can be understood as integer multiples of the mode $\vec{k}_{0}$ (i.e., $\pm\vec{k}_{0}$, $\pm2\vec{k}_{0}$, $\pm3\vec{k}_{0}$,... $\pm n\vec{k}_{0}$,...). 

This motivates us to propose the following ansatz for the form of the state $|\xi^{(II)}\rangle$;

\begin{eqnarray}
&|\xi^{(II)}\rangle =& ..|0^{(II)}_{ -n\vec{k}_{n} }\rangle..|0^{(II)}_{ -n\vec{k}_{2} }\rangle\otimes..|0^{(II)}_{ -\vec{k}_{2} }\rangle  
...\otimes|0^{(II)}_{ -n\vec{k}_{1} }\rangle\otimes..|0^{(II)}_{ -2\vec{k}_{1} }\rangle\otimes|0^{(II)}_{ -\vec{k}_{1} }\rangle.. 
\otimes|\xi^{(II)}_{ -n\vec{k}_{0} }\rangle.. \otimes|\xi^{(II)}_{ -2\vec{k}_{0} }\rangle \otimes| \xi^{(II)}_{ -\vec{k}_{0} }\rangle \otimes \nonumber \\
&&|\xi^{(II)}_{0}\rangle \otimes |\xi^{(II)}_{ \vec{k}_{0}}\rangle \otimes|\xi^{(II)}_{ 2\vec{k}_{0} }\rangle.. \otimes|\xi^{(II)}_{ n\vec{k}_{0} }\rangle 
.. \otimes |0^{(II)}_{ \vec{k}_{1} }\rangle\otimes |0^{(II)}_{ 2\vec{k}_{1} }\rangle.. \otimes|0^{(II)}_{n\vec{k}_{1}}\rangle..  \otimes|0^{(II)}_{ \vec{k}_{2} }\rangle..\otimes|0^{(II)}_{ n\vec{k}_{2} }\rangle..\otimes |0^{(II)}_{ n\vec{k}_{n} }\rangle,.. 
\end{eqnarray}
Up to second order in perturbations, it is convenient  to write:

\begin{eqnarray}
&|\xi^{(II)}\rangle =& ..|0^{(II)}_{ -n\vec{k}_{n} }\rangle..|0^{(II)}_{ -n\vec{k}_{2} }\rangle\otimes..|0^{(II)}_{ -\vec{k}_{2} }\rangle  ...\otimes|0^{(II)}_{ -n\vec{k}_{1} }\rangle\otimes..|0^{(II)}_{ -2\vec{k}_{1} }\rangle 
\otimes|0^{(II)}_{ -\vec{k}_{1} }\rangle.. \otimes|0^{(II)}_{ -3\vec{k}_{0} }\rangle\otimes|\xi^{(II)}_{ -2\vec{k}_{0} }\rangle \otimes| \xi^{(II)}_{ -\vec{k}_{0} }\rangle \otimes|\xi^{(II)}_{0}\rangle \otimes \nonumber \\
&&
 |\xi^{(II)}_{ \vec{k}_{0}}\rangle \otimes|\xi^{(II)}_{ 2\vec{k}_{0} }\rangle\otimes |0^{(II)}_{ 3\vec{k}_{0} }\rangle.. \otimes|0^{(II)}_{ n\vec{k}_{0} }\rangle 
.. \otimes |0^{(II)}_{ \vec{k}_{1} }\rangle\otimes |0^{(II)}_{ 2\vec{k}_{1} }\rangle.. \otimes|0^{(II)}_{n\vec{k}_{1}}\rangle..  \otimes|0^{(II)}_{ \vec{k}_{2} }\rangle..\otimes|0^{(II)}_{ n\vec{k}_{2} }\rangle..\otimes |0^{(II)}_{ n\vec{k}_{n} }\rangle,.. 
\end{eqnarray}

For this reason, working to second order in $\varepsilon$, it is possible to consider that the general form of the state $|\xi^{(II)}\rangle$ is such, that only the modes $\vec{k}=0$, $\pm\vec{k}_{0}$ and $\pm2\vec{k}_{0}$
are excited. Whereas the rest of the modes $\vec{k}$ are found in their corresponding base states.

Now, assuming that such a state $|\xi^{(II)}\rangle$ is a highly coherent one, it will be possible to write it as :

\begin{equation}\label{EstadoexcitadoNoHI}
|\xi^{(II)}\rangle =  
\mathcal{F}(\xi^{(II)}_{-2\vec{k}_{0}} \hat{a}^{(II)\dag}_{-2\vec{k}_{0}})\mathcal{F}(\xi^{(II)}_{-\vec{k}_{0}} \hat{a}^{(II)\dag}_{-\vec{k}_{0}})  
\mathcal{F}(\xi^{(II)}_{0} \hat{a}^{(II)\dag}_{0}) \mathcal{F}(\xi^{(II)}_{\vec{k}_{0}} \hat{a}^{(II)\dag}_{\vec{k}_{0}}) \mathcal{F}(\xi^{(II)}_{2\vec{k}_{0}} \hat{a}^{(II)\dag}_{2\vec{k}_{0}})   
|0^{(II)}\rangle, 
\end{equation}
where $\mathcal{F}(\hat{X}) \propto e^{ \hat{X}}$. In this way, the state $|\xi^{(II)}\rangle$ will be determined once the numbers  $\Big\{ \xi^{(II)}_{0},\hskip.03cm \xi^{(II)}_{\pm\vec{k}_{0}},\hskip.03cm \xi^{(II)}_{\pm2\vec{k}_{0}} \Big\}$ are specified. These values are in time given by  $\xi^{(II)}_{0} = \langle\xi^{(II)}|\hat{a}^{(II)}_{0}|\xi^{(II)}\rangle$, $\xi^{(II)}_{\pm\vec{k}_{0}} =  \langle\xi^{(II)}|\hat{a}^{(II)}_{\pm\vec{k}_{0}}|\xi^{(II)}\rangle$ and $\xi^{(II)}_{\pm2\vec{k}_{0}} =  \langle\xi^{(II)}|\hat{a}^{(II)}_{\pm2\vec{k}_{0}}|\xi^{(II)}\rangle$.

Additionally,  given that according to the ansatz (\ref{ordpertHas2Phi}), (\ref{ordpertHas2Psi}), (\ref{ordpertHas2Ht}) and (\ref{Uk2eps}), along with the fact that at zeroth order in perturbations, the only excited  mode present in the problem is $\vec{k}=0$, there will be no significant deviation from an almost perfectly homogeneous and isotropic spacetime. For this reason, at zeroth order in $\varepsilon$, our setting is completely  equivalent to the construction of the SSC-I. Therefore, performing equivalent steps to those taken to find 
(\ref{expHIcomparar5}),    we can obtain the result for $\xi^{(II)}_{0}$, that is:

\begin{equation}\label{IIexpHIcomparar5}
 \xi^{(II)}_{0} \approx \sqrt{\frac{  L^{3}H^{(II)}_{0} }{ 16\pi G \epsilon^{(II)}} } = \sqrt{\frac{  L^{3}H^{(II)}_{0} }{ 2 \epsilon^{(II)} t_{p}^{2} } }, 
\end{equation}
whereas $^{(0)}\phi^{(II)}_{\xi,0}(\eta)$, will be analogous to $\phi^{(I)}_{\xi,0}(\eta)$ found in
(\ref{expHIcomparar}), but with the label $^{(II)}$ instead of $^{(I)}$, that is;

\begin{equation}\label{expHIcompararII} 
^{(0)}\phi^{(II)}_{\xi,0}(\eta) = \frac{2 \xi^{(II)}_{0} }{L^{3/2}}\sqrt{ \frac{1}{ H^{(II)}_{0} } }\Big( -H_{0}^{(II)}\eta \Big)^{ m^{2}/(3H_{0}^{2(II)}) }.
\end{equation}

Now, in order to determine the rest of the numbers $\xi^{(II)}_{\pm\vec{k}_{0}}$ and $\xi^{(II)}_{\pm2\vec{k}_{0}}$ to second order in $\varepsilon$, we will continue with the construction of the state  $|\xi^{(II)}\rangle$ to higher orders in this parameter.

Therefore, calculating $\langle\xi^{(II)}|\hat{\phi}^{(II)}(\boldsymbol{x})|\xi^{(II)}\rangle$, we find that at higher orders $\varepsilon^{n}$ (with $n\in\mathbb{N}-\{0\}$) the deviation from a spatially homogenous  and isotropic space time manifests itself. We obtain that : $\phi^{(II)}_{\xi}(\boldsymbol{x})\equiv\langle\xi^{(II)}|\hat{\phi}^{(II)}(\boldsymbol{x})|\xi^{(II)}\rangle$. That is, the expectation value of (\ref{ucuanticoU7ante}) in the state 
$|\xi^{(II)}\rangle$, can be written as:

\begin{eqnarray}\label{expectacionh2} 
\phi^{(II)}_{\xi}(\boldsymbol{x}) &=& \langle\xi^{(II)}| \sum_{\vec{k}} \Big[u_{\vec{k}}(\boldsymbol{x})\hat{a}_{\vec{k}} + u^{\ast}_{\vec{k}}(\boldsymbol{x})\hat{a}^{\dag}_{\vec{k}} \Big]|\xi^{(II)}\rangle , \hskip.2cm \textup{given that: } \xi^{(II)}_{n\vec{k}_{0}} =  \langle\xi^{(II)}|\hat{a}^{(II)}_{n\vec{k}_{0}}|\xi^{(II)}\rangle \hskip.1cm 
\nonumber\\ 
&& \textup{with} \hskip.1cm n=0,\pm1,\pm2. 
\textup{ and since the rest of the modes } 
\textup{ $\vec{k}$, such that  $\vec{k} \neq \hskip.05cm p\vec{k}_{0}$ with $p\in\{0,\pm1,\pm2\}$} \nonumber\\ 
&& \textup{are found in their corresponding base states, we obtain;}
\nonumber\\     
&&
\nonumber\\ 
\phi^{(II)}_{\xi}(\boldsymbol{x}) &=& \Big[ u_{0}(\boldsymbol{x}) \xi^{(II)}_{0} + c.c \Big] + \Big[ u_{\vec{k}_{0}}(\boldsymbol{x}) \xi^{(II)}_{\vec{k}_{0}} + c.c \Big] + \Big[ u_{2\vec{k}_{0}}(\boldsymbol{x}) \xi^{(II)}_{2\vec{k}_{0}} + c.c \Big] 
\nonumber\\ 
&+& \Big[ u_{-\vec{k}_{0}}(\boldsymbol{x}) \xi^{(II)}_{-\vec{k}_{0}} + c.c \Big] + \Big[ u_{-2\vec{k}_{0}}(\boldsymbol{x}) \xi^{(II)}_{-2\vec{k}_{0}} + c.c \Big] 
\end{eqnarray}
substituting in the previous equation $u_{\vec{k}}(\boldsymbol{x})$ from (\ref{Uk2eps})  and rearranging the coefficients of  $e^{ i\vec{k}\cdot\vec{x} }$, we find;

\begin{eqnarray}\label{Aexpectacionhcontinu2}
\phi^{(II)}_{\xi}(\boldsymbol{x}) 
&=& \phi^{(II)}_{\xi,0}(\eta) + \Big\{\hskip.1cm\Big[ \phi^{(II)}_{\xi,\vec{k}_{0}}(\eta) e^{ i\vec{k}_{0}\cdot\vec{x} } + \phi^{(II)}_{\xi,2\vec{k}_{0}}(\eta) e^{ 2i\vec{k}_{0}\cdot\vec{x} } \Big] 
\hskip.1cm + \hskip.1cm  c.c \Big\}, 
\end{eqnarray}
where the auxiliary functions $\phi^{(II)}_{\xi,0}(\eta)$, $\phi^{(II)}_{\xi,\vec{k}_{0}}(\eta)$ and $\phi^{(II)}_{\xi,2\vec{k}_{0}}(\eta)$ are given respectively by:

\begin{eqnarray}
L^{3/2} \phi^{(II)}_{\xi,0}\!(\eta) &=& \Big[\Big( \delta^{(0)}v^{(II)}_{0}\!(\eta) + \varepsilon^{2}\theta^{(II)}_{0}\!(\eta) \Big)\xi^{(II)}_{0} + \varepsilon \Big( \delta^{(1)}v^{(II)-}_{ \vec{k}_{0} }\!(\eta)\xi^{(II)}_{ \vec{k}_{0} }  + \delta^{(1)}v^{(II)+}_{ -\vec{k}_{0} }\!(\eta)\xi^{(II)}_{ -\vec{k}_{0} } \Big) + O(\varepsilon^{3})  \Big] + c.c, \label{separaexpectacionh2subcasoB0} \\
L^{3/2} \phi^{(II)}_{\xi,\vec{k}_{0}}\!(\eta) &=&  \delta^{(0)}v^{(II)}_{ \vec{k}_{0} }\!(\eta)\xi^{(II)}_{ \vec{k}_{0} } +  \delta^{(0)}v^{(II) \ast }_{ -\vec{k}_{0} }\!(\eta)\xi^{(II)\ast}_{ -\vec{k}_{0} } + \varepsilon \Big[ \delta^{(1)}v^{(II)+}_{0}\!(\eta)\xi^{(II)}_{0} + (\delta^{(1)}v^{(II)-}_{0}\!(\eta))^{\ast}\xi_{0}^{(II)\ast} \Big] + O(\varepsilon^{3}), \label{Aseparaexpectacionh2subcasoB0} \\
L^{3/2} \phi^{(II)}_{\xi,2\vec{k}_{0}}\!(\eta) &=&  \delta^{(0)}v^{(II)}_{ 2\vec{k}_{0} }\!(\eta)\xi^{(II)}_{ 2\vec{k}_{0} } + \delta^{(0)}v^{(II) \ast }_{ -2\vec{k}_{0} }\!(\eta)\xi^{(II)\ast}_{ -2\vec{k}_{0} } + \varepsilon \Big[ \delta^{(1)}v^{(II)+}_{\vec{k}_{0}}\!(\eta)\xi^{(II)}_{\vec{k}_{0}} + (\delta^{(1)}v^{(II)-}_{-\vec{k}_{0}}\!(\eta))^{\ast}\xi_{-\vec{k}_{0}}^{(II)\ast} \Big] + \nonumber \\ 
&&  \hskip.1cm \varepsilon^{2} \Big[ \delta^{(2)}\tilde{v}^{(II)+}_{0}\!(\eta)\xi^{(II)}_{0} + (\delta^{(2)}\tilde{v}^{(II)-}_{0}\!(\eta))^{\ast}\xi_{0}^{(II)\ast} \Big] + O(\varepsilon^{3}). \label{Bseparaexpectacionh2subcasoB0} 
\end{eqnarray}

Since at zeroth order in $\varepsilon$ only the $\vec{k}=0$ mode contributes to our calculation, at first order it is the  $\vec{k}=0$ and $\vec{k}_{0}$ modes that contribute, whereas at second order the $\vec{k}=0$, $\vec{k}_{0}$, $2\vec{k}_{0}$ modes contribute, means that  $\xi^{(II)}_{ \pm \vec{k}_{0} } = \varepsilon \xi^{ (II) }_{(1), \pm \vec{k}_{0}}$ (that is, $\xi^{(II)}_{ \pm \vec{k}_{0}}$ is first order in $\varepsilon$), and $\xi^{ (II) }_{ \pm 2\vec{k}_{0}} = \varepsilon^{2} \xi^{ (II) }_{(2), \pm 2\vec{k}_{0}}$ (i.e., $\xi^{(II)}_{ \pm 2\vec{k}_{0}}$ is second order in $\varepsilon$). 

For this reason, a second-order expansion in $\varepsilon$ implies that the equations (\ref{separaexpectacionh2subcasoB0}), (\ref{Aseparaexpectacionh2subcasoB0}) and (\ref{Bseparaexpectacionh2subcasoB0}), can be written respectively as:

\begin{eqnarray}
\phi^{(II)}_{\xi,0}(\eta) &=& \hskip.05cm ^{(0)}\phi^{(II)}_{\xi,0}(\eta)\varepsilon^{0} + \hskip.05cm ^{(2)}\phi^{(II)}_{\xi,0}(\eta)\varepsilon^{2} + O(\varepsilon^{3}),\label{separaexpectacionh2subcasoB} \\
\phi^{(II)}_{\xi,\vec{k}_{0}}(\eta) &=&\hskip.05cm ^{(1)}\phi^{(II)}_{\xi,\vec{k}_{0}}(\eta)\varepsilon + O(\varepsilon^{3}), \label{Aseparaexpectacionh2subcasoB} \\
\phi^{(II)}_{\xi,2\vec{k}_{0}}(\eta) &=& \hskip.05cm ^{(2)}\phi^{(II)}_{\xi,2\vec{k}_{0}}(\eta)\varepsilon^{2} + O(\varepsilon^{3}). \label{Bseparaexpectacionh2subcasoB} 
\end{eqnarray}

Therefore, Eq.~(\ref{expectacionhcontinu2}) at second order can be written as

\begin{eqnarray}\label{expectacionhcontinu2}
\phi^{(II)}_{\xi}(\boldsymbol{x}) 
&=&  \hskip.05cm ^{(0)}\phi^{(II)}_{\xi,0}(\eta)\varepsilon^{0} +  \hskip.05cm ^{(2)}\phi^{(II)}_{\xi,0}(\eta)\varepsilon^{2} + \Big\{\hskip.1cm\Big[ \hskip.05cm ^{(1)} \phi^{(II)}_{\xi,\vec{k}_{0}}(\eta) e^{ i\vec{k}_{0}\cdot\vec{x} }\varepsilon + \hskip.05cm ^{(2)}\phi^{(II)}_{\xi,2\vec{k}_{0}}(\eta) e^{ 2i\vec{k}_{0}\cdot\vec{x} }\varepsilon^{2} \Big] 
\hskip.1cm + \hskip.1cm  c.c \Big\}. 
\end{eqnarray}

\begin{center}
\textbf{ First-order expansion in $\varepsilon$ }
\end{center}
At first order in $\varepsilon$ the semiclassical Einstein equations; 

\begin{eqnarray}
&&\delta^{(1)}G_{l}\hskip.02cm^{\eta}[\hskip.1cm^{[1]}\!g^{(II)}(\boldsymbol{x})] = 8\pi G \langle\xi^{(II)}|\delta^{(1)}\hat{T}_{l}\hskip.02cm^{\eta}[\hskip.1cm^{[1]}\!g^{(II)}(\boldsymbol{x}),\hskip.1cm^{[1]}\!\hat{\phi}^{(II)}(\boldsymbol{x}),\hskip.1cm^{[1]}\!\hat{\pi}^{(II)}(\boldsymbol{x})]|\xi^{(II)}\rangle,\nonumber \\
&&\delta^{(1)}G_{\eta}\hskip.02cm^{l}[\hskip.1cm^{[1]}\!g^{(II)}(\boldsymbol{x})] = 8\pi G \langle\xi^{(II)}|\delta^{(1)}\hat{T}_{\eta}\hskip.02cm^{l}[\hskip.1cm^{[1]}\!g^{(II)}(\boldsymbol{x}),\hskip.1cm^{[1]}\!\hat{\phi}^{(II)}(\boldsymbol{x}),\hskip.1cm^{[1]}\!\hat{\pi}^{(II)}(\boldsymbol{x})]|\xi^{(II)}\rangle,\nonumber \\
&&\textup{ with  } l= x, y, z.
\end{eqnarray}

Are reduced to the semiclassical version of Eq.~(32) of  \cite{Acquaviva}, permuting $\delta^{(1)}\varphi(\boldsymbol{x})$ (meaning a first-order perturbation of the scalar field $\varphi$) by $^{(1)}\phi^{(II)}_{\xi}(\boldsymbol{x})$ (meaning the first-order contribution in $\varepsilon$ of the expectation value of the field $\hat{\phi}^{(II)}$ in the state $|\xi^{(II)}\rangle$), these equations correspond to:

\begin{equation}\label{Eze}
2\varepsilon\partial_{l}\tilde{\Psi}'_{(1)}(\boldsymbol{x}) + 2\varepsilon\mathcal{H}\partial_{l}\tilde{\Phi}_{(1)}(\boldsymbol{x}) = 8\pi G\hskip.05cm  ^{(0)}\!\phi^{'(II)}_{\xi}\partial_{l}\hskip.02cm ^{(1)}\!\phi^{(II)}_{\xi}(\boldsymbol{x}).  
\end{equation}
Using (\ref{expectacionhcontinu2}) in order to calculate the right-hand side of the previous equation. Taking into account that, without loss of generality, we have fixed $\vec{k}_{0}=(0,0,k_{0})$, 

\begin{equation}
 \partial_{l}\hskip.03cm^{(1)}\!\phi^{(II)}_{\xi}(\boldsymbol{x}) = \delta_{lz}\hskip.05cm\partial_{z}\hskip.03cm^{(1)}\!\phi^{(II)}_{\xi}(\boldsymbol{x}) = \varepsilon \Big( ik_{0z} \hskip.03cm^{(1)}\!\phi^{(II)}_{\xi,\vec{k}_{0}}(\eta) e^{ i\vec{k}_{0}\cdot\vec{x} } + c.c \Big) \delta_{lz}, 
\end{equation}
whereas using (\ref{ordpertHas2Phi}), (\ref{ordpertHas2Psi}) and (\ref{ordpertHas2Ht}) to expand the left-hand side, we finally find that equation~ (\ref{Eze}), can be written as:

\begin{eqnarray}\label{EzeFin}
&&\Big[i\Big( P'_{(1)}(\eta) + \mathcal{H}^{(II)}P_{(1)}(\eta) \Big)e^{ i\vec{k}_{0}\cdot\vec{x} } + c.c\Big]  = 4\pi G\hskip.2cm ^{(0)}\!\phi^{'(II)}_{\xi,0}\Big[i\hskip.1cm ^{(1)}\!\phi^{'(II)}_{\xi,\vec{k}_{0}}(\eta)e^{ i\vec{k}_{0}\cdot\vec{x} } + c.c\Big]. 
\end{eqnarray}
Using that the functions $e^{ i\vec{k}_{0}\cdot\vec{x} }$ and $e^{- i\vec{k}_{0}\cdot\vec{x} }$ are linearly independent, on the previous equation, we find that the set of equations.

\begin{eqnarray}
&& P'_{(1)}(\eta) + \mathcal{H}^{(II)}P_{(1)}(\eta) = 4\pi G\hskip.2cm ^{(0)}\!\phi^{'(II)}_{\xi,0}\hskip.05cm ^{(1)}\!\phi^{(II)}_{\xi,\vec{k}_{0}}(\eta),  \label{SisEzeFin}
\\
&& P^{'\ast}_{(1)}(\eta) + \mathcal{H}^{(II)}P^{\ast}_{(1)}(\eta) = 4\pi G\hskip.2cm ^{(0)}\!\phi^{'(II)}_{\xi,0}\hskip.05cm ^{(1)}\!\phi^{(II)\ast}_{\xi,\vec{k}_{0}}(\eta),
\label{ASisEzeFin} 
\end{eqnarray}
 given that $\hskip.2cm ^{(0)}\phi^{'(II)}_{\xi,0}(\eta), P_{(1)}(\eta)\in \mathbb{R}$, from the previous equation, we can conclude that $^{(1)}\phi^{'(II)}_{ \xi,\vec{k}_{0} }\in\mathbb{R}.$ No  analogous requirement appeared  in  \cite{DiezTaSudarD} working  to first order in perturbations as the  overall phase  was physically  meaningless.   However, at  second order in perturbation theory some phases  become physically  relevant.

We consider now the $jl$ components of the semiclassical Einstein equations:

\begin{equation*}
\delta^{(1)}G_{j}\hskip.02cm^{l}[\hskip.03cm^{[1]}g^{(II)}(\boldsymbol{x})] = 8\pi G \langle\xi^{(II)}|\delta^{(1)}\hat{T}_{j}\hskip.02cm^{l}[\hskip.03cm^{[1]}g^{(II)}(\boldsymbol{x}),\hskip.02cm^{[1]}\hat{\phi}^{(II)}(\boldsymbol{x}),\hskip.02cm^{[1]}\hat{\pi}^{(II)}(\boldsymbol{x})]|\xi^{(II)}\rangle, \hskip.07cm \textup{with $j\neq l$.}  
\end{equation*}
they get reduced to the semiclassical version of Equation~(34) in reference  \cite{Acquaviva},  that is:

\begin{equation}
\partial_{j}\partial^{l}(\tilde{\Psi}_{(1)}(\boldsymbol{x}) - \tilde{\Phi}_{(1)}(\boldsymbol{x}) ) = 0.
\end{equation}
Now, since the metric potentials:  $\tilde{\Psi}_{(1)}(\boldsymbol{x})$ and $\tilde{\Phi}_{(1)}(\boldsymbol{x})$ depend only on 
$(\eta,z)$, we conclude that the previous equation is trivially satisfied. As it can be understood  as a mixed partial derivative (in the spatial coordinates) from first-order contributions of the metric potentials. Therefore, mixed partial derivatives (of the spatial coordinates): $\partial_{j}\partial_{l}\tilde{\Psi}_{(1)}$ and $\partial_{j}\partial_{l}\tilde{\Phi}_{(1)}$, will be identically zero independently from one another : $\partial_{j}\partial_{l}\tilde{\Psi}_{(1)} = \partial_{j}\partial_{l}\tilde{\Phi}_{(1)} = 0$, for $j\neq l$ with $j,l = x,y,z$.

On the other hand, we now consider the $\eta\eta$ components of the semiclassical Einstein equations. This corresponds to the equation:

\begin{equation}
\delta^{(1)}G_{\eta}\hskip.02cm^{\eta}[\hskip.03cm^{[1]}g^{(II)}(\boldsymbol{x})] = 8\pi G \langle\xi^{(II)}|\delta^{(1)}\hat{T}_{\eta}\hskip.02cm^{\eta}[\hskip.03cm^{[1]}g^{(II)}(\boldsymbol{x}),\hskip.03cm^{[1]}\hat{\phi}^{(II)}(\boldsymbol{x}),\hskip.03cm^{[1]}\hat{\pi}^{(II)}(\boldsymbol{x})]|\xi^{(II)}\rangle, 
\end{equation}
this equation is analogous to the semiclassical version of equation~(30) in \cite{Acquaviva};

\begin{eqnarray}
&&2\nabla^{2} \tilde{\Psi}_{(1)}(\boldsymbol{x}) - 6\mathcal{H}^{(II)}\tilde{\Psi}'_{(1)}(\boldsymbol{x}) = 8\pi G\Big\{\hskip.2cm ^{(0)}\!\phi^{'(II)}_{\xi,0}\Big[\hskip.1cm ^{(1)}\!\phi^{'(II)}_{\xi,\vec{k}_{0}}(\eta)e^{ i\vec{k}_{0}\cdot\vec{x} } + c.c  \Big]  + \nonumber \\ 
&&  \hskip.05cm a^{2(II)} m^{2} \hskip.09cm ^{(0)}\!\phi^{2(II)}_{\xi,0} \tilde{\Phi}_{(1)}(\boldsymbol{x}) + a^{2(II)} m^{2}   \hskip.09cm ^{(0)}\!\phi^{(II)}_{\xi,0} \Big(  \hskip.05cm ^{(1)}\!\phi^{(II)}_{\xi,\vec{k}_{0}}(\eta)e^{ i\vec{k}_{0}\cdot\vec{x} } + c.c \Big) \Big\}.
\end{eqnarray}
By using (\ref{ordpertHas2Phi}), (\ref{ordpertHas2Psi}) and (\ref{ordpertHas2Ht}) to expand the previous equation, we find:

\begin{eqnarray}
&&-k_{0}^{2}\Big( P_{(1)}(\eta) e^{ i\vec{k}_{0}\cdot\vec{x} } + c.c\Big)  - 3\mathcal{H}^{(II)} \Big( P'_{(1)}(\eta) e^{ i\vec{k}_{0}\cdot\vec{x} } + c.c\Big) = 8\pi G\Big\{\hskip.05cm ^{(0)}\!\phi^{'(II)}_{\xi,0} \Big[\hskip.05cm ^{(1)}\!\phi^{'(II)}_{\xi,\vec{k}_{0}}(\eta)e^{ i\vec{k}_{0}\cdot\vec{x} } + c.c  \Big] \nonumber \\ 
&&\hskip.05cm + a^{2(II)} m^{2}   \hskip.09cm ^{(0)}\!\phi^{2(II)}_{\xi,0} \Big( P_{(1)}(\eta) e^{ i\vec{k}_{0}\cdot\vec{x} } + c.c\Big) + a^{2(II)} m^{2}   \hskip.09cm ^{(0)}\!\phi^{(II)}_{\xi,0}\Big(  \hskip.05cm ^{(1)}\!\phi^{(II)}_{\xi,\vec{k}_{0}}(\eta)e^{ i\vec{k}_{0}\cdot\vec{x} } + c.c \Big) \Big\}.  
\end{eqnarray}
By regrouping the coefficients of the functions $\{ e^{ i\vec{k}_{0}\cdot\vec{x} }, e^{ - i\vec{k}_{0}\cdot\vec{x} } \}$, and taking into account that the functions $e^{ \pm i\vec{k}_{0}\cdot\vec{x} }$ are linearly independent, we then obtain respectively that the coefficients of  $e^{ i\vec{k}_{0}\cdot\vec{x} }$ and $e^{-i\vec{k}_{0}\cdot\vec{x}}$ imply that;

\begin{eqnarray}
&& 3\mathcal{H}^{(II)}P'_{(1)}(\eta) + k_{0}^{2}P_{(1)}(\eta) = -4\pi G \Big[\hskip.01cm ^{(0)}\!\phi^{'(II)}_{\xi,0} \hskip.04cm ^{(1)}\!\phi^{'(II)}_{\xi,\vec{k}_{0}}(\eta) + a^{2(II)} m^{2} \Big( \hskip.02cm ^{(0)}\!\phi^{2(II)}_{\xi,0}P_{(1)}(\eta) + \hskip.02cm ^{(0)}\!\phi^{(II)}_{\xi,0}\hskip.04cm ^{(1)}\!\phi^{(II)}_{\xi,\vec{k}_{0}}(\eta) \Big) \Big], \label{SisEzeFnn}\\
&&3\mathcal{H}^{(II)}P'^{\ast}_{(1)}(\eta) + k_{0}^{2}P^{\ast}_{(1)}(\eta) = -4\pi G \Big[ \hskip.005cm ^{(0)}\!\phi^{'(II)}_{\xi,0} \hskip.005cm ^{(1)}\!\phi^{'(II)\ast}_{\xi,\vec{k}_{0}}(\eta) + a^{2(II)} m^{2} \Big( \hskip.005cm ^{(0)}\!\phi^{2(II)}_{\xi,0}P^{\ast}_{(1)}(\eta) + \hskip.005cm ^{(0)}\!\phi^{(II)}_{\xi,0}\hskip.005cm ^{(1)}\!\phi^{(II)\ast}_{\xi,\vec{k}_{0}}(\eta) \Big) \Big].\label{ASisEzeFnn} 
\end{eqnarray}
We conclude that these equations are the complex conjugate of each other, therefore, they correspond to the same equation.

\begin{center}
\textbf{Dynamical equation for the function $P_{(1)}(\eta)$ }
\end{center}
Considering the $ll$ components of the semiclassical Einstein equations,

\begin{equation}\label{Dinprima}
\delta^{(1)}G_{l}\hskip.02cm^{l}[\hskip.03cm^{[1]}\!g^{(II)}(\boldsymbol{x})] = 8\pi G \langle\xi^{(II)}|\delta^{(1)}\hat{T}_{l}\hskip.02cm ^{l}[\hskip.03cm^{[1]}\!g^{(II)}(\boldsymbol{x}),\hskip.03cm^{[1]}\hat{\phi}^{(II)}(\boldsymbol{x}),\hskip.03cm^{[1]}\hat{\pi}^{(II)}(\boldsymbol{x})]|\xi^{(II)}\rangle. 
\end{equation}
 
Despite the fact that the symmetries of the problem are such, that the metric potentials $\tilde{\Phi}_{(1)}(\boldsymbol{x})$ and $\tilde{\Psi}_{(1)}(\boldsymbol{x})$ only depend on the $(\eta,z)$ variables, the components $xx$, $yy$, $zz$ of equation~(\ref{Dinprima}) are identical to each other and get reduced to the semiclassical version of equation~(34) of  \cite{Acquaviva}. That is :

\begin{eqnarray}
&&\tilde{\Psi}''_{(1)}(\boldsymbol{x}) + \Big(2\frac{ a^{''(II)} }{ a^{(II)} } - \mathcal{H}^{2} \Big)( \tilde{\Psi}_{(1)}(\boldsymbol{x}) + \tilde{\Phi}_{(1)}(\boldsymbol{x}) )  + \mathcal{H}( 2\tilde{\Psi}'_{(1)}(\boldsymbol{x}) + \tilde{\Phi}'_{(1)}(\boldsymbol{x}) ) = 4\pi G \Big[ \hskip.001cm -\hskip.001cm^{(0)}\!\phi^{'2(II)}_{\xi,0}\tilde{\Phi}_{(1)}(\boldsymbol{x}) + \hskip.005cm ^{(0)}\!\phi^{'(II)}_{\xi,0} \Big( \nonumber\\ && ^{(1)}\!\phi^{'(II)}_{ \xi,\vec{k}_{0} }(\eta)e^{ i\vec{k}_{0}\cdot\vec{x} } 
+ c.c \Big)  -  \Big( \hskip.005cm ^{(0)}\!\phi^{'2(II)}_{ \xi,0} - a^{2(II)}m^{2}\hskip.01cm ^{(0)}\!\phi^{2(II)}_{ \xi,0} \Big)\tilde{\Psi}_{(1)}(\boldsymbol{x}) 
- a^{2(II)}m^{2}\hskip.005cm ^{(0)}\!\phi^{(II)}_{ \xi,0}\Big( \hskip.005cm ^{(1)}\!\phi^{(II)}_{ \xi,\vec{k}_{0} }(\eta)e^{ i\vec{k}_{0}\cdot\vec{x} } + c.c \Big) \Big],
\end{eqnarray}
after substituting (\ref{ordpertHas2Phi}), (\ref{ordpertHas2Psi}) and (\ref{ordpertHas2Ht}) on the previous equation and keeping terms to first order in $\varepsilon$ we find:

\begin{eqnarray}\label{dinamica1e}
&& \Big[ P''_{(1)}(\eta) + 3\mathcal{H}P'_{(1)}(\eta) + 2\Big( 2 \frac{ a^{''(II)} }{ a^{(II)} } - \mathcal{H}^{2} \Big)P_{(1)}(\eta) \Big]e^{ i\vec{k}_{0}\cdot\vec{x} } + c.c = 4\pi G \bigg[ \hskip.05cm-\hskip.05cm ^{(0)}\!\phi^{'2(II)}_{\xi,0} \Big( P_{(1)}(\eta)e^{ i\vec{k}_{0}\cdot\vec{x} } + c.c \Big) \nonumber\\
&&  + \hskip.05cm ^{(0)}\!\phi^{'(II)}_{\xi,0} \Big( \hskip.05cm ^{(1)}\!\phi^{'(II)}_{ \xi,\vec{k}_{0} }(\eta)e^{ i\vec{k}_{0}\cdot\vec{x} } + c.c \Big) 
- \Big( \hskip.05cm ^{(0)}\!\phi^{'2(II)}_{\xi,0} - a^{2(II)}m^{2}\hskip.09cm ^{(0)}\!\phi^{2(II)}_{\xi,0} \Big)\Big( P_{(1)}(\eta)e^{ i\vec{k}_{0}\cdot\vec{x} } + c.c \Big)  \nonumber \\
&& - a^{2(II)}m^{2}\hskip.05cm ^{(0)}\!\phi^{(II)}_{ \xi,0}\Big( \hskip.05cm ^{(1)}\!\phi^{(II)}_{ \xi,\vec{k}_{0} }(\eta)e^{ i\vec{k}_{0}\cdot\vec{x} } + c.c \Big) \bigg]. 
\end{eqnarray}
By using the linear independence of the functions $e^{ i\vec{k}_{0}\cdot\vec{x} }$ and $e^{ -i\vec{k}_{0}\cdot\vec{x} }$, we obtain the following equations which are the complex conjugate of each other:

\begin{center}
\textbf{The coefficient of $e^{i\vec{k}_{0}\cdot\vec{x}}$}
\end{center}
Regrouping the coefficients of $e^{i\vec{k}_{0}\cdot\vec{x}}$ in the previous equation, we find:

\begin{eqnarray}
&&P''_{(1)}(\eta) + 3\mathcal{H}P'_{(1)}(\eta) + 2\Big( 2 \frac{ a^{''(II)} }{ a^{(II)} } - \mathcal{H}^{2} \Big)P_{(1)}(\eta) = 4\pi G \bigg[ \hskip.041cm-\hskip.041cm^{(0)}\!\phi^{'2(II)}_{\xi,0}P_{(1)}(\eta) + \hskip.09cm ^{(0)}\!\phi^{'(II)}_{\xi,0}\hskip.09cm ^{(1)}\!\phi^{'(II)}_{ \xi,\vec{k}_{0} }(\eta) \nonumber \\
&& - \Big( \hskip.05cm ^{(0)}\!\phi^{'2(II)}_{ \xi,0} - a^{2(II)}m^{2}\hskip.09cm ^{(0)}\!\phi^{2(II)}_{ \xi,0} \Big) P_{(1)}(\eta) - a^{2(II)}m^{2}\hskip.09cm ^{(0)}\!\phi^{(II)}_{ \xi,0}\hskip.09cm ^{(1)}\!\phi^{(II)}_{ \xi,\vec{k}_{0} }(\eta) \bigg]. \end{eqnarray}
By using the definition $\mathcal{H}^{(II)} = \frac{ a^{'(II)} }{ a^{(II)} },$ we can see that $2\mathcal{H}^{'(II)} + \mathcal{H}^{2(II)} = 2\frac{ a^{''(II)} }{ a^{(II)} } - \mathcal{H}^{2}$. Hence the previous equation can be written as

\begin{eqnarray}\label{1Pmasdin}
&&  P''_{(1)}(\eta) + 3\mathcal{H}P'_{(1)}(\eta) + 2(2\mathcal{H}^{'(II)} + \mathcal{H}^{2(II)})P_{(1)}(\eta) 
= 4\pi G \bigg[ \hskip.041cm-2\hskip.041cm^{(0)}\!\phi^{'2(II)}_{\xi,0}P_{(1)}(\eta) + \hskip.05cm ^{(0)}\!\phi^{'(II)}_{\xi,0}\hskip.05cm ^{(1)}\!\phi^{'(II)}_{ \xi,\vec{k}_{0} }(\eta)  \nonumber \\
&& \hskip.05cm + \hskip.05cm a^{2(II)}m^{2}\hskip.05cm ^{(0)}\!\phi^{(II)}_{ \xi,0} \Big( \hskip.05cm ^{(0)}\!\phi^{(II)}_{ \xi,0}P_{(1)}(\eta)  - \hskip.05cm ^{(1)}\!\phi^{(II)}_{ \xi,\vec{k}_{0} }(\eta) \Big)  \bigg]. 
\end{eqnarray}

\begin{center}
\textbf{Coefficient of $e^{-i\vec{k}_{0}\cdot\vec{x}}$}
\end{center}
Proceeding in a similar manner as in the previous case, that is, working with the coefficient of  $e^{-i\vec{k}_{0}\cdot\vec{x}}$ we find; 

\begin{eqnarray}\label{1Pmendin}
  &&P^{''\ast}_{(1)}(\eta) + 3\mathcal{H}P^{'\ast}_{(1)}(\eta) + 2(2\mathcal{H}^{'(II)} + \mathcal{H}^{2(II)})P^{\ast}_{(1)}(\eta) 
= 4\pi G \bigg[ \hskip.05cm-2\hskip.05cm^{(0)}\!\phi^{'2(II)}_{\xi,0}P^{\ast}_{(1)}(\eta) + \hskip.05cm ^{(0)}\!\phi^{'(II)}_{\xi,0}\hskip.09cm ^{(1)}\!\phi^{'(II)\ast}_{ \xi,\vec{k}_{0} }(\eta)  \nonumber \\
&& \hskip.05cm + \hskip.05cm  a^{2(II)}m^{2}\hskip.09cm ^{(0)}\!\phi^{(II)}_{ \xi,0} \Big( \hskip.05cm ^{(0)}\!\phi^{(II)}_{ \xi,0}P^{\ast}_{(1)}(\eta)  - \hskip.05cm ^{(1)}\!\phi^{(II)\ast}_{ \xi,\vec{k}_{0} }(\eta)  \Big)  \bigg],
\end{eqnarray}
which is the complex conjugate of equation~(\ref{1Pmasdin}).

\underline{\textbf{Remark:}} The pair of equations~(\ref{SisEzeFin})-(\ref{ASisEzeFin}) are the complex conjugate of each other. In the same way, equations:  (\ref{SisEzeFnn})-(\ref{ASisEzeFnn}) and (\ref{1Pmasdin})-(\ref{1Pmendin}) are also one the complex conjugate of the other. Now, since we have defined the functions  $P_{(1)}(\eta)$ and $^{(0)}\phi^{(II)}_{\xi,0}(\eta)$, as functions taking values over the set of real numbers, the compatibility of these equations is guaranteed if the quantity $^{(1)}\phi^{(II)}_{ \xi,\vec{k}_{0} }(\eta)$ is defined over the set of real numbers.
 
In order to verify the requirement $\hskip.1cm ^{(1)}\phi^{(II)}_{ \xi,\vec{k}_{0} }(\eta) \in\mathbb{R}$, we proceed in the following manner: in the setting of this problem, we have constructed $\hskip.1cm^{(0)}\phi^{(II)}_{\xi,0}(\eta)$ as a function giving real values, (see equation~(\ref{separaexpectacionh2subcasoB})). Whereas condition $\hskip.2cm ^{(1)}\phi^{(II)}_{ \xi,\vec{k}_{0} }(\eta) \in\mathbb{R}$, arises as a requirement so that each pair of equations forming the systems (\ref{SisEzeFin})-(\ref{ASisEzeFin}),  (\ref{SisEzeFnn})-(\ref{ASisEzeFnn}) and (\ref{1Pmasdin})-(\ref{1Pmendin}), are compatible.

Therefore, we ought to verify that our construction of $\hskip.01cm ^{(1)}\phi^{(II)}_{ \xi,\vec{k}_{0} }(\eta)$, (see equation (\ref{Aseparaexpectacionh2subcasoB})), admits that this function is real. Working with the definition 
$\hskip.002cm ^{(1)}\phi^{(II)}_{ \xi,\vec{k}_{0} }(\eta) \varepsilon = \phi^{(II)}_{ \xi,\vec{k}_{0} }(\eta)$;

\begin{eqnarray}\label{Imagenreal}
L^{3/2}\hskip.04cm ^{(1)}\phi^{(II)}_{\xi,\vec{k}_{0}}\!(\eta)\varepsilon = \varepsilon \delta^{(0)}v^{(II)}_{ \vec{k}_{0} }\!(\eta)\xi^{(II)}_{(1), \vec{k}_{0} }  +  \varepsilon \delta^{(0)}v^{(II) \ast }_{ -\vec{k}_{0} }\!(\eta)\xi^{(II)\ast}_{(1), -\vec{k}_{0} }   + \varepsilon \Big[ \delta^{(1)}v^{(II)+}_{0}\!(\eta)\xi^{(II)}_{0} + (\delta^{(1)}v^{(II)-}_{0}\!(\eta))^{\ast}\xi_{0}^{(II)\ast} \Big], 
\end{eqnarray}
we observe that in principle, this variable takes on complex values.
However, the second term in (\ref{Imagenreal}) can be analyzed in the following manner: we had previously found in  (\ref{primperturIIceromodo}) and (\ref{primperturIIceromodo2}) that the equations of motion for the quantities $\delta^{(1)}v^{(II)+}_{0}(\eta)$ and $\delta^{(1)}v^{(II)-}_{0}(\eta)$ are identical. Whereas in (\ref{condicionIn}) there was an initial condition in $\eta=\eta_{c}$ for $\delta^{(1)}v^{(II)+}_{\vec{k}}(\eta)$, $\delta^{(1)}v^{'(II)+}_{\vec{k}}(\eta)$, $\delta^{(1)}v^{(II)-}_{\vec{k}}(\eta)$ and $\delta^{(1)}v^{'(II)-}_{\vec{k}}(\eta)$, 
compatible with the contribution at first order in $\varepsilon$ of the normalization condition induced by the symplectic product. Using (\ref{condicionIn}) for the mode $\vec{k}=0$ 
we find; $\delta^{(1)}v^{(II)+}_{0}(\eta_{c}) = \delta^{(1)}v^{(II)-}_{0}(\eta_{c}) = 4\delta^{(0)}v^{(II)}_{0}(\eta_{c})P_{(1)}(\eta_{c})$, along with $\delta^{(1)}v^{'(II)+}_{0}(\eta_{c}) = \delta^{(1)}v^{'(II)-}_{0}(\eta_{c}) = 0$. As consequence, the solution $\{$ $\delta^{(1)}v^{(II)+}_{0}(\eta)$,  $\delta^{(1)}v^{(II)-}_{0}(\eta)$ $\}$ of the systems (\ref{primperturIIceromodo}) and (\ref{primperturIIceromodo2}) that satisfy the previous initial conditions by the unitarity and existence theorem, we find that $\delta^{(1)}v^{(II)+}_{0}(\eta) = \delta^{(1)}v^{(II)-}_{0}(\eta)$. Therefore the second term in (\ref{Imagenreal}) can be written as;

\begin{eqnarray}\label{Resultado1}
 \delta^{(1)}v^{(II)+}_{0}(\eta)\xi_{0} + (\delta^{(1)}v^{(II)-}_{0}(\eta))^{\ast}\xi_{0}^{\ast} &=&  \delta^{(1)}v^{(II)+}_{0}(\eta)\xi_{0} + ( \delta^{(1)}v^{(II)+}_{0}(\eta)\xi_{0} )^{\ast}
 = 2\mathcal{R}e[ \delta^{(1)}v^{(II)+}_{0}(\eta)\xi_{0} ] \nonumber \\
 &=& 2\mathcal{R}e[(\delta^{(1)}v^{(II)-}_{0}(\eta))^{\ast}\xi_{0}^{\ast}] \in \mathbb{R}, 
\end{eqnarray}
whereas the first term, the function  $\delta^{(0)}v^{(II)}_{ \vec{k}_{0} }$ derived in (\ref{ceroperturBDII}) is;

\begin{equation}\label{ceroperturBDII1}
\delta^{(0)}v^{(II)}_{\vec{k}}(\eta) \approx \sqrt{ \frac{1}{2k} }\Big(-H_{0}^{(II)}\eta\Big) \Big( 1 - \frac{i}{k\eta} \Big) e^{-ik\eta}, \hskip.3cm \textup{ for }  k\neq0, 
\end{equation}
we can see that the change $\vec{k} \rightarrow -\vec{k}$, in the previous equation corresponds to:

\begin{equation}\label{ceroperturBDII1a}
\delta^{(0)}v^{(II)}_{-\vec{k}}(\eta) \approx \sqrt{ \frac{1}{2k} }\Big(-H_{0}^{(II)}\eta\Big) \Big( 1 - \frac{i}{k\eta} \Big) e^{-ik\eta} \Rightarrow \delta^{(0)}v^{(II)}_{\vec{k}}(\eta) = \delta^{(0)}v^{(II)}_{-\vec{k}}(\eta),
\end{equation}
that is; given that $\delta^{(0)}v^{(II)}_{\vec{k}}(\eta)$ only depends on the module of the vector $\vec{k}$, it is then invariant with the change  $\vec{k}$ by $-\vec{k}$. Now, by imposing that $\xi^{(II)}_{(1), \vec{k}_{0} } = \xi^{(II)}_{(1), -\vec{k}_{0} }$, we obtain that the first term in  (\ref{Imagenreal}) is also real. This  condition  was   used in dealing with  equations (58a) and (58b) in \cite{DiezTaSudarD}, although it is not   required   due to the   un-physical nature of overall phases in that treatment. Here on the other hand,  one  needs to  take  proper care of these aspects   in order to achieve  a   self-consistent   construction of the SSC-II, to second order. That is:

\begin{eqnarray}\label{Resultado2}
\delta^{(0)}v^{(II)}_{ \vec{k}_{0} }(\eta)\xi^{(II)}_{(1), \vec{k}_{0} } +  \delta^{(0)}v^{(II) \ast }_{ -\vec{k}_{0} }(\eta)\xi^{(II)\ast}_{(1), -\vec{k}_{0} }  &=& \delta^{(0)}v^{(II)}_{ \vec{k}_{0} }(\eta)\xi^{(II)}_{(1), \vec{k}_{0} } + ( \delta^{(0)}v^{(II)}_{ \vec{k}_{0} }(\eta)\xi^{(II)}_{(1), \vec{k}_{0} } )^{\ast} = 2\mathcal{R}e[ \delta^{(0)}v^{(II)}_{ \vec{k}_{0} }(\eta)\xi^{(II)}_{(1), \vec{k}_{0} } ] \nonumber\\
&=& 2\mathcal{R}e[ \delta^{(0)}v^{(II) \ast }_{ -\vec{k}_{0} }(\eta)\xi^{(II)\ast}_{(1), -\vec{k}_{0} } ] \in \mathbb{R}. 
\end{eqnarray}
Finally, by rearranging the results  (\ref{Resultado1}) and (\ref{Resultado2}), we can see that the expression of $^{(1)}\phi^{(II)}_{\xi,\vec{k}_{0}}(\eta)$ presented in (\ref{Imagenreal}) can be understood as the sum of quantities defined over the set of real numbers. Therefore, $^{(1)}\phi^{(II)}_{\xi,\vec{k}_{0}}(\eta)\in\mathbb{R}$. 

In summary, the system of equations governing the evolution of the function $P_{(1)}$ is given by (\ref{SisEzeFin}), (\ref{SisEzeFnn}) and (\ref{1Pmasdin}), that is;

\begin{eqnarray}
&&P'_{(1)}(\eta) + \mathcal{H}^{(II)}P_{(1)}(\eta) = 4\pi G\hskip.05cm ^{(0)}\!\phi^{'(II)}_{\xi,0}\hskip.09cm ^{(1)}\!\phi^{(II)}_{\xi,\vec{k}_{0}}(\eta), \label{constricc1} 
\\
&&3\mathcal{H}^{(II)}P'_{(1)}(\eta) + k_{0}^{2} P_{(1)}(\eta) = -4\pi G \Big[\hskip.005cm ^{(0)}\!\phi^{'(II)}_{\xi,0} \hskip.01cm ^{(1)}\!\phi^{'(II)}_{\xi,\vec{k}_{0}}(\eta) + a^{2(II)} m^{2} \Big( \hskip.005cm ^{(0)}\!\phi^{2(II)}_{\xi,0}P_{(1)}(\eta) + \hskip.005cm ^{(0)}\!\phi^{(II)}_{\xi,0}\hskip.01cm ^{(1)}\!\phi^{(II)}_{\xi,\vec{k}_{0}}(\eta) \Big) \Big], \label{constricc2} 
\\
&&P''_{(1)}(\eta) + 3\mathcal{H}P'_{(1)}(\eta) + 2(2\mathcal{H}^{'(II)} + \mathcal{H}^{2(II)})P_{(1)}(\eta) = 4\pi G \bigg[ \hskip.05cm - 2\hskip.05cm ^{(0)}\!\phi^{'2(II)}_{\xi,0}P_{(1)}(\eta) + \hskip.05cm ^{(0)}\!\phi^{'(II)}_{\xi,0}\hskip.05cm ^{(1)}\!\phi^{'(II)}_{ \xi,\vec{k}_{0} }(\eta)  \nonumber \\
&& + a^{2(II)}m^{2}\hskip.01cm ^{(0)}\!\phi^{(II)}_{ \xi,0} \Big( \hskip.05cm ^{(0)}\!\phi^{(II)}_{ \xi,0}P_{(1)}(\eta)  - \hskip.05cm ^{(1)}\!\phi^{(II)}_{ \xi,\vec{k}_{0} }(\eta) \Big)  \bigg]. \label{movimientoNH} 
\end{eqnarray}

Equations (\ref{constricc1}) and (\ref{constricc2}) are restrictions from the theory on the quantities $P_{(1)}$, $^{(1)}\phi^{(II)}_{\xi,\vec{k}_{0}}$ and its first derivatives with respect to conformal time.  Whilst (\ref{movimientoNH}), is the dynamical equation for the quantity $P_{(1)}$. Restrictions (\ref{constricc1}), (\ref{constricc2}) can be used to write $\hskip.2cm ^{(1)}\phi^{(II)}_{ \xi,\vec{k}_{0} }\hskip.1cm$ and $\hskip.1cm ^{(1)}\phi^{'(II)}_{ \xi,\vec{k}_{0} }$ as functions of $P_{(1)}$ and $P'_{(1)}$. The resulting expressions can be in time, substituted into equation (\ref{movimientoNH}). In this manner, the dynamical equation for  $P_{(1)}$ can be written as a homogeneous equation:

\begin{eqnarray}\label{movimiento1}
&&P''_{(1)}(\eta) + 2 \bigg( 3\mathcal{H}^{(II)} + \frac{ a^{ 2(II) } m^{2}\hskip.05cm ^{(0)}\!\phi^{ (II) }_{ \xi,0 } }{ ^{(0)}\!\phi^{'(II)}_{\xi,0} } \bigg) P'_{(1)}(\eta) + \bigg( k_{0}^{2} + 4\mathcal{H}^{ '(II) } + 2\mathcal{H}^{2(II)} + 8\pi G \hskip.05cm ^{(0)}\!\phi^{'2(II)}_{\xi,0} \nonumber\\ 
&& + \frac{  2a^{ 2(II) } m^{ 2 }\hskip.05cm ^{(0)}\!\phi^{(II)}_{\xi,0}  }{ ^{(0)}\!\phi^{'(II)}_{\xi,0} }\mathcal{H}^{(II)} \bigg)P_{(1)}(\eta) = 0.
\end{eqnarray}
The previous system of equations is valid for $\eta>\eta_{c}.$ However, starting from (\ref{constricc1}) and (\ref{constricc2}) evaluated at $\eta = \eta_{c}$,  we obtain a system of two algebraic linear equations with two unknown quantities $P_{(1)}(\eta_{c})$ and $P'_{(1)}(\eta_{c})$. The solution for this system renders $P_{(1)}(\eta_{c})$ and $P'_{(1)}(\eta_{c})$ in terms of the initial conditions $^{(1)}\phi^{(II)}_{\xi,\vec{k}_{0}}(\eta_{c})$ and $^{(1)}\phi^{'(II)}_{\xi,\vec{k}_{0}}(\eta_{c})$, that is;

\begin{eqnarray}
-\frac{P_{(1)}(\eta_{c})}{4\pi G} &=& \frac{ \hskip.001cm ^{(0)}\!\phi^{'(II)}_{\xi,0}(\eta_{c}) \hskip.01cm ^{(1)}\!\phi^{'(II)}_{\xi,\vec{k}_{0}}(\eta_{c})  + \Big( a^{2(II)}(\eta_{c}) m^{2}\hskip.01cm ^{(0)}\!\phi^{(II)}_{\xi,0}(\eta_{c}) + 3\mathcal{H}^{ (II) }(\eta_{c})\hskip.01cm ^{(0)}\!\phi^{'(II)}_{\xi,0}(\eta_{c})  \Big)\hskip.01cm ^{(1)}\!\phi^{(II)}_{\xi,\vec{k}_{0}}(\eta_{c})   }{ k_{0}^{2} + 4\pi G a^{2(II)}(\eta_{c})m^{2} \hskip.01cm ^{(0)}\!\phi^{2(II)}_{\xi,0}(\eta_{c}) - 3\mathcal{H}^{ 2(II) }(\eta_{c})  },  
\label{condicionesPyPdA} \\
P'_{(1)}(\eta_{c})&=& \frac{ \hskip.001cm ^{(0)}\!\phi^{'(II)}_{\xi,0}(\eta_{c}) \hskip.01cm ^{(1)}\!\phi^{'(II)}_{\xi,\vec{k}_{0}}(\eta_{c})  + \Big( a^{2(II)}(\eta_{c}) m^{2}\hskip.01cm ^{(0)}\!\phi^{(II)}_{\xi,0}(\eta_{c}) + 3\mathcal{H}^{ (II) }(\eta_{c})\hskip.01cm ^{(0)}\!\phi^{'(II)}_{\xi,0}(\eta_{c})  \Big)\hskip.01cm ^{(1)}\!\phi^{(II)}_{\xi,\vec{k}_{0}}(\eta_{c})   }{ (4\pi G \mathcal{H}^{ (II) }(\eta_{c}))^{-1} \Big(k_{0}^{2} + 4\pi G a^{2(II)}(\eta_{c})m^{2} \hskip.01cm ^{(0)}\!\phi^{2(II)}_{\xi,0}(\eta_{c}) - 3\mathcal{H}^{ 2(II) }(\eta_{c}) \Big) } \nonumber\\
&& + \hskip.05cm 4\pi G\hskip.05cm ^{(0)}\!\phi^{'(II)}_{\xi,0}\hskip.09cm ^{(1)}\!\phi^{(II)}_{\xi,\vec{k}_{0}}(\eta). \label{condicionesPyPd}
\end{eqnarray}

Conditions (\ref{condicionesPyPdA}) and (\ref{condicionesPyPd}) determine a unique particular solution of ecuation (\ref{movimiento1}).
Now, regarding the numbers $\xi^{(II)}_{(1),\pm\vec{k}_{0}}$, which at linear order in $\varepsilon$ 
determine the state $|\xi^{(II)}\rangle$, they can be fixed through the following recipe:
\\

\textbf{Step (1):} We know the initial conditions $^{(1)}\phi^{(II)}_{\xi,\vec{k}_{0}}(\eta_{c})$ and $^{(1)}\phi^{'(II)}_{\xi,\vec{k}_{0}}(\eta_{c})$, then by (\ref{condicionesPyPdA}) and (\ref{condicionesPyPd}) the initial values $P_{(1)}(\eta_{c})$ and $P'_{(1)}(\eta_{c})$ for the equation (\ref{movimiento1}) are generated. In this way, we determine the function $P_{(1)}(\eta)$ uniquely.
\\

\textbf{Step (2):} From  $P_{(1)}(\eta_{c})$ the system (\ref{condicionIn}) allows to determine: 
$\delta^{(1)}v^{(II)\pm}_{\vec{k}}(\eta_{c})$ and $\delta^{(1)}v^{'(II)\pm}_{\vec{k}}(\eta_{c})$, that is; 

\begin{eqnarray}\label{AlgcondicionIn} 
\delta^{(1)}v^{'(II)\pm}_{\vec{k}}(\eta_{c}) = 0, \hskip.7cm  
\delta^{(1)}v^{(II)\pm}_{\vec{k}}(\eta_{c}) = 4\delta^{(0)}v^{(II)}_{\vec{k}}(\eta_{c})P_{(1)}(\eta_{c}). 
\end{eqnarray}
Which give us necessary and sufficient information as initial conditions to find unique solutions $\delta^{(1)}v^{(II)+}_{\vec{k}}(\eta)$ of Eq.~(\ref{Klv1masII}), and $\delta^{(1)}v^{(II)-}_{\vec{k}}(\eta)$ of  Eq.~(\ref{Klv1menII}).
\\

\textbf{Step (3):} Finally, starting from equation~(\ref{Imagenreal}) along with its first derivative with respect to $\eta$, we evaluate these two equations at $\eta_{c}$.  Taking into account the initial data $\delta^{(1)}v^{(II)\pm}_{\vec{k}}(\eta_{c})$, $\delta^{(1)}v^{'(II)\pm}_{\vec{k}}(\eta_{c})$, $^{(1)}\phi^{(II)}_{ \xi,\vec{k}_{0} }(\eta_{c})$ and $^{(1)}\phi^{'(II)}_{ \xi,\vec{k}_{0} }(\eta_{c})$. We obtain a system of two algebraic linear equations with two unknowns; $\xi^{(II)}_{(1),\vec{k}_{0}}$ and $\xi^{(II)\ast}_{(1),-\vec{k}_{0}}$. This system is:

\begin{eqnarray}
&& L^{3/2}\hskip.1cm ^{(1)}\!\phi^{(II)}_{\xi,\vec{k}_{0}}(\eta_{c}) =  \delta^{(0)}v^{(II)}_{ \vec{k}_{0} }(\eta)\xi^{(II)}_{(1), \vec{k}_{0} } +  \delta^{(0)}v^{(II) \ast }_{ -\vec{k}_{0} }(\eta)\xi^{(II)\ast}_{(1), -\vec{k}_{0} }   +  \Big[ \delta^{(1)}v^{(II)+}_{0}(\eta)\xi^{(II)}_{0} + (\delta^{(1)}v^{(II)-}_{0}(\eta))^{\ast}\xi_{0}^{(II)\ast} \Big] \hskip.02cm\Bigg|_{\eta=\eta_{c}} \\
&&L^{3/2}\hskip.05cm ^{(1)}\!\phi^{'(II)}_{\xi,\vec{k}_{0}}(\eta_{c}) = \delta^{(0)}v^{'(II)}_{ \vec{k}_{0} }(\eta)\xi^{(II)}_{(1), \vec{k}_{0} } +  \delta^{(0)}v^{'(II) \ast }_{ -\vec{k}_{0} }(\eta)\xi^{(II)\ast}_{ (1),-\vec{k}_{0} }   + \Big[ \delta^{(1)}v^{'(II)+}_{0}(\eta)\xi^{(II)}_{0} + (\delta^{(1)}v^{'(II)-}_{0}(\eta))^{\ast}\xi_{0}^{(II)\ast} \Big] \hskip.01cm \Bigg|_{\eta=\eta_{c}} 
\end{eqnarray}
Hence, by solving the previous system we find $\xi^{(II)}_{(1),\pm\vec{k}_{0}}$ as function of initial data $^{(1)}\phi^{(II)}_{\xi,\vec{k}_{0}}(\eta_{c})$, $^{(1)}\phi^{'(II)}_{\xi,\vec{k}_{0}}(\eta_{c})$. And also in terms of quantities 
that were determined at zeroth-order in $\varepsilon$.
\begin{center}
\textbf{Expansion at second order in $\varepsilon$}
\end{center}
Now we continue with the description at second order in $\varepsilon$. 

Equations $\delta^{(2)}G_{\eta}\hskip.1cm^{l}[\hskip.05cm^{[2]}g^{(II)}(\boldsymbol{x})] = 8\pi G \langle\xi^{(II)}|\delta^{(1)}\hat{T}_{\eta}\hskip.1cm^{l}[\hskip.05cm^{[2]}g^{(II)}(\boldsymbol{x}),\hskip.05cm^{[2]}\hat{\phi}^{(II)}(\boldsymbol{x}),\hskip.05cm^{[2]}\hat{\pi}^{(II)}(\boldsymbol{x})]|\xi^{(II)}\rangle$ and $\delta^{(2)}G_{l}\hskip.1cm^{\eta}[\hskip.05cm^{[2]}g^{(II)}(\boldsymbol{x})] = 8\pi G \langle\xi^{(II)}|\delta^{(1)}\hat{T}_{l}\hskip.1cm^{\eta}[\hskip.05cm^{[2]}g^{(II)}(\boldsymbol{x}),\hskip.05cm^{[2]}\hat{\phi}^{(II)}(\boldsymbol{x}),\hskip.05cm^{[2]}\hat{\pi}^{(II)}(\boldsymbol{x})]|\xi^{(II)}\rangle$, with $l=x,y,z$ 
get reduce to an equation analogous to the semiclassical version of equation~(6.49) of
\cite{secondorderCPTMEJOR}, that is;

\begin{eqnarray}\label{649semiclasica}
2\varepsilon^{2}\partial_{l}\tilde{\Psi}'_{(2)}(\boldsymbol{x}) + 2\varepsilon^{2}\mathcal{H}\partial_{l}\tilde{\Phi}_{(2)}(\boldsymbol{x}) - 8 \pi G \hskip.05cm ^{(0)}\!\phi^{'(II)}_{\xi}\hskip.05cm \partial_{l}\hskip.01cm  ^{(2)}\!\phi^{(II)}_{\xi}(\boldsymbol{x}) = -8\varepsilon^{2}\tilde{\Psi}_{(1)}(\boldsymbol{x})\partial_{l}\tilde{\Psi}'_{(1)}(\boldsymbol{x}) \nonumber\\
 + 8\varepsilon^{2}\mathcal{H}^{(II)}\tilde{\Psi}_{(1)}(\boldsymbol{x})\partial_{l}\tilde{\Psi}_{(1)}(\boldsymbol{x}) - 4\varepsilon^{2}\tilde{\Psi}'_{(1)}(\boldsymbol{x})\partial_{l}\tilde{\Psi}_{(1)}(\boldsymbol{x}) + 16 \pi G \hskip.05cm ^{(1)}\!\phi^{'(II)}_{\xi}(\boldsymbol{x}) \hskip.01cm \partial_{l}\hskip.01cm ^{(1)}\!\phi^{(II)}_{\xi}(\boldsymbol{x}) 
\end{eqnarray}
using $\phi^{(II)}_{\xi}(\boldsymbol{x})$ given by (\ref{expectacionhcontinu2}) in order to expand the matter part of the equation up to second order in $\varepsilon$ , we obtain;
\begin{eqnarray}\label{Eze2}
&&2\partial_{l}\tilde{\Psi}'_{(2)}(\boldsymbol{x}) + 2\mathcal{H}\partial_{l}\tilde{\Phi}_{(2)}(\boldsymbol{x}) + 8\tilde{\Psi}_{(1)}(\boldsymbol{x})\partial_{l}\tilde{\Psi}'_{(1)}(\boldsymbol{x}) - 8\mathcal{H}^{(II)}\tilde{\Psi}_{(1)}(\boldsymbol{x})\partial_{l}\tilde{\Psi}_{(1)}(\boldsymbol{x}) + 4\tilde{\Psi}'_{(1)}(\boldsymbol{x})\partial_{l}\tilde{\Psi}_{(1)}(\boldsymbol{x}) = 8\pi G \Big\{ \hskip.02cm^{(0)}\!\phi^{'(II)}_{\xi,0} \nonumber\\ 
&& \times \Big[ 2ik_{0l}\hskip.05cm^{(2)}\!\phi^{(II)}_{\xi,2\vec{k}_{0}}(\eta)e^{2i\vec{k}_{0}\cdot\vec{x}} + c.c \Big] + 2 \Big[ \hskip.05cm^{(1)}\!\phi^{'(II)}_{\xi,\vec{k}_{0}}(\eta)e^{i\vec{k}_{0}\cdot\vec{x}}  + c.c \Big] \Big[ i k_{0l}\hskip.05cm^{(1)}\!\phi^{(II)}_{\xi,\vec{k}_{0}}(\eta)e^{i\vec{k}_{0}\cdot\vec{x}}  + c.c \Big] \Big\}, 
\end{eqnarray} 

where $k_{0l}$, corresponds to the $l$-th $(l : x,y,z)$ component of the vector $\vec{k}_{0}$. Using (\ref{ordpertHas2Phi}), (\ref{ordpertHas2Psi}) and (\ref{ordpertHas2Ht}) to expand the left-hand side of the
previous equation to second order in $\varepsilon$, we find that;

\begin{eqnarray}\label{Eze2ecuacionl}
&& \Big[ 4ik_{0l}\hskip.03cm \Big( P'_{(2)}(\eta) + \mathcal{H}^{(II)}P_{(2)}(\eta)  \Big) e^{2i\vec{k}_{0}\cdot\vec{x}} 
+ c.c \Big] + 4 \Big( P_{(1)}(\eta)e^{i\vec{k}_{0}\cdot\vec{x}} + c.c  \Big)\Big( i k_{0l} P'_{(1)}(\eta)e^{i\vec{k}_{0}\cdot\vec{x}} + c.c  \Big) -
\nonumber \\
&&  4\mathcal{H}^{(II)}\Big( P_{(1)}(\eta)e^{i\vec{k}_{0}\cdot\vec{x}} + c.c  \Big)\Big( i k_{0l} P_{(1)}(\eta)e^{i\vec{k}_{0}\cdot\vec{x}} + c.c  \Big) + 2\Big( P'_{(1)}(\eta)e^{i\vec{k}_{0}\cdot\vec{x}} + c.c  \Big)\Big( i k_{0l} P_{(1)}(\eta)e^{i\vec{k}_{0}\cdot\vec{x}} + c.c  \Big) 
\nonumber \\
&& = 4\pi G \Big\{ \hskip.05cm^{(0)}\!\phi^{'(II)}_{\xi,0}  
\Big[ 2ik_{0l}\hskip.05cm^{(2)}\!\phi^{(II)}_{\xi,2\vec{k}_{0}}(\eta)e^{2i\vec{k}_{0}\cdot\vec{x}}  
+ c.c \Big] + 2 \Big[ \hskip.05cm^{(1)}\!\phi^{'(II)}_{\xi,\vec{k}_{0}}(\eta)e^{i\vec{k}_{0}\cdot\vec{x}}  + c.c \Big] \Big[ ik_{0l}\hskip.05cm ^{(1)}\!\phi^{(II)}_{\xi,\vec{k}_{0}}(\eta)e^{i\vec{k}_{0}\cdot\vec{x}}  + c.c \Big] \Big\},
\end{eqnarray}
after rearranging and regrouping coefficients of $e^{i\vec{k}\cdot\vec{x}}$ in the previous equation, we find that it takes the form:

\begin{equation}\label{C2k0k1y1}
C_{-2\vec{k}_{0}}(\eta)e^{-2i\vec{k}_{0}\cdot\vec{x}} + C_{0}(\eta) + C_{2\vec{k}_{0}}(\eta)e^{2i\vec{k}_{0}\cdot\vec{x}}  = 0.
\end{equation}
Because of the linear independence of the set of functions $\{1, e^{-2i\vec{k}_{0}\cdot\vec{x}}, e^{2i\vec{k}_{0}\cdot\vec{x}} \}$, we deduce the following equations:

For the coefficient $C_{0}(\eta)$, by linear independence, equation~(\ref{C2k0k1y1}) implies that  $C_{0}(\eta)=0$, where we find: 

\begin{eqnarray}\label{problematica1}
&& i\vec{k}_{0}\!\cdot\!\hat{u}\Big\{ -4 P_{(1)}(\eta) P^{'\ast}_{(1)}(\eta) + 4P^{\ast}_{(1)}(\eta) P^{'}_{(1)}(\eta) + 4\mathcal{H}^{(II)} P_{(1)}(\eta) P^{\ast}_{(1)}(\eta) - 4\mathcal{H}^{(II)} P^{\ast}_{(1)}(\eta) P_{(1)}(\eta) - 2 P'_{(1)}(\eta) P^{\ast}_{(1)}(\eta) \nonumber \\
&& + 2 P^{'\ast}_{(1)}(\eta)P_{(1)}(\eta) + 4\pi G \hskip.1cm \Big( 2 \hskip.05cm ^{(1)}\!\phi^{'(II)}_{\xi,\vec{k}_{0}}(\eta) \hskip.05cm^{(1)}\!\phi^{(II)\ast}_{\xi,\vec{k}_{0}}(\eta) - 2\hskip.05cm^{(1)}\!\phi^{'(II)\ast}_{\xi,\vec{k}_{0}}(\eta) \hskip.05cm ^{(1)}\!\phi^{(II)}_{\xi,\vec{k}_{0}}(\eta)\hskip.01cm \Big) \Big\} = 0,
\end{eqnarray}
with $\hskip.1cm\hat{k}_{0}\|\hat{z}$. 
We can see that in the previous equation, there only appear the quantities $P_{(1)}(\eta)$, $P'_{(1)}(\eta)$, $^{(1)}\phi^{(II)\ast}_{\xi,\vec{k}_{0}}(\eta)$, and $^{(1)}\phi^{'(II)\ast}_{\xi,\vec{k}_{0}}(\eta)$ which were already determined at first order in $\varepsilon$. However, given that $P_{(1)}(\eta)$, $P'_{(1)}(\eta)$, $^{(1)}\phi^{(II)\ast}_{\xi,\vec{k}_{0}}(\eta)$ and $^{(1)}\phi^{'(II)\ast}_{\xi,\vec{k}_{0}}(\eta)$, are real valued, we find then that (\ref{problematica1}) is trivially satisfied.  At this point the fact  that, at first order we have imposed reality  the conditions on  $P_{(1)}(\eta)$, $^{(1)}\phi^{(II)}_{\xi,\vec{k}_{0}} $ and  thus   we need  $\xi^{(II)}_{(1),\vec{k}_{0}} = \xi^{(II)}_{(1),-\vec{k}_{0}}$, for   the consistency of  Equation~(\ref{problematica1}) .


We reconsider now equation~(\ref{C2k0k1y1}). The coefficients $C_{\pm2\vec{k}_{0}}$ of (\ref{C2k0k1y1}) lead to a pair of equations which are the complex conjugate of each other, since $C_{2\vec{k}_{0}} = C_{-2\vec{k}_{0}}^{\ast}$. Therefore,  $C_{2\vec{k}_{0}} = 0$, implies:

\begin{equation}\label{casodos1}
P'_{(2)}(\eta) + \mathcal{H}^{(II)}P_{(2)}(\eta) + \frac{3}{2} P_{(1)}(\eta)P'_{(1)}(\eta) - \mathcal{H}^{(II)} P^{2}_{(1)}(\eta) = 2\pi G \Big\{ \hskip.05cm ^{(0)}\!\phi^{'(II)}_{\xi,0}\hskip.1cm^{(2)}\!\phi^{(II)}_{\xi,2\vec{k}_{0}}(\eta) +  \hskip.05cm ^{(1)}\!\phi^{'(II)}_{\xi,\vec{k}_{0}}(\eta) \hskip.1cm ^{(1)}\!\phi^{(II)}_{\xi,\vec{k}_{0}}(\eta) \Big\}.
\end{equation}
In summary,  (\ref{casodos1}) corresponds to a restriction from the theory imposed on the functions 
$P_{(2)}(\eta)$, $P^{'}_{(2)}(\eta)$, $^{(2)}\phi^{(II)}_{\xi,2\vec{k}_{0}}(\eta)$ and $^{(2)}\phi^{'(II)}_{\xi,2\vec{k}_{0}}(\eta)$. We have found that equation~(\ref{C2k0k1y1}), resulting from expanding  (\ref{Eze2}), does not lead to restrictions on $a_{(2)}(\eta)$, $a'_{(2)}(\eta)$, $^{(2)}\phi^{(II)}_{\xi,0}(\eta)$ and $^{(2)}\phi^{'(II)}_{\xi,0}(\eta)$. Nevertheless, a restriction of this kind would arise from working directly with the homogenous part\footnote{Any function $f(\eta,\vec{x})$ can be decomposed as the sum of a homogeneous part $\textrm{P}_{hom}\{f(\eta,\vec{x})\}$ and an inhomogeneous part $\textrm{P}_{inhom}\{f(\eta,\vec{x})\}$, i.e., $f(\eta,\vec{x}) = \textrm{P}_{hom}\{f(\eta,\vec{x})\} + \textrm{P}_{inhom}\{f(\eta,\vec{x})\}$, where the function $\textrm{P}_{hom}\{f(\eta,\vec{x})\}$ it comes from grouping all the contributions to $f(\eta,\vec{x})$ that are independent of spatial coordinates (then, in general $\textrm{P}_{hom}\{f(\eta,\vec{x})\}$ will have the form; $\textrm{P}_{hom}\{f(\eta,\vec{x})\} = constant + c(\eta)$), whereas the function $\textrm{P}_{inhom}\{f(\eta,\vec{x})\}$ is the part that depends on the spatial coordinates, i.e., $\textrm{P}_{inhom}\{f(\eta,\vec{x})\}\neq constant + c(\eta)$.} ($\textrm{P}_{hom}$) from (\ref{649semiclasica}). In other words: the restriction would arise from working with the semiclassical version of (6.51) in \cite{secondorderCPTMEJOR}. 

We can observe that the homogeneous part on the left-hand side of (\ref{649semiclasica}) is zero. This comes from:
\begin{eqnarray}\label{PartEescalar}
&&\textrm{P}_{hom}\Big\{ -8\varepsilon^{2}\tilde{\Psi}_{(1)}(\eta,\vec{x})\partial_{l}\tilde{\Psi}'_{(1)}(\eta,\vec{x}) + 8\varepsilon^{2}\mathcal{H}^{(II)}\tilde{\Psi}_{(1)}(\eta,\vec{x})\partial_{l}\tilde{\Psi}_{(1)}(\eta,\vec{x}) - 4\varepsilon^{2}\tilde{\Psi}'_{(1)}(\eta,\vec{x})\partial_{l}\tilde{\Psi}_{(1)}(\eta,\vec{x})
\nonumber\\
&& 
+ 16 \pi G \hskip.05cm ^{(1)}\!\phi^{'(II)}_{\xi}(\eta,\vec{x}) \hskip.1cm \partial_{l}\hskip.05cm ^{(1)}\!\phi^{(II)}_{\xi}(\eta,\vec{x})  \Big\}  = \varepsilon^{2}\textrm{P}_{hom}\Big\{ 
-8\Big[P_{(1)}(\eta)e^{i\vec{k}_{0}\cdot\vec{x}}  + c.c \Big]\Big[ik_{0l}P'_{(1)}(\eta)e^{i\vec{k}_{0}\cdot\vec{x}}  + c.c \Big] \nonumber\\
&& + 8\mathcal{H}^{(II)}\Big[P_{(1)}(\eta)e^{i\vec{k}_{0}\cdot\vec{x}}  + c.c \Big]\Big[ik_{0l}P_{(1)}(\eta)e^{i\vec{k}_{0}\cdot\vec{x}}  + c.c \Big] 
- 4\Big[P'_{(1)}(\eta)e^{i\vec{k}_{0}\cdot\vec{x}}  + c.c \Big]\Big[ik_{0l}P_{(1)}(\eta)e^{i\vec{k}_{0}\cdot\vec{x}}  \nonumber\\
&& + c.c \Big] + 16 \pi G \Big[ \hskip.05cm^{(1)}\!\phi^{'(II)}_{\xi,\vec{k}_{0}}(\eta) e^{i\vec{k}_{0}\cdot\vec{x}}  + c.c \Big] \Big[ i k_{0l}\hskip.05cm^{(1)}\!\phi^{(II)}_{\xi,\vec{k}_{0}}(\eta) e^{i\vec{k}_{0}\cdot\vec{x}}  + c.c \Big] \Big\}, 
\end{eqnarray} 
which is equal to zero given the form of the mathematical expression, and along with the fact that: $P_{(1)}(\eta)$, $P'_{(1)}(\eta)$, $^{(1)}\phi^{(II)}_{\xi,\vec{k}_{0}}(\eta)$ and $^{(1)}\phi^{(II)}_{\xi,\vec{k}_{0}}(\eta)$ $\in\mathbb{R}$. Consequently, the right-hand side of (\ref{649semiclasica}) implies that; 

\begin{equation}\label{PhomPsi}
\textrm{P}_{hom}\Big\{ 2\varepsilon^{2}\partial_{l}\tilde{\Psi}'_{(2)}(\eta,\vec{x}) + 2\varepsilon^{2}\mathcal{H}\partial_{l}\tilde{\Phi}_{(2)}(\eta,\vec{x}) - 8 \pi G \hskip.1cm ^{(0)}\!\phi^{'(II)}_{\xi}(\eta) \hskip.1cm \partial_{l}\hskip.001cm ^{(2)}\!\phi^{(II)}_{\xi}(\eta,\vec{x}) \Big\} = 0.
\end{equation}
The argument of $\textrm{P}_{hom}$ in (\ref{PhomPsi}) can be written as;

\begin{equation}
 \textrm{P}_{hom}\Big\{ \partial_{l}\Big[ \varepsilon^{2} \tilde{\Psi}'_{(2)}(\eta,\vec{x}) + \varepsilon^{2} \mathcal{H}\tilde{\Phi}_{(2)}(\eta,\vec{x}) - 4\pi G \hskip.1cm ^{(0)}\!\phi^{'(II)}_{\xi}(\eta) \hskip.1cm ^{(2)}\!\phi^{(II)}_{\xi}(\eta,\vec{x}) - \varepsilon^{2} \hskip.1cm ^{(2)}\!\mu(\eta) \Big] \Big\}   = 0
\end{equation}
with $^{(2)}\mu(\eta)$ being an unknown function which only depends on the time coordinate $\eta$. Therefore, the above equation implies that;

\begin{equation}
  \partial_{l}\Big[ \varepsilon^{2} \tilde{\Psi}'_{(2)}(\eta,\vec{x}) + \varepsilon^{2} \mathcal{H}\tilde{\Phi}_{(2)}(\eta,\vec{x}) - 4\pi G \hskip.1cm ^{(0)}\!\phi^{'(II)}_{\xi}(\eta) \hskip.1cm ^{(2)}\!\phi^{(II)}_{\xi}(\eta,\vec{x}) - \varepsilon^{2} \hskip.1cm ^{(2)}\!\mu(\eta) \Big]  = \varepsilon^{2} \hskip.1cm ^{(2)}\!\Gamma(\eta,\vec{x}),
\end{equation}
with $ ^{(2)}\Gamma(\eta,\vec{x})$ being an unknown function such that $\textrm{P}_{hom}\{ \hskip.005cm ^{(2)}\Gamma(\eta,\vec{x}) \} = 0$ which can be determined through equation~(\ref{649semiclasica}). After applying 
$\partial_{l}^{-1}$ on both sides of the previous equation, we obtain:

\begin{equation}\label{homogeq}
 \varepsilon^{2}\tilde{\Psi}'_{(2)}(\eta,\vec{x}) + \varepsilon^{2}\mathcal{H}\tilde{\Phi}_{(2)}(\eta,\vec{x}) - 4\pi G \hskip.1cm ^{(0)}\!\phi^{'(II)}_{\xi}(\eta) \hskip.1cm ^{(2)}\!\phi^{(II)}_{\xi}(\eta,\vec{x}) - \varepsilon^{2}\hskip.1cm ^{(2)}\!\tilde{\mu}(\eta)  = \varepsilon^{2}\hskip.1cm ^{(2)}\!\tilde{\Gamma}(\eta,\vec{x}),
\end{equation}
where $ ^{(2)}\tilde{\Gamma}(\eta,\vec{x})$ satisfies the conditions $\textrm{P}_{hom}\{ \hskip.05cm ^{(2)}\!\tilde{\Gamma}(\eta,\vec{x}) \} = 0$ and $\partial_{l} \hskip.05cm ^{(2)}\!\tilde{\Gamma}(\eta,\vec{x}) = \hskip.05cm ^{(2)}\!\Gamma(\eta,\vec{x})$, whereas $ ^{(2)}\tilde{\mu}(\eta)$ corresponds to the sum of $^{(2)}\mu(\eta)$ and the function arising from  $\partial_{l}^{-1} \hskip.05cm ^{(2)}\!\Gamma(\eta,\vec{x})$ that only depends on the time component. Lastly, using (\ref{ordpertHas2Phi}) and (\ref{ordpertHas2Psi}) to expand (\ref{homogeq}) to second order in $\varepsilon$ taking afterwards $\textrm{P}_{hom}$ from the expansion, we obtain;

\begin{equation}\label{restricciona}  
 a'_{(2)}(\eta) - \mathcal{H}a_{(2)}(\eta) = 4\pi G \hskip.1cm ^{(0)}\!\phi^{'(II)}_{\xi,0}(\eta) \hskip.05cm ^{(2)}\!\phi^{(II)}_{\xi,0}(\eta) + \hskip.05cm ^{(2)}\!\tilde{\mu}(\eta).
\end{equation}
In this way, we find the restriction equation for $a_{(2)}(\eta)$, $a'_{(2)}(\eta)$ and $^{(2)}\phi^{(II)}_{\xi,0}(\eta)$. 

The other restrictions of the theory come from the second-order contribution in $\varepsilon$ of the 
$\eta\eta$ component of the semiclassical Einstein equations.;\\

The equation; $\delta^{(2)}G_{\eta}\hskip.1cm^{\eta}[\hskip.05cm^{[2]}g^{(II)}(\boldsymbol{x})] = 8\pi G \hskip.05cm \langle\xi^{(II)}|\delta^{(2)}\hat{T}_{\eta}\hskip.1cm^{\eta}[\hskip.05cm^{[2]}g^{(II)}(\boldsymbol{x}),\hskip.05cm^{[2]}\hat{\phi}^{(II)}(\boldsymbol{x}),\hskip.05cm^{[2]}\hat{\pi}^{(II)}(\boldsymbol{x})]|\xi^{(II)}\rangle,$ which gets reduced to the semiclassical version of (6.53) in \cite{secondorderCPTMEJOR}, corresponds to:

\begin{eqnarray}\label{Eee2}
&&-3\mathcal{H}^{(II)}\tilde{\Psi}'_{(2)}(\boldsymbol{x}) + \nabla^{2}\tilde{\Psi}_{(2)}(\boldsymbol{x}) - \mathcal{H}^{'(II)}\tilde{\Phi}_{(2)}(\boldsymbol{x}) - 2\mathcal{H}^{2(II)}\tilde{\Phi}_{(2)}(\boldsymbol{x})  + 2\tilde{\Phi}_{(1)}(\boldsymbol{x})\tilde{\Phi}''_{(1)}(\boldsymbol{x}) + 3 \tilde{\Phi}^{'2}_{(1)}(\boldsymbol{x}) + 3 (\vec{\nabla}\tilde{\Phi}_{(1)}(\boldsymbol{x}))^{2} \nonumber\\ 
&& + 10\tilde{\Phi}_{(1)}(\boldsymbol{x})\nabla^{2}\tilde{\Phi}_{(1)}(\boldsymbol{x}) + 4(\mathcal{H}^{'(II)} + 2\mathcal{H}^{(II)2} )\tilde{\Phi}^{2}_{(1)}(\boldsymbol{x}) = 4 \pi G \Big\{ \hskip.1cm ^{(0)}\!\phi^{'(II)}_{\xi,0}\Big[ \hskip.1cm ^{(2)}\!\phi^{'(II)}_{\xi,0}(\eta) + \Big( \hskip.1cm ^{(2)}\!\phi^{'(II)}_{\xi,2\vec{k}_{0}}(\eta) e^{ 2i\vec{k}_{0}\cdot\vec{x} }   + c.c \Big) \Big] \nonumber \\
&&
+ \Big(\hskip.1cm ^{(1)}\!\phi^{'(II)}_{\xi,\vec{k}_{0}}(\eta) e^{ i\vec{k}_{0}\cdot\vec{x} }  + c.c \Big)^{2} + a^{2(II)}m^{2} \Big[ \hskip.1cm ^{(0)}\!\phi^{(II)}_{\xi,0}\hskip.1cm ^{(2)}\!\phi^{(II)}_{\xi,0}(\eta) + \hskip.1cm ^{(0)}\!\phi^{(II)}_{\xi,0}\Big( \hskip.1cm ^{(2)}\!\phi^{(II)}_{\xi,2\vec{k}_{0}}(\eta) e^{ 2i\vec{k}_{0}\cdot\vec{x} } + c.c \Big) \nonumber \\
&&
+ \Big(\hskip.1cm^{(1)}\!\phi^{(II)}_{\xi,\vec{k}_{0}}(\eta) e^{ i\vec{k}_{0}\cdot\vec{x} }  + c.c \Big)^{2} \Big]  + \Big(i \vec{k}_{0}\hskip.1cm ^{(1)}\!\phi^{(II)}_{\xi,\vec{k}_{0}}(\eta) e^{ i\vec{k}_{0}\cdot\vec{x} }  + c.c \Big)^{2}  \Big\}. 
\end{eqnarray}
Using  (\ref{ordpertHas2Phi}) and (\ref{ordpertHas2Psi}) in order to expand the left-hand side of the previous equation at second order in $\varepsilon$, we find:

\begin{eqnarray}\label{Eee2fin}
&& - 6\mathcal{H}^{(II)}\Big[ a'_{(2)}(\eta) + \Big( P'_{(2)}(\eta)e^{ 2i\vec{k}_{0}\cdot\vec{x} } + c.c \Big) \Big]  - 8\Big( k_{0}^{2}P_{(2)}(\eta) e^{ 2i\vec{k}_{0}\cdot\vec{x} } + c.c  \Big)  - 2\Big( \mathcal{H}^{'(II)} + \nonumber\\ 
&& 2\mathcal{H}^{2(II)} \Big)\Big[ -a_{(2)}(\eta) + \Big( P_{(2)}(\eta)e^{ 2i\vec{k}_{0}\cdot\vec{x} } + c.c \Big) \Big] + 2\Big( P_{(1)}(\eta)e^{ i\vec{k}_{0}\cdot\vec{x} } + c.c \Big)\Big( P''_{(1)}(\eta)e^{ i\vec{k}_{0}\cdot\vec{x} } + c.c \Big) \nonumber\\ 
&&
+ 3\Big( P'_{(1)}(\eta)e^{ i\vec{k}_{0}\cdot\vec{x} } + c.c \Big)^{2} + 3\Big( i\vec{k}_{0}P_{(1)}(\eta)e^{ i\vec{k}_{0}\cdot\vec{x} } + c.c \Big)^{2} - 10k_{0}^{2} \Big(P_{(1)}(\eta)e^{ i\vec{k}_{0}\cdot\vec{x} } + c.c \Big)^{2}  \nonumber\\ 
&&
+ 4\Big( \mathcal{H}^{'(II)} + 2\mathcal{H}^{2(II)} \Big)\Big( P_{(1)}(\eta)e^{ i\vec{k}_{0}\cdot\vec{x} } + c.c \Big)^{2} = 4\pi G \Big\{ \hskip.1cm ^{(0)}\!\phi^{'(II)}_{\xi,0}\Big[ \hskip.1cm ^{(2)}\!\phi^{'(II)}_{\xi,0}(\eta) + \nonumber \\
&&
\Big( \hskip.1cm ^{(2)}\!\phi^{'(II)}_{\xi,2\vec{k}_{0}}(\eta) e^{ 2i\vec{k}_{0}\cdot\vec{x} } 
+ c.c \Big)   \Big]
+ \Big(\hskip.1cm ^{(1)}\!\phi^{'(II)}_{\xi,\vec{k}_{0}}(\eta) e^{ i\vec{k}_{0}\cdot\vec{x} }  + c.c \Big)^{2} + a^{2(II)}m^{2} \Big[ \hskip.1cm ^{(0)}\!\phi^{(II)}_{\xi,0}\hskip.1cm^{(2)}\!\phi^{(II)}_{\xi,0}(\eta) + \nonumber \\
&&
 \hskip.1cm ^{(0)}\!\phi^{(II)}_{\xi,0}\Big( \hskip.1cm ^{(2)}\!\phi^{(II)}_{\xi,2\vec{k}_{0}}(\eta)e^{ 2i\vec{k}_{0}\cdot\vec{x} }  
+ c.c \Big) + 
\Big(\hskip.1cm^{(1)}\!\phi^{(II)}_{\xi,\vec{k}_{0}}(\eta) e^{ i\vec{k}_{0}\cdot\vec{x} }  + c.c \Big)^{2} \Big]  + \Big(i \vec{k}_{0}\hskip.1cm ^{(1)}\!\phi^{(II)}_{\xi,\vec{k}_{0}}(\eta) e^{ i\vec{k}_{0}\cdot\vec{x} }  + c.c \Big)^{2}  \Big\}, 
\end{eqnarray}
after manipulating the coefficients of the functions $\{e^{\pm2i\vec{k}_{0}\cdot\vec{x}},\hskip.05cm 1\}$,
we find that the previous equation takes the form:

\begin{equation}\label{k0k1indist}
\boldsymbol{C}_{-2\vec{k}_{0}}(\eta)e^{-2i\vec{k}_{0}\cdot\vec{x}} + \boldsymbol{C}_{0}(\eta) + \boldsymbol{C}_{2\vec{k}_{0}}(\eta)e^{2i\vec{k}_{0}\cdot\vec{x}} = 0. 
\end{equation}
Due to the linear independence of the functions $\{ e^{\pm2i\vec{k}_{0}\cdot\vec{x}}, 1\}$, we conclude that
$\boldsymbol{C}_{\pm2\vec{k}_{0}}(\eta) = \boldsymbol{C}_{0}(\eta) = 0.$ 

The equation $\boldsymbol{C}_{2\vec{k}_{0}}(\eta) = 0$,  corresponds to;

\begin{eqnarray}\label{Eee2finmas}
&& - 6\mathcal{H}^{(II)} P'_{(2)}(\eta)  - 8 k_{0}^{2}P_{(2)}(\eta)   - 2 \Big( \mathcal{H}^{'(II)}  + 2\mathcal{H}^{2(II)} \Big) P_{(2)}(\eta) + 2P_{(1)}(\eta)P''_{(1)}(\eta) + 3 (P'_{(1)}(\eta))^{2} \nonumber\\ 
&&
- 13k_{0}^{2} P_{(1)}^{2}(\eta) + 4\Big( \mathcal{H}^{'(II)} + 2\mathcal{H}^{2(II)} \Big)P_{(1)}^{2}(\eta) 
= 4\pi G \Big\{ \hskip.03cm ^{(0)}\!\phi^{'(II)}_{\xi,0} \hskip.03cm ^{(2)}\!\phi^{'(II)}_{\xi,2\vec{k}_{0}}(\eta) + \Big( \hskip.02cm ^{(1)}\!\phi^{'(II)}_{\xi,\vec{k}_{0}}(\eta)\Big)^{2}  
\nonumber \\
&&  
+ a^{ 2(II) } m^{ 2 } \Big[ \hskip.05cm^{(0)}\!\phi^{(II)}_{\xi,0}\hskip.05cm ^{(2)}\!\phi^{(II)}_{\xi,2\vec{k}_{0}}(\eta)  + \Big(\hskip.05cm ^{(1)}\!\phi^{(II)}_{\xi,\vec{k}_{0}}(\eta)\Big)^{2} \Big] 
- k_{0}^{2} \Big(\hskip.05cm^{(1)}\!\phi^{(II)}_{\xi,\vec{k}_{0}}(\eta)\Big)^{2}  \Big\}. 
\end{eqnarray}

On the other hand, proceeding in an equivalent way, we find that the equation $\boldsymbol{C}_{-2\vec{k}_{0}}(\eta) = 0$ corresponds to the complex conjugate of the previous expression. Whereas equation $\boldsymbol{C}_{0}(\eta) = 0$, corresponds to;

\begin{eqnarray}\label{Eee2fincero}
&& - 6\mathcal{H}^{(II)} a^{'}_{(2)}(\eta) + 2\Big( \mathcal{H}^{'(II)}  + 2\mathcal{H}^{2(II)} \Big) a_{(2)}(\eta) + 2(P_{(1)}(\eta)P^{''\ast}_{(1)}(\eta) + P^{\ast}_{(1)}(\eta)P^{''}_{(1)}(\eta)) + 6 (P^{'}_{(1)}(\eta))^{2} \nonumber\\ 
&&
- 14k_{0}^{2} (P_{(1)}(\eta))^{2} + 8\Big( \mathcal{H}^{'(II)} + 2\mathcal{H}^{2(II)} \Big)(P_{(1)}(\eta))^{2} = 4\pi G \Big\{ \hskip.1cm ^{(0)}\!\phi^{'(II)}_{\xi,0} \hskip.1cm^{(2)}\phi^{'(II)}_{\xi,0} + \nonumber \\ 
&&
2(\hskip.05cm^{(1)}\!\phi^{'(II)}_{\xi,\vec{k}_{0}}(\eta))^{2}
+ a^{2(II)}m^{2} \Big( \hskip.05cm^{(0)}\!\phi^{(II)}_{\xi,0}\hskip.1cm ^{(2)}\!\phi^{(II)}_{\xi,0}(\eta)  +  2 (\hskip.02cm^{(1)}\!\phi^{(II)}_{\xi,\vec{k}_{0}}(\eta))^{2} \Big) 
+ 2k_{0}^{2} \hskip.01cm (\hskip.01cm ^{(1)}\!\phi^{(II)}_{\xi,\vec{k}_{0}}(\eta))^{2} \Big\}. 
\end{eqnarray}
We find that the equation $\boldsymbol{C}_{2\vec{k}_{0}}(\eta) = 0$ corresponds to a restriction of the 
functions $P_{(2)}(\eta)$, $^{(2)}\phi^{(II)}_{\xi,2\vec{k}_{0}}(\eta)$ and its time derivatives. Whereas equation
$\boldsymbol{C}_{0}(\eta) = 0$ provides a restriction to the functions $a_{(2)}(\eta)$, $^{(2)}\phi^{(II)}_{\xi,0}(\eta)$ and its time derivatives.\\

Continuing with the expansion, the dynamic equations for the contributions at second-order in
$\varepsilon$ of the metric potentials come from the equations  $\delta^{(2)}G_{l}\hskip.1cm^{j}[\hskip.05cm^{[2]}g^{(II)}(\boldsymbol{x})] = 8\pi G \hskip.04cm \langle\xi^{(II)}|\delta^{(2)}\hat{T}_{l}\hskip.1cm^{j}[\hskip.05cm^{[2]}g^{(II)}(\boldsymbol{x}),\hskip.05cm^{[2]}\hat{\phi}^{(II)}(\boldsymbol{x}),\hskip.05cm^{[2]}\hat{\pi}^{(II)}(\boldsymbol{x})]|\xi^{(II)}\rangle$, with 
 $l,j:x,y,z$. These can be understood as the semiclassical version of equation~(6.55) in reference \cite{secondorderCPTMEJOR}, which we are going to write as:
 
\begin{eqnarray}\label{Eij2}
&&\partial_{l}\partial_{j}( \tilde{\Psi}_{(2)}(\boldsymbol{x}) - \tilde{\Phi}_{(2)}(\boldsymbol{x}) ) + \Big\{ -\nabla^{2}\tilde{\Psi}_{(2)}(\boldsymbol{x}) + 2\tilde{\Psi}''_{(2)}(\boldsymbol{x}) + 4\mathcal{H}^{(II)}\tilde{\Psi}'_{(2)}(\boldsymbol{x}) + 2\mathcal{H}^{(II)}\tilde{\Phi}'_{(2)}(\boldsymbol{x}) + ( 2\mathcal{H}^{'(II)} + 4\mathcal{H}^{(II)2} ) \tilde{\Phi}_{(2)}(\boldsymbol{x}) \nonumber\\
&& + \nabla^{2}\tilde{\Phi}_{(2)}(\boldsymbol{x}) \Big\}\delta_{lj} + \frac{1}{2}( \partial^{2}_{\eta} + 2\mathcal{H}^{(II)}\partial_{\eta} - \nabla^{2})\hskip.01cm ^{(2)}h_{lj}(\boldsymbol{x}) + 4 \partial_{l}\tilde{\Phi}_{(1)}(\boldsymbol{x})\partial_{j}\tilde{\Phi}_{(1)}(\boldsymbol{x}) + 8\tilde{\Phi}_{(1)}(\boldsymbol{x})\partial_{l}\partial_{j}\tilde{\Phi}_{(1)}(\boldsymbol{x}) -2\Big\{ \tilde{\Phi}^{'2}_{(1)}(\boldsymbol{x}) + \nonumber\\
&& 8\mathcal{H}^{(II)}\tilde{\Phi}_{(1)}(\boldsymbol{x})\tilde{\Phi}'_{(1)}(\boldsymbol{x}) - 2\tilde{\Phi}_{(1)}(\boldsymbol{x})\tilde{\Phi}''_{(1)}(\boldsymbol{x}) + 3(\vec{\nabla}\tilde{\Phi}_{(1)}(\boldsymbol{x}))^{2} + 
2\tilde{\Phi}_{(1)}(\boldsymbol{x})\nabla^{2}\tilde{\Phi}_{(1)}(\boldsymbol{x})  + 4(\mathcal{H}^{'(II)} + 2\mathcal{H}^{(II)2} ) (\tilde{\Phi}_{(1)}(\boldsymbol{x}))^{2}  \Big\} \delta_{lj} \nonumber \\
&& = 8 \pi G\Big\{ \hskip.1cm^{(0)}\!\phi^{'(II)}_{\xi,0} \Big[ \hskip.1cm ^{(2)}\!\phi^{'(II)}_{\xi,0}(\eta) + \Big(\hskip.1cm ^{(2)}\!\phi^{'(II)}_{\xi,2\vec{k}_{0}}(\eta) e^{2i\vec{k}_{0}\cdot\vec{x}}  + c.c \Big)\Big] + \Big(\hskip.1cm^{(1)}\!\phi^{'(II)}_{\xi,\vec{k}_{0}}(\eta)e^{i\vec{k}_{0}\cdot\vec{x}}  + c.c \Big)^{2} + \nonumber \\ && - \Big(i\vec{k}_{0}\hskip.1cm ^{(1)}\!\phi^{(II)}_{\xi,\vec{k}_{0}}(\eta)e^{i\vec{k}_{0}\cdot\vec{x}}  + c.c \Big)^{2} - a^{2(II)} m^{2} \Big[ \hskip.1cm ^{(0)}\!\phi^{(II)}_{\xi,0}\hskip.1cm ^{(2)}\!\phi^{(II)}_{\xi,0}(\eta) + \hskip.1cm ^{(0)}\!\phi^{(II)}_{\xi,0}\Big( \hskip.1cm ^{(2)}\!\phi^{(II)}_{\xi,2\vec{k}_{0}}(\eta) e^{2i\vec{k}_{0}\cdot\vec{x}}  +  c.c  \Big) \nonumber \\ 
&& + \Big( \hskip.1cm ^{(1)}\!\phi^{(II)}_{\xi,\vec{k}_{0}}(\eta) e^{i\vec{k}_{0}\cdot\vec{x}}  +  c.c  \Big)^{2} \Big] \Big\} \delta_{lj} + 16\pi G \Big(ik_{0l}\hskip.1cm ^{(1)}\!\phi^{(II)}_{\xi,\vec{k}_{0}}(\eta)e^{i\vec{k}_{0}\cdot\vec{x}}  + c.c \Big)\Big(ik_{0j}\hskip.1cm ^{(1)}\!\phi^{(II)}_{\xi,\vec{k}_{0}}(\eta) e^{i\vec{k}_{0}\cdot\vec{x}}  + c.c \Big). 
\end{eqnarray}
In order to examine these equations, we start with the $lj$ components such that $l\neq j$. Given the symmetry of the problem, the only non trivial component  $l\neq j$ of (\ref{Eij2}) corresponds to $l=x,j=y$, or equivalently
$l=y,j=x$.\\

Hence, the equations;
$\delta^{(2)}G_{x}\hskip.1cm^{y}[\hskip.05cm^{[2]}g^{(II)}(\boldsymbol{x})] = 8\pi G \langle\xi^{(II)}|\delta^{(2)}\hat{T}_{x}\hskip.1cm^{y}[\hskip.05cm^{[2]}g^{(II)}(\boldsymbol{x}),\hskip.05cm^{[2]}\hat{\phi}^{(II)}(\boldsymbol{x}),\hskip.05cm^{[2]}\hat{\pi}^{(II)}(\boldsymbol{x})]|\xi^{(II)}\rangle$  
and 
$\delta^{(2)}G_{y}\hskip.1cm^{x}[\hskip.05cm^{[2]}g^{(II)}(\boldsymbol{x})] = 8\pi G \langle\xi^{(II)}|\delta^{(2)}\hat{T}_{y}\hskip.1cm^{x}[\hskip.05cm^{[2]}g^{(II)}(\boldsymbol{x}),\hskip.05cm^{[2]}\hat{\phi}^{(II)}(\boldsymbol{x}),\hskip.05cm^{[2]}\hat{\pi}^{(II)}(\boldsymbol{x})]|\xi^{(II)}\rangle$ are equivalent and get simplified to;
 
\begin{equation}\label{ondaT}
^{(2)}\!h''_{xy}(\eta,z) + 2\mathcal{H}^{(II)}\hskip.05cm^{(2)}\!h'_{xy}(\eta,z) = \partial^{2}_{z} \hskip.05cm^{(2)}\!h_{xy}(\eta,z). 
\end{equation} 
Recalling our considerations for the tensor perturbations at the start of this analysis, that is: $^{(2)}h_{xy}(\eta,z) = \hskip.05cm ^{(2)}h_{yx}(\eta,z) = \hskip.05cm ^{(2)}h_{T}(\eta,z)$, (the transverse mode of the gravitational wave). By using (\ref{ordpertHas2Ht}) in order to expand the previous equation to second order in $\varepsilon$, we find:

\begin{eqnarray}\label{ondaTMezcla}
&&\Big( \hskip.02cm ^{(2)}\!h''_{T}(\eta) + 2\mathcal{H}^{(II)} \hskip.02cm ^{(2)}\!h'_{T}(\eta) \Big) + \Big[ \Big( \hskip.02cm ^{(2)}\!H''_{T}(\eta) + 2\mathcal{H}^{(II)} \hskip.02cm ^{(2)}\!H'_{T}(\eta) \Big)e^{2i\vec{k}_{0}\cdot\vec{x}} + c.c \Big] 
= \Big( \partial^{2}_{z} \hskip.05cm ^{(2)}\!H_{T}(\eta) e^{2i\vec{k}_{0}\cdot\vec{x}} + c.c \Big),
\end{eqnarray}
making use here of the linear independence of the functions $\{e^{\pm 2i\vec{k}_{0}\cdot\vec{x}},1\}$, we find that the functions $\hskip.02cm^{(2)}h_{T}(\eta)$, $\hskip.02cm^{(2)}H_{T}(\eta)$ satisfy the dynamical equations;

\begin{eqnarray}
&& \hskip.2cm^{(2)}\!h''_{T}(\eta) + 2\mathcal{H}^{(II)} \hskip.051cm^{(2)}\!h'_{T}(\eta) = 0, \label{ondah} \\
&& \hskip.2cm^{(2)}\!H''_{T}(\eta) + 2\mathcal{H}^{(II)} \hskip.051cm^{(2)}\!H'_{T}(\eta)  =  - 2k_{0z}^{2}\hskip.001cm^{(2)}\!H_{T}(\eta). \label{ondaH}
\end{eqnarray}
The pair of previous equations correspond to dynamical equations for the functions $^{(2)}h_{T}(\eta)$ and $^{(2)}H_{T}(\eta)$ which through equation (\ref{ordpertHas2Ht})  contribute at second order in $\varepsilon$
to the transverse mode of tensor perturbations. \\

Focusing on equation~(\ref{Eij2}) for $l=j=x$, same as for $l=j=y$, and taking into account that for the present setting: $\tilde{\Phi}_{(1)}(\eta,z)$, $\tilde{\Psi}_{(1)}(\eta,z)$, $\tilde{\Phi}_{(2)}(\eta,z)$, $\tilde{\Psi}_{(2)}(\eta,z)$, $^{(2)}h_{xx}(\eta,z) = \hskip.05cm^{(2)}h_{yy}(\eta,z) = -\frac{1}{2}\hskip.05cm^{(2)}h_{zz}(\eta,z) = \hskip.05cm^{(2)}h_{L}(\eta,z)$ (where $h_{L}(\boldsymbol{x})$ represents the longitudinal mode of the gravitational wave).  Equation~(\ref{Eij2}) is reduced to:

\begin{eqnarray}\label{Exxyy2}
&& -\nabla^{2}\tilde{\Psi}_{(2)}(\boldsymbol{x}) + 2\tilde{\Psi}''_{(2)}(\boldsymbol{x}) + 4\mathcal{H}^{(II)}\tilde{\Psi}'_{(2)}(\boldsymbol{x}) + 2\mathcal{H}^{(II)}\tilde{\Phi}'_{(2)}(\boldsymbol{x}) + ( 2\mathcal{H}^{'(II)} + 4\mathcal{H}^{(II)2} ) \tilde{\Phi}_{(2)}(\boldsymbol{x})  + \nabla^{2}\tilde{\Phi}_{(2)}(\boldsymbol{x}) + \nonumber\\
&& 
\frac{1}{2}( \partial^{2}_{\eta} + 2\mathcal{H}^{(II)}\partial_{\eta} - \nabla^{2})\hskip.05cm ^{(2)}h_{L}(\boldsymbol{x}) - 2\Big\{ \tilde{\Phi}^{'2}_{(1)}(\boldsymbol{x}) 
+ 8\mathcal{H}^{(II)}\tilde{\Phi}_{(1)}(\boldsymbol{x})\tilde{\Phi}'_{(1)}(\boldsymbol{x}) - 2\tilde{\Phi}_{(1)}(\boldsymbol{x})\tilde{\Phi}''_{(1)}(\boldsymbol{x}) + 3(\vec{\nabla}\tilde{\Phi}_{(1)}(\boldsymbol{x}))^{2} + \nonumber\\
&& 
2\tilde{\Phi}_{(1)}(\boldsymbol{x})\nabla^{2}\tilde{\Phi}_{(1)}(\boldsymbol{x})  + 4(\mathcal{H}^{'(II)} + 2\mathcal{H}^{(II)2} )(\tilde{\Phi}_{(1)}(\boldsymbol{x}))^{2}  \Big\}
= 8 \pi G\Big\{ \hskip.05cm^{(0)}\!\phi^{'(II)}_{\xi,0} \Big[ \hskip.05cm^{(2)}\!\phi^{'(II)}_{\xi,0}(\eta) + \Big(\hskip.05cm^{(2)}\!\phi^{'(II)}_{\xi,2\vec{k}_{0}}(\eta)e^{2i\vec{k}_{0}\cdot\vec{x}} \nonumber \\
&& + c.c \Big)\Big] + 
\Big(\hskip.05cm^{(1)}\!\phi^{'(II)}_{\xi,\vec{k}_{0}}(\eta)e^{i\vec{k}_{0}\cdot\vec{x}}  + c.c \Big)^{2} 
- \Big(i\vec{k}_{0}\hskip.05cm^{(1)}\!\phi^{(II)}_{\xi,\vec{k}_{0}}(\eta)e^{i\vec{k}_{0}\cdot\vec{x}}  + c.c \Big)^{2} - a^{2(II)} m^{2} \Big[ \hskip.05cm^{(0)}\!\phi^{(II)}_{\xi,0}\hskip.05cm ^{(2)}\!\phi^{(II)}_{\xi,0}(\eta)
\nonumber \\
&&
+ \hskip.05cm^{(0)}\!\phi^{(II)}_{\xi,0}\Big( \hskip.2cm^{(2)}\!\phi^{(II)}_{\xi,2\vec{k}_{0}}(\eta) e^{2i\vec{k}_{0}\cdot\vec{x}}  +  c.c  \Big) + 
\Big( \hskip.2cm^{(1)}\!\phi^{(II)}_{\xi,\vec{k}_{0}}(\eta) e^{i\vec{k}_{0}\cdot\vec{x}}  +  c.c  \Big)^{2} \Big] \Big\}.
\end{eqnarray}

On the other hand, working with equation (\ref{Exxyy2}) along with the $l=j=z$ component of the system of equations  (\ref{Eij2}), we find;

\begin{eqnarray}\label{waveLongitudinal}
&&\frac{3}{2} ( \partial^{2}_{\eta} + 2\mathcal{H}^{(II)}\partial_{\eta} - \nabla^{2})\hskip.05cm^{(2)}h_{L}(\boldsymbol{x}) = \partial^{2}_{z} (\tilde{\Psi}_{(2)}(\boldsymbol{x}) - \tilde{\Phi}_{(2)}(\boldsymbol{x})) + 4 ( \partial_{z}\tilde{\Phi}_{(1)}(\boldsymbol{x}) )^{2} + 8\tilde{\Phi}_{(1)}(\boldsymbol{x}) \partial^{2}_{z}\tilde{\Phi}_{(1)}(\boldsymbol{x}) \nonumber \\ 
&&-\hskip.05cm 16\pi G\Big( ik_{0z}\hskip.001cm^{(1)}\!\phi^{(II)}_{\xi,\vec{k}_{0}}(\eta) e^{i\vec{k}_{0}\cdot\vec{x}}  +  c.c  \Big)^{2}.
\end{eqnarray}

Now substituting (\ref{waveLongitudinal}) in (\ref{Exxyy2}), we obtain;

\begin{eqnarray}\label{Exxyy2nenwave}
&& -\frac{2}{3}\partial^{2}_{z}( \tilde{\Psi}_{(2)}(\boldsymbol{x}) - \tilde{\Phi}_{(2)}(\boldsymbol{x}) ) + 2\tilde{\Psi}''_{(2)}(\boldsymbol{x}) + 2\mathcal{H}^{(II)}(2\tilde{\Psi}'_{(2)}(\boldsymbol{x}) + \tilde{\Phi}'_{(2)}(\boldsymbol{x}) ) + ( 2\mathcal{H}^{'(II)} + 4\mathcal{H}^{(II)2} ) \tilde{\Phi}_{(2)}(\boldsymbol{x}) - \frac{14}{3}\Big(\partial_{z}\tilde{\Phi}_{(1)}(\boldsymbol{x})\Big)^{2} \nonumber \\
&& - \frac{4}{3}\tilde{\Phi}_{(1)}(\boldsymbol{x})\partial^{2}_{z}\tilde{\Phi}_{(1)}(\boldsymbol{x}) - 16\mathcal{H}^{(II)}\tilde{\Phi}_{(1)}(\boldsymbol{x})\tilde{\Phi}'_{(1)}(\boldsymbol{x}) + 4\tilde{\Phi}_{(1)}(\boldsymbol{x})\tilde{\Phi}''_{(1)}(\boldsymbol{x}) - 2(\tilde{\Phi}'_{(1)}(\boldsymbol{x}))^{2} - 8 (\mathcal{H}^{'(II)} + 2\mathcal{H}^{(II)2})(\tilde{\Phi}_{(1)}(\boldsymbol{x}))^{2}
 \nonumber \\
&& = 8 \pi G\Big\{ \hskip.01cm^{(0)}\!\phi^{'(II)}_{\xi,0} \Big[ \hskip.01cm ^{(2)}\!\phi^{'(II)}_{\xi,0}(\eta) + \Big(\hskip.01cm ^{(2)}\!\phi^{'(II)}_{\xi,2\vec{k}_{0}}(\eta) e^{2i\vec{k}_{0}\cdot\vec{x}}  + c.c \Big)\Big] + \Big(\hskip.01cm^{(1)}\!\phi^{'(II)}_{\xi,\vec{k}_{0}}(\eta)e^{i\vec{k}_{0}\cdot\vec{x}}  + c.c \Big)^{2}  
- \Big(i\vec{k}_{0}\hskip.02cm^{(1)}\phi^{(II)}_{\xi,\vec{k}_{0}}(\eta)e^{i\vec{k}_{0}\cdot\vec{x}}  + c.c \Big)^{2} \nonumber \\ 
&&
- a^{2(II)} m^{2} \Big[ \hskip.02cm^{(0)}\!\phi^{(II)}_{\xi,0}\hskip.041cm ^{(2)}\!\phi^{(II)}_{\xi,0}(\eta) +     \hskip.041cm ^{(0)}\!\phi^{(II)}_{\xi,0}\Big( \hskip.01cm ^{(2)}\!\phi^{(II)}_{\xi,2\vec{k}_{0}}(\eta) e^{2i\vec{k}_{0}\cdot\vec{x}}  +  c.c  \Big) 
+ \Big( \hskip.01cm ^{(1)}\!\phi^{(II)}_{\xi,\vec{k}_{0}}(\eta)e^{i\vec{k}_{0}\cdot\vec{x}}  +  c.c  \Big)^{2} \Big] \Big\} \nonumber \\ 
&& + \frac{16\pi G}{3}\Big( ik_{0z}\hskip.01cm ^{(1)}\!\phi^{(II)}_{\xi,\vec{k}_{0}}(\eta) e^{i\vec{k}_{0}\cdot\vec{x}}  +  c.c  \Big)^{2}.  
\end{eqnarray}
Using (\ref{ordpertHas2Phi}) and (\ref{ordpertHas2Psi}) in order to expand the left-hand side of the previous equation to second order in  $\varepsilon$, we arrive to;

\begin{eqnarray}\label{finalExxyy2nenwave}
&& 4a''_{(2)}(\eta) + 4\Big( P''_{(2)}(\eta)e^{2i\vec{k}_{0}\cdot\vec{x}}  + c.c \Big) 
+ 4\mathcal{H}^{(II)} \Big\{ a'_{(2)}(\eta) + \Big( 3P'_{(2)}(\eta)  e^{2i\vec{k}_{0}\cdot\vec{x}} + c.c \Big)\Big\} + 4\Big( \mathcal{H}^{'(II)} + 2\mathcal{H}^{(II)2} \Big)\times\nonumber \\
&&\Big( P_{(2)}(\eta)e^{2i\vec{k}_{0}\cdot\vec{x}}  + c.c \Big)  
 - \frac{14}{3}\Big(i\vec{k}_{0}P_{(1)}(\eta)e^{i\vec{k}_{0}\cdot\vec{x}} + c.c \Big)^{2} + \frac{4}{3}\Big( P_{(1)}(\eta)e^{i\vec{k}_{0}\cdot\vec{x}}  + c.c  \Big)\Big(k_{0}^{2}P_{(1)}(\eta)e^{i\vec{k}_{0}\cdot\vec{x}}  + c.c  \Big) \nonumber \\
&&  
- 16\mathcal{H}^{(II)}\Big( P_{(1)}(\eta)e^{i\vec{k}_{0}\cdot\vec{x}}  + c.c  \Big)\Big( P'_{(1)}(\eta)e^{i\vec{k}_{0}\cdot\vec{x}} + c.c  \Big)
+ 4\Big( P_{(1)}(\eta)e^{i\vec{k}_{0}\cdot\vec{x}}  + c.c  \Big)\Big( P''_{(1)}(\eta)e^{i\vec{k}_{0}\cdot\vec{x}}  + c.c  \Big)  
\nonumber \\
&&
- 2\Big( P'_{(1)}(\eta)e^{i\vec{k}_{0}\cdot\vec{x}}  + c.c  \Big)^{2} - 8 (\mathcal{H}^{'(II)} + 2\mathcal{H}^{(II)2})\Big(P_{(1)}(\eta)e^{i\vec{k}_{0}\cdot\vec{x}}  + c.c  \Big)^{2}
= 8 \pi G\Big\{ \hskip.1cm^{(0)}\!\phi^{'(II)}_{\xi,0} \Big[ \hskip.1cm^{(2)}\!\phi^{'(II)}_{\xi,0}(\eta)\nonumber\\
&& 
+ \Big(\hskip.1cm^{(2)}\!\phi^{'(II)}_{\xi,2\vec{k}_{0}}(\eta)e^{2i\vec{k}_{0}\cdot\vec{x}}  + c.c \Big)\Big] + 
\Big(\hskip.01cm^{(1)}\!\phi^{'(II)}_{\xi,\vec{k}_{0}}(\eta)e^{i\vec{k}_{0}\cdot\vec{x}}  + c.c \Big)^{2}  
- \Big(i\vec{k}_{0}\hskip.05cm^{(1)}\!\phi^{(II)}_{\xi,\vec{k}_{0}}(\eta)e^{i\vec{k}_{0}\cdot\vec{x}}  + c.c \Big)^{2} \nonumber \\
&& 
- a^{2(II)} m^{2} \Big[ \hskip.1cm^{(0)}\!\phi^{(II)}_{\xi,0}\hskip.1cm^{(2)}\!\phi^{(II)}_{\xi,0}(\eta) + \hskip.01cm^{(0)}\!\phi^{(II)}_{\xi,0}\Big(\hskip.02cm^{(2)}\!\phi^{(II)}_{\xi,2\vec{k}_{0}}(\eta) e^{2i\vec{k}_{0}\cdot\vec{x}}  +  c.c  \Big) + 
\Big( \hskip.01cm^{(1)}\!\phi^{(II)}_{\xi,\vec{k}_{0}}(\eta) e^{i\vec{k}_{0}\cdot\vec{x}}  +  c.c  \Big)^{2} \Big] \Big\} \nonumber \\ 
&& + \frac{16\pi G}{3}\Big( i\vec{k}_{0}\hskip.1cm ^{(1)}\!\phi^{(II)}_{\xi,\vec{k}_{0}}(\eta) e^{i\vec{k}_{0}\cdot\vec{x}}  +  c.c  \Big)^{2}, 
\end{eqnarray}
which can be written as;

\begin{equation}\label{situacion12iguales} 
\mathcal{C}_{-2\vec{k}_{0}}(\eta)e^{-2i\vec{k}_{0}\cdot\vec{x}} + \mathcal{C}_{0}(\eta) + \mathcal{C}_{2\vec{k}_{0}}(\eta)e^{2i\vec{k}_{0}\cdot\vec{x}} = 0. 
\end{equation}

\begin{center}
\textbf{Equation for $\mathcal{C}_{2\vec{k}_{0}}(\eta) = 0$}
\end{center}
\begin{eqnarray}\label{DinamicaP2}
&& 4 P''_{(2)}(\eta)  + 12\mathcal{H}^{(II)}P'_{(2)}(\eta) + 4\Big( \mathcal{H}^{'(II)} + 2\mathcal{H}^{(II)2} \Big) P_{(2)}(\eta) - 16\mathcal{H}^{(II)}P_{(1)}(\eta)P'_{(1)}(\eta) + 4P_{(1)}(\eta) P''_{(1)}(\eta) - 2 (P'_{(1)}(\eta))^{2} \nonumber\\
&& - 8 \Big(\mathcal{H}^{'(II)} - \frac{3}{4}k_{0}^{2}  + 2\mathcal{H}^{(II)2} \Big) P_{(1)}^{2}(\eta) = 8 \pi G\Big\{ \hskip.01cm^{(0)}\!\phi^{'(II)}_{\xi,0} \hskip.01cm ^{(2)}\!\phi^{'(II)}_{\xi,2\vec{k}_{0}}(\eta) + \Big(\hskip.05cm^{(1)}\!\phi^{'(II)}_{\xi,\vec{k}_{0}}(\eta)\Big)^{2} 
+ \frac{1}{3}k_{0}^{2} \Big(\hskip.01cm^{(1)}\!\phi^{(II)}_{\xi,\vec{k}_{0}}(\eta)\Big)^{2} \nonumber \\ 
&&
-a^{2(II)} m^{2} \Big[ \hskip.02cm^{(0)}\!\phi^{(II)}_{\xi,0}\hskip.05cm^{(2)}\!\phi^{(II)}_{\xi,2\vec{k}_{0}}(\eta) + \Big( \hskip.05cm^{(1)}\!\phi^{(II)}_{\xi,\vec{k}_{0}} (\eta) \Big)^{2} \Big] \Big\}.  
\end{eqnarray}
On the other hand, the equation obtained from $\mathcal{C}_{-2\vec{k}_{0}}(\eta) = 0$ corresponds to the complex conjugate of equation~(\ref{DinamicaP2}).

A convenient way to write the previous equation is; starting from definitions $\tilde{\mathrm{Q}}^{2}_{(1)}(\eta) = \frac{3}{2} P_{(1)}(\eta)P'_{(1)}(\eta)  - \mathcal{H}^{(II)} P^{2}_{(1)}(\eta) - 2\pi G\hskip.05cm ^{(1)}\phi^{'(II)}_{\xi,\vec{k}_{0}}(\eta)\hskip.05cm ^{(1)}\phi^{(II)}_{\xi,\vec{k}_{0}}(\eta)
$ and $\tilde{\mathrm{S}}_{(1)}^{2}(\eta) = 2P_{(1)}(\eta)P''_{(1)}(\eta) + 3 (P'_{(1)}(\eta))^{2} - 13k_{0}^{2} P_{(1)}^{2}(\eta) + 4\Big( \mathcal{H}^{'(II)} + 2\mathcal{H}^{2(II)} \Big)P_{(1)}^{2}(\eta) - 4\pi G \Big\{ \Big( \hskip.1cm ^{(1)}\phi^{'(II)}_{\xi,\vec{k}_{0}}(\eta) \Big)^{2} + a^{2(II)}m^{2} \Big(\hskip.1cm ^{(1)}\phi^{(II)}_{\xi,\vec{k}_{0}}(\eta) \Big)^{2} - k_{0}^{2} \Big(\hskip.1cm ^{(1)}\phi^{(II)}_{\xi,\vec{k}_{0}}(\eta) \Big)^{2} \Big\}$. Using also (\ref{casodos1}) and (\ref{Eee2finmas}) in order to find $^{(2)}\phi^{(II)}_{\xi,2\vec{k}_{0}}(\eta)$ and $^{(2)}\phi^{'(II)}_{\xi,2\vec{k}_{0}}(\eta)$, we obtain:

\begin{eqnarray}
^{(2)}\phi^{(II)}_{\xi,2\vec{k}_{0}}(\eta) &=&\frac{ P'_{(2)}(\eta)  + \mathcal{H}^{(II)}\hskip.04cm P_{(2)}(\eta) + \tilde{\mathrm{Q}}^{2}_{(1)}(\eta) }{2\pi G\hskip.05cm^{(0)}\phi^{'(II)}_{\xi,0}(\eta) }, \label{evasuenno1} \\
^{(2)}\phi^{'(II)}_{\xi,2\vec{k}_{0}}(\eta)&=& l_{(0)} P'_{(2)}(\eta)  + \tilde{q}_{(0)} P_{(2)}(\eta) + \tilde{\mathrm{W}}^{2}_{(1)}(\eta), \label{evasuenno2}
\end{eqnarray}
with $l_{(0)} = -\frac{1}{ 4 \pi G\hskip.05cm ^{(0)}\phi^{'(II)}_{\xi,0} }\Big( 6\mathcal{H}^{(II)} +  \frac{ 2 a^{2(II)}m^{2}\hskip.05cm ^{(0)}\phi^{(II)}_{\xi,0} }{ ^{(0)}\phi^{'(II)}_{\xi,0} } \Big)$, $\tilde{q}_{(0)} = -\frac{1}{ 4 \pi G\hskip.05cm ^{(0)}\phi^{'(II)}_{\xi,0} }\Big( 8||\vec{k}_{0}||^{2} + 2\mathcal{H}^{'(II)}\mathcal{A} + 4\mathcal{H}^{2(II)}\mathcal{A} +  \frac{2 a^{2(II)}m^{2}\hskip.05 cm ^{(0)}\phi^{(II)}_{\xi,0} \mathcal{H}^{(II)} \mathcal{A} }{ ^{(0)}\phi^{'(II)}_{\xi,0} } \Big)\hskip.08cm$ and $\hskip.08cm\tilde{\mathrm{W}}^{2}_{(1)} = \frac{1}{ 4 \pi G\hskip.05cm ^{(0)}\phi^{'(II)}_{\xi,0} }\Big( \tilde{\mathrm{S}}^{2}_{(1)} - \frac{ 2 a^{2(II)}m^{2}\hskip.05cm ^{(0)}\phi^{(II)}_{\xi,0} \tilde{\mathrm{Q}}^{2}_{(1)} }{ ^{(0)}\phi^{'(II)}_{\xi,0} } \Big).$ 
\\
Defining $\tilde{\mathrm{Z}}^{2}_{(1)}(\eta)$ as;

\begin{eqnarray}\label{ruidoZ1Definicion}
\tilde{\mathrm{Z}}^{2}_{(1)}(\eta) &=&\frac{3||\vec{k}_{0}||^{2}}{2} P^{2}_{(1)}(\eta) - 4\mathcal{H}^{(II)}P_{(1)}(\eta)P'_{(1)}(\eta) + P_{(1)}(\eta)P''_{(1)}(\eta) - \frac{1}{2}(P'_{(1)}(\eta))^{2} -2 \Big( \mathcal{H}^{'(II)} + 2\mathcal{H}^{(II)2} \Big) P_{(1)}^{2}(\eta) \nonumber \\
&& - 2 \pi G\Big\{ \Big(\hskip.2cm^{(1)}\phi^{'(II)}_{\xi,\vec{k}_{0}}(\eta)\Big)^{2} + ||\vec{k}_{0}||^{2} \Big(\hskip.2cm^{(1)}\phi^{(II)}_{\xi,\vec{k}_{0}}(\eta)\Big)^{2} 
- a^{2(II)} m^{2}\Big( \hskip.2cm^{(1)}\phi^{(II)}_{\xi,\vec{k}_{0}}(\eta) \Big)^{2}  \Big\}, \end{eqnarray}
now, applying the previous definition and substituting (\ref{evasuenno1}) and (\ref{evasuenno2}) into (\ref{DinamicaP2}) we find;

\begin{eqnarray}\label{ruidoNewDinamicaP2CasA}
&& P''_{(2)}(\eta) + \Bigg[ 3\mathcal{H}^{(II)} - 2 \pi G \hskip.03cm^{(0)}\!\phi^{'(II)}_{\xi,0}l_{(0)} + \frac{a^{2(II)}m^{2}\hskip.03cm^{(0)}\!\phi^{(II)}_{\xi,0}}{ \hskip.03cm^{(0)}\!\phi^{'(II)}_{\xi,0} }\Bigg]P'_{(2)}(\eta) 
+ \Bigg[  \mathcal{H}^{'(II)}  
 + 2\mathcal{H}^{(II)2} - 2\pi G \hskip.03cm ^{(0)}\!\phi^{'(II)}_{\xi,0} \tilde{q}_{(0)}  \nonumber \\ 
&& + \hskip.03cm \frac{ a^{2(II)}m^{2}\hskip.03cm^{(0)}\!\phi^{(II)}_{\xi,0} \mathcal{H}^{(II)} }{ \hskip.03cm^{(0)}\!\phi^{'(II)}_{\xi,0} }  \Bigg] P_{(2)}(\eta) 
= 2\pi G \hskip.03cm ^{(0)}\!\phi^{'(II)}_{\xi,0}\tilde{\mathrm{W}}^{2}_{(1)}(\eta) - \frac{ a^{2(II)}m^{2}\hskip.03cm^{(0)}\!\phi^{(II)}_{\xi,0} }{ \hskip.03cm^{(0)}\!\phi^{'(II)}_{\xi,0} }\tilde{\mathrm{Q}}^{2}_{(1)}(\eta) - \tilde{\mathrm{Z}}^{2}_{(1)}(\eta), 
\end{eqnarray}
which is a non-homogeneous linear differential equation of second order (dynamical equation for $P_{(2)}(\eta)$), where the non-homogeneous term $2\pi G \hskip.03cm ^{(0)}\phi^{'(II)}_{\xi,0}\tilde{\mathrm{W}}^{2}_{(1)}(\eta) - \frac{ a^{2(II)}m^{2}\hskip.03cm ^{(0)}\phi^{(II)}_{\xi,0}  }{ \hskip.03cm ^{(0)}\phi^{'(II)}_{\xi,0} }\tilde{\mathrm{Q}}^{2}_{(1)}(\eta) - \tilde{\mathrm{Z}}^{2}_{(1)}(\eta)$, corresponds to a set of quantities that were determined at zeroth and first order in $\varepsilon$.


\begin{center}
\textbf{Equation $\mathcal{C}_{0}(\eta) = 0$}
\end{center}

\begin{eqnarray}\label{Dinamicaa2}
&& 4a''_{(2)}(\eta) + 4\mathcal{H}^{(II)}a'_{(2)}(\eta)  
- 16\mathcal{H}^{(II)}\Big( P_{(1)}(\eta)P^{'\ast}_{(1)}(\eta)  + P^{\ast}_{(1)}(\eta)P'_{(1)}(\eta)  \Big) + 4\Big( P_{(1)}(\eta)P^{''\ast}_{(1)}(\eta)  + P^{\ast}_{(1)}(\eta)P''_{(1)}(\eta)  \Big) 
- 4(P'_{(1)}(\eta))^{2} \nonumber \\
&& 
- 16 \Big( \mathcal{H}^{'(II)} + \frac{5}{12}k_{0}^{2} + 2\mathcal{H}^{(II)2} \Big)(P_{(1)}(\eta))^{2}
= 8\pi G\Big\{ \hskip.01cm^{(0)}\!\phi^{'(II)}_{\xi,0} \hskip.01cm ^{(2)}\!\phi^{'(II)}_{\xi,0}(\eta) + 2 (\hskip.01cm ^{(1)}\!\phi^{'(II)}_{\xi,\vec{k}_{0}}(\eta))^{2} - \frac{2}{3} k_{0}^{2} (\hskip.01cm ^{(1)}\!\phi^{(II)}_{\xi,\vec{k}_{0}}(\eta))^{2} \nonumber \\
&& 
- a^{2(II)} m^{2} \Big[ \hskip.02cm ^{(0)}\!\phi^{(II)}_{\xi,0}\hskip.02cm ^{(2)}\!\phi^{(II)}_{\xi,0}(\eta) + \hskip.01cm 2 (\hskip.01cm ^{(1)}\!\phi^{(II)}_{\xi,\vec{k}_{0}}(\eta))^{2} \Big] \Big\},  
\end{eqnarray}
which, with the aid of  equations~(\ref{restricciona}) and (\ref{Eee2fincero}) can be written as; 

\begin{eqnarray}\label{Dinamicaa2desacoplada}
&& a''_{(2)}(\eta) + 4\mathcal{H}^{(II)}a'_{(2)}(\eta) - ( \mathcal{H}^{'(II)}  + 2\mathcal{H}^{2(II)} ) a_{(2)}(\eta) + \frac{ a^{2(II)} m^{2} \hskip.01cm ^{(0)}\!\phi^{(II)}_{\xi,0} }{2\hskip.03cm ^{(0)}\!\phi^{'(II)}_{\xi,0} }( a'_{(2)}(\eta) - \mathcal{H}^{(II)} a_{(2)}(\eta) )
= 8\mathcal{H}^{(II)}P_{(1)}(\eta)P'_{(1)}(\eta) \nonumber \\
&& + 8\Big(\frac{2}{3}k_{0}^{2} + \mathcal{H}^{'(II)} + \mathcal{H}^{(II)2} \Big)(P_{(1)}(\eta))^{2} + 4 (P'_{(1)}(\eta))^{2} 
- 8 \pi G \Big( \frac{2}{3} k_{0}^{2} +  a^{2(II)} m^{2} \Big) (\hskip.05cm ^{(1)}\!\phi^{(II)}_{\xi,\vec{k}_{0}}(\eta))^{2}.
\end{eqnarray}

This equation governs the dynamics of $a_{(2)}$ and is analogous to (\ref{ruidoNewDinamicaP2CasA}). That is, 
the non-homogenous term is determined completely at first order in $\varepsilon$.  The initial data $a_{(2)}(\eta_{c})$ and $a'_{(2)}(\eta_{c})$, are known by using the initial conditions $^{(2)}\phi^{(II)}_{\xi,0}(\eta_{c})$ and $^{(2)}\phi^{'(II)}_{\xi,0}(\eta_{c})$ through equations (\ref{restricciona}) and (\ref{Eee2fincero}) evaluated in $\eta_{c}$. 

The dynamical equations for the functions $^{(2)}\!h_{L}(\eta)$, $^{(2)}\!H_{L}(\eta)$ are obtained by substituting (\ref{ordpertHas2Ht}) into equation~(\ref{waveLongitudinal})  and expanding up to second order in 
$\varepsilon$:

\begin{eqnarray}\label{waveLongitudinalexp}
&&
3\Big( \hskip.05cm ^{(2)}\!h''_{L}(\eta) + 2\mathcal{H}^{(II)} \hskip.05cm ^{(2)}\!h'_{L}(\eta) \Big) + 3\Big[ \Big( \hskip.01cm ^{(2)}\!H''_{L}(\eta) + 2\mathcal{H}^{(II)} \hskip.05cm ^{(2)}\!H'_{L}(\eta) \Big)e^{2i\vec{k}_{0}\cdot\vec{x}} + c.c \Big] 
- 3\Big(\partial^{2}_{z}\hskip.01cm ^{(2)}\!H_{L}(\eta) e^{2i\vec{k}_{0}\cdot\vec{x}} + c.c \Big) \nonumber \\
&&  =  4k_{0}^{2} \Big(iP_{(1)}(\eta)e^{i\vec{k}_{0}\cdot\vec{x}} + c.c \Big)^{2} - 8k_{0}^{2} \Big(P_{(1)}(\eta)e^{i\vec{k}_{0}\cdot\vec{x}} + c.c \Big)^{2} - \hskip.05cm 16\pi G\Big( ik_{0z}\hskip.031cm ^{(1)}\!\phi^{(II)}_{\xi,\vec{k}_{0}}(\eta) e^{i\vec{k}_{0}\cdot\vec{x}}  +  c.c  \Big)^{2},
\end{eqnarray}
which has the form; 

\begin{equation}\label{situacion12diferentes} 
\boldsymbol{\mathcal{C}}_{-2\vec{k}_{0}}(\eta)e^{-2i\vec{k}_{0}\cdot\vec{x}} + \boldsymbol{\mathcal{C}}_{0}(\eta) + \boldsymbol{\mathcal{C}}_{2\vec{k}_{0}}(\eta)e^{2i\vec{k}_{0}\cdot\vec{x}}  
= 0. 
\end{equation}\\

\begin{center}
\textbf{Equation for $\boldsymbol{\mathcal{C}}_{0}(\eta) = 0$}
\end{center}
\begin{equation}\label{CCero}
^{(2)}\!h''_{L}(\eta) + 2\mathcal{H}^{(II)} \hskip.01cm ^{(2)}\!h'_{L}(\eta)  =  - \frac{16}{3}k_{0}^{2} \Big( (P_{(1)}(\eta))^{2} - 4 \pi G (\hskip.05cm ^{(1)}\!\phi^{(II)}_{\xi,\vec{k}_{0}}(\eta))^{2} \Big), 
\end{equation}
this equation governs the dynamics of $^{(2)}h_{L}(\eta)$. 
\\

\begin{center}
\textbf{Equation $\boldsymbol{\mathcal{C}}_{ 2\vec{k}_{0} }(\eta) = (\boldsymbol{\mathcal{C}}_{ -2\vec{k}_{0} }(\eta) )^{\ast} = 0$} 
\end{center}
\begin{equation}\label{3waveLongitudinalexp}
^{(2)}\!H''_{L}(\eta) + 2\mathcal{H}^{(II)} \hskip.05cm ^{(2)}\!H'_{L}(\eta) + 4 k_{0}^{2} \hskip.05cm ^{(2)}\!H_{L}(\eta) 
= -4 k_{0}^{2}\Big( P_{(1)}^{2}(\eta) - \pi G \hskip.01cm (\hskip.01cm ^{(1)}\!\phi^{(II)}_{\xi,\vec{k}_{0}}(\eta) )^{2} \Big). 
\end{equation}
In this way, we obtain the dynamical equation for the function $^{(2)}H_{L}(\eta)$, which in the same manner as 
(\ref{CCero}), is an inhomogeneous equation where the non-homogeneous part is determined at first order in 
$\varepsilon$. 

\subsection{Matching of the two space times at the collapse hypersurface}

Let us suppose that on certain Cauchy hypersurface $\eta=\eta_{c}$ the state $|\xi^{(I)} \rangle \in\mathscr{H}^{(I)}$, has only one mode $\vec{k}$ excited, which is $\vec{k}=0$. All other modes 
$\vec{k}\neq0$ are in their corresponding base states (those of a harmonic oscillator). This hypersurface undergoes a self-induced collapse in such a way that, in general, the post-collapse state could have the 
modes $\vec{k}=n\vec{k}_{0}$ excited, where $n\in\mathbb{Z}$. Along with the excitation on the mode $\vec{k}=0$.

Henceforth, we will define the state  $|\zeta^{(I)}_{g}\rangle\in\mathscr{H}^{(I)}$ as the post-collapse state after previously having the state $|\xi^{(I)}\rangle\in\mathscr{H}^{(I)}$. The transition from $|\xi^{(I)} \rangle$ to  $|\zeta^{(I)}_{g}\rangle$ is denoted as:

\begin{equation}
|\xi^{(I)} \rangle \rightarrow |\zeta^{(I)}_{g} \rangle. 
\end{equation}

It is worth mentioning that the set $\{g^{(I)}(\boldsymbol{x}), \hat{\phi}^{(I)}(\boldsymbol{x}), \hat{\pi}^{(I)}(\boldsymbol{x}), |\zeta^{(I)}_{g}\rangle\in \mathscr{H}^{(I)} \}$ does not represent a SSC-I. This is a consequence that the state $|\zeta^{(I)}_{g}\rangle$ has more excited modes than the state $|\xi^{(I)}\rangle$. 
Therefore, in general it is fulfilled that:

$$\langle\xi^{(I)}|\hat{T}_{\mu}\hskip.02cm^{\nu}[g^{(I)}(\boldsymbol{x}),\hat{\phi}^{(I)}(\boldsymbol{x}),\hat{\pi}^{(I)}(\boldsymbol{x})]|\xi^{(I)}\rangle \neq \langle\zeta^{(I)}_{g}|\hat{T}_{\mu}\hskip.02cm^{\nu}[g^{(I)}(\boldsymbol{x}),\hat{\phi}^{(I)}(\boldsymbol{x}),\hat{\pi}^{(I)}(\boldsymbol{x})]|\zeta^{(I)}_{g}\rangle,$$ 
since: $G_{\mu}\hskip.02cm^{\nu}[g^{(I)}] = 8\pi G \hskip.1cm\langle\xi^{(I)}|\hat{T}_{\mu}\hskip.02cm^{\nu}[g^{(I)}(\boldsymbol{x}),\hat{\phi}^{(I)}(\boldsymbol{x}),\hat{\pi}^{(I)}(\boldsymbol{x})]|\xi^{(I)}\rangle
$ we conclude;

\begin{equation}\label{noSSCeinsU1}
G_{\mu}\hskip.02cm^{\nu}[g^{(I)}] \neq 8\pi G \hskip.1cm\langle\zeta^{(I)}_{g}|\hat{T}_{\mu}\hskip.02cm^{\nu}[g^{(I)}(\boldsymbol{x}),\hat{\phi}^{(I)}(\boldsymbol{x}),\hat{\pi}^{(I)}(\boldsymbol{x})]|\zeta^{(I)}_{g}\rangle.
\end{equation}
However, we will relate the post-collapse state $|\zeta^{(I)}_{g} \rangle\in\mathscr{H}^{(I)}$ with a state
$|\xi\rangle \in\mathscr{H}$, that shares the same excited modes as the state  $|\zeta^{(I)}_{g}\rangle$, and be at the same time, the state used to construct a SSC. With this clarified, we see that the state $|\xi^{(II)}\rangle \in\mathscr{H}^{(II)}$, that we use to construct SSC-II, fulfills the requirements demanded by our hypothesis by being related with the state $|\zeta^{(I)}_{g}\rangle$. We argue that, on certain Cauchy hypersurface $\eta=\eta_{c}$, that is the collapse hypersurface, the states $|\zeta^{(I)}_{g}\rangle$ and $|\xi^{(II)}\rangle$ are related through:

\begin{equation}\label{juntura}
\langle\zeta^{(I)}_{g}|\hat{T}_{\mu}\hskip.02cm^{\nu}[g^{(I)}(\boldsymbol{x}),\hat{\phi}^{(I)}(\boldsymbol{x}),\hat{\pi}^{(I)}(\boldsymbol{x})]|\zeta^{(I)}_{g} \rangle |_{{ \scriptstyle\eta=\eta_{c}}} = \langle\xi^{(II)}|\hat{T}_{\mu}\hskip.02cm^{\nu}[g^{(II)}(\boldsymbol{x}),\hat{\phi}^{(II)}(\boldsymbol{x}),\hat{\pi}^{(II)}(\boldsymbol{x})]|\xi^{(II)} \rangle |_{{ \scriptstyle \eta=\eta_{c}}}, \nonumber \\
\end{equation}
moreover, on this hypersurface, the metric, in going from  $g^{(I)}(\boldsymbol{x})$ to $g^{(II)}(\boldsymbol{x})$, will do so in a continuous manner. That is, $g^{(I)}(\eta_{c},\vec{x}) = g^{(II)}(\eta_{c},\vec{x})$. The set built from the two previous conditions will be denominated, matching condition.
\\

We should  remind the reader that our proposal for the matching conditions envisions  the  gluing  the two space times resulting from the two SSC's constructions. Our approach  is motivated  by  the analogy with a particle that undergoes a sudden impact. In the limit in which the duration of action of the force generating the impulse can be taken to zero, it is natural to assume continuity of the position but sudden jump in the velocity. Following this, we take the induced 3-metric of the collapse hypersurface to be continuous as one approaches that hypersurface from the future and past. However, the instantaneous "rate of change" of the induced  metric  on  that  hypersurface  when viewed {\it  a la}  ADM  (and corresponding to the extrinsic curvature), is assumed to  change discontinuously  (i.e. to undergo a sudden  jump). 
We are aware that this  must be taken   to represent an effective description of  something  much more complicated   going on at the level of the   true underlying degrees of freedom  of a quantum gravity theory  which  we   take as  completely unknown  at this point. This  general   framwork is reflected in Equations~(\ref{empalemresul0}), (\ref{empalemresul1A}), (\ref{empalemresul1C}), (\ref{empalemresul1D}), (\ref{primerordenPetaC}), (\ref{carcomple}), (\ref{2condicionjuntura}), (\ref{edez}), (\ref{2potencialEnColap}), (\ref{solnuevaultima2}), (\ref{condicionB2}),  below.
\\

On the other hand, following the reasoning about the post-collapse state, we will write:

\begin{eqnarray}\label{Guiapostcolapso}
|\zeta^{(I)}_{g} \rangle &=&  ..|0^{(I)}_{ -\vec{k}_{m} } \rangle \otimes..|0^{(I)}_{ -2\vec{k}_{1} } \rangle \otimes|0^{(I)}_{ -\vec{k}_{1} } \rangle \otimes..|\zeta^{(I)}_{ -3\vec{k}_{0} } \rangle \otimes |\zeta^{(I)}_{ -2\vec{k}_{0} } \rangle \otimes |\zeta^{(I)}_{ -\vec{k}_{0} } \rangle \otimes |\zeta^{(I)}_{ 0 } \rangle \otimes |\zeta^{(I)}_{ \vec{k}_{0} } \rangle \otimes |\zeta^{(I)}_{ 2\vec{k}_{0} } \rangle  \nonumber \\
&&
\otimes|\zeta^{(I)}_{ 3\vec{k}_{0} } \rangle\otimes..|0^{(I)}_{ \vec{k}_{1} } \rangle\otimes |0^{(I)}_{ 2\vec{k}_{1} } \rangle.. \otimes|0^{(I)}_{\vec{k}_{m} } \rangle ..., \textup{ siendo} \hskip.3cm  |\zeta^{(I)}_{ \vec{k} } \rangle = \mathcal{F}(\zeta^{(I)}_{\vec{k}} \hat{a}^{(II)\dag}_{ \vec{k} }) |0^{(II)}_{\vec{k}}\rangle, 
\end{eqnarray}
where all other modes $\vec{k}$ different from $n\vec{k}_{0}$ with $n\in\mathbb{Z}$, are in their corresponding base states. However, despite all the modes  $n\vec{k}_{0}$ being excited on the collapse hypersurface, we saw during the construction of the SSC-II, that the excitation of these modes occurs at different orders in 
$\varepsilon$. Therefore, the numbers $\{ \zeta^{(I)}_{ n\vec{k}_{0} } \}_{n\in\mathbb{Z}}$, along with the parameters $\{ \xi^{(II)}_{ n\vec{k}_{0} } \}_{n\in\mathbb{Z}}$, appear in our treatment at different orders in $\varepsilon$. This takes place in such a way that: $\zeta^{(I)}_{ n\vec{k}_{0} } = \varepsilon^{n} \zeta^{(I)}_{(n), n\vec{k}_{0} }$.

Hence, we note that at second order in $\epsilon$, only $\zeta^{(I)}_{0}$, $\zeta^{(I)}_{ \pm\vec{k}_{0} }$ and $\zeta^{(I)}_{ \pm2\vec{k}_{0} }$ intervene in the expansion. Therefore, by simplicity, it is convenient to write the post-collapse state as a state with only $\vec{k}=0$, $\pm\vec{k}_{0}$ and $\pm2\vec{k}_{0},$ modes excited.

\begin{eqnarray}\label{Guiapostcolapsoaprox}
|\zeta^{(I)}_{g} \rangle &=&  ..|0^{(I)}_{ -\vec{k}_{m} } \rangle \otimes..|0^{(I)}_{ -2\vec{k}_{1} } \rangle \otimes|0^{(I)}_{ -\vec{k}_{1} } \rangle \otimes..|0^{(I)}_{ -3\vec{k}_{0} } \rangle \otimes |\zeta^{(I)}_{ -2\vec{k}_{0} } \rangle \otimes |\zeta^{(I)}_{ -\vec{k}_{0} } \rangle \otimes |\zeta^{(I)}_{ 0 } \rangle \nonumber \\
&& \otimes |\zeta^{(I)}_{ \vec{k}_{0} } \rangle \otimes |\zeta^{(I)}_{ 2\vec{k}_{0} } \rangle \otimes|0^{(I)}_{ 3\vec{k}_{0} } \rangle \otimes ..|0^{(I)}_{ \vec{k}_{1} } \rangle\otimes |0^{(I)}_{ 2\vec{k}_{1} } \rangle.. \otimes|0^{(I)}_{\vec{k}_{m} } \rangle ...
\end{eqnarray}

On the other hand, in the same way\footnote{The procedure is analogous but instead of using $u^{(II)}_{\vec{k}}(\boldsymbol{x})$ given by  (\ref{Uk2eps}), we use  $u^{(I)}_{\vec{k}}(\boldsymbol{x})$ given by (\ref{modoshynh}). }  as (\ref{expectacionhcontinu2}) was found. We calculate $\langle\zeta^{(I)}_{g}|\hat{\phi}^{(I)}(\boldsymbol{x})|\zeta^{(I)}_{g}\rangle$, finding :

\begin{eqnarray}\label{carcexpectacionh2}
\phi^{(I)}_{\zeta_{g}}(\boldsymbol{x}) &\equiv& \langle\zeta^{(I)}_{g}|\hat{\phi}^{(I)}(\boldsymbol{x})|\zeta^{(I)}_{g} \rangle 
\nonumber\\
&=& \phi^{(I)}_{\zeta_{g},0}(\eta) + \Big\{\Big[ \phi^{(I)}_{\zeta_{g},\vec{k}_{0}}(\eta) e^{ i\vec{k}_{0}\cdot\vec{x} } + \phi^{(I)}_{\zeta_{g},2\vec{k}_{0}}(\eta) e^{ 2i\vec{k}_{0}\cdot\vec{x} } \Big] + c.c \Big\}, 
\end{eqnarray}
where the auxiliary functions $\phi^{(I)}_{\zeta_{g},0}(\eta)$, $\phi^{(I)}_{\zeta_{g},\vec{k}_{0}}(\eta)$ y $\phi^{(I)}_{\zeta_{g},2\vec{k}_{0}}(\eta)$, correspond to;

\begin{eqnarray}
L^{3/2} \phi^{(I)}_{\zeta_{g},0}(\eta) &=& 
v^{(I)}_{0}(\eta) \zeta_{0}^{(I)} + c.c \label{Aseparaexpectacionh2} 
\\
L^{3/2} \phi^{(I)}_{\zeta_{g},\vec{k}_{0}}(\eta) &=&  v^{(I)}_{ \vec{k}_{0} }(\eta) \zeta^{(I)}_{ \vec{k}_{0} } +  v^{(I) \ast }_{ -\vec{k}_{0} }(\eta) \zeta^{(I)\ast}_{ -\vec{k}_{0} } 
\label{Bseparaexpectacionh2} \\
L^{3/2} \phi^{(I)}_{\zeta_{g},2\vec{k}_{0}}(\eta) &=&  v^{(I)}_{ 2\vec{k}_{0} }(\eta) \zeta^{(I)}_{ 2\vec{k}_{0} } +  v^{(I) \ast }_{ -2\vec{k}_{0} }(\eta) \zeta^{(I)\ast}_{ -2\vec{k}_{0} } \label{cultoseparaexpectacionh2} 
\end{eqnarray}

At zeroth order in $\varepsilon$, the only non zero equations from the matching condition (\ref{juntura}) are: 
$\langle(\delta^{(0)}\hat{T}_{\eta}\hskip.02cm^{ \eta}(\eta_{c}))^{(I)} \rangle_{\zeta_{g}} = \langle(\delta^{(0)}\hat{T}_{\eta}\hskip.02cm^{\eta}(\eta_{c}))^{(II)}\rangle_{\xi}$, and   $\langle(\delta^{(0)}\hat{T}_{l}\hskip.02cm^{l}(\eta_{c}))^{(I)}\rangle_{\zeta_{g}} = \langle(\delta^{(0)}\hat{T}_{l}\hskip.02cm^{l}(\eta_{c}))^{(II)}\rangle_{\xi}$. 
These get reduced to:

\begin{eqnarray}
&&[ (^{(0)}\phi^{'(I)}_{\zeta_{g},0})^{2} + a^{(I)2}m^{2}(^{(0)}\phi^{(I)}_{\zeta_{g},0})^{2} ]\Big|_{\eta_{c}} =  [(^{(0)}\phi^{'(II)}_{\xi,0})^{2} + a^{(II)2}m^{2}(^{(0)}\phi^{(II)}_{\xi,0})^{2}]\Big|_{\eta_{c}}, \nonumber\\ 
\\ 
&&
[(^{(0)}\phi^{'(I)}_{\zeta_{g},0})^{2} - a^{(I)2}m^{2}(^{(0)}\phi^{(I)}_{\zeta_{g},0})^{2} ]\Big|_{\eta_{c}} = [(^{(0)}\phi^{'(II)}_{\xi,0})^{2} - a^{(II)2}m^{2}(^{(0)}\phi^{(II)}_{\xi,0})^{2}]\Big|_{\eta_{c}}.\nonumber\\ 
\end{eqnarray}
The previous system of equations has as solution ; $(^{(0)}\phi^{'(I)}_{\zeta_{g},0}(\eta_{c}))^{2} = (^{(0)}\phi^{'(II)}_{\xi,0}(\eta_{c}))^{2}$, and  $a^{(I)2}(\eta_{c})(^{(0)}\phi^{(I)}_{\zeta_{g},0}(\eta_{c}))^{2} = a^{(II)2}(\eta_{c})(^{(0)}\phi^{(II)}_{\xi,0}(\eta_{c}))^{2}$. Now given the continuity on the metric at the collapse hypersurface : $a^{(I)}(\eta_{c}) = a^{(II)}(\eta_{c})$, we obtain:

\begin{equation}\label{empalemresul0}
^{(0)}\phi^{(I)}_{\zeta_{g},0}(\eta_{c}) \hskip.2cm = \hskip.2cm^{(0)}\phi^{(II)}_{\xi,0}(\eta_{c}), \hskip.2cm \textup{and} \hskip.2cm  ^{(0)}\phi^{'(I)}_{\zeta_{g},0}(\eta_{c}) \hskip.2cm = \hskip.2cm^{(0)}\phi^{'(II)}_{\xi,0}(\eta_{c}). 
\end{equation}

We now proceed to discuss the matching at $\eta=\eta_{c}$ at first order in $\varepsilon$. 
To this order, the only non zero equations derived from the matching condition (\ref{juntura}) are: $\langle(\delta^{(1)}\hat{T}_{\eta}\hskip.02cm^{\eta}(\eta_{c},\vec{x}))^{(I)}\rangle_{\zeta_{g}} = \langle( \delta^{(1)} \hat{T}_{\eta}\hskip.02cm^{\eta}(\eta_{c},\vec{x}) )^{(II)}\rangle_{\xi}$,    $\langle(\delta^{(1)}\hat{T}_{\eta}\hskip.02cm^{l}(\eta_{c},\vec{x}))^{(I)} \rangle_{\zeta_{g}} = \langle(\delta^{(1)}\hat{T}_{\eta}\hskip.02cm^{l}(\eta_{c},\vec{x}))^{(II)}\rangle_{\xi}$, $\langle(\delta^{(1)}\hat{T}_{l}\hskip.02cm^{\eta}(\eta_{c},\vec{x}))^{(I)} \rangle_{\zeta_{g}} = \langle(\delta^{(1)}\hat{T}_{l}\hskip.02cm^{\eta}(\eta_{c},\vec{x}))^{(II)}\rangle_{\xi}$, and   $\langle(\delta^{(1)}\hat{T}_{l}\hskip.02cm^{l}(\eta_{c},\vec{x}))^{(I)} \rangle_{\zeta_{g}} = \langle(\delta^{(1)}\hat{T}_{l}\hskip.02cm^{l}(\eta_{c},\vec{x}))^{(II)}\rangle_{\xi}$. These equations reduce respectively to:  

\begin{eqnarray}
&&^{(0)}\!\phi^{'(I)}_{\zeta_{g},0}(\eta_{c}) \hskip.02cm ^{(1)}\!\phi^{'(I)}_{\zeta_{g},\vec{k}_{0}}(\eta_{c}) + a^{(I)2}(\eta_{c}) m^{2} \hskip.02cm ^{(0)}\!\phi^{(I)}_{\zeta_{g},0}(\eta_{c}) \hskip.02cm ^{(1)}\!\phi^{(I)}_{\zeta_{g},\vec{k}_{0}}(\eta_{c})  
= \hskip.05cm ^{(0)}\!\phi^{'(II)}_{\xi,0}(\eta_{c}) \hskip.02cm ^{(1)}\!\phi^{'(II)}_{\xi,\vec{k}_{0}}(\eta_{c}) \nonumber \\
&&
- [ \hskip.01cm ^{(0)}\!\phi^{'(II)}_{\xi,0}(\eta_{c}) \hskip.01cm]^{2}P_{(1)}(\eta_{c}) + a^{(II)2}(\eta_{c}) m^{2}\hskip.02cm ^{(0)}\!\phi^{(II)}_{\xi,0}(\eta_{c}) \hskip.02cm ^{(1)}\!\phi^{(II)}_{\xi,\vec{k}_{0}}(\eta_{c}),
\label{empalmeP1}\\
&& ^{(0)}\!\phi^{'(I)}_{\zeta_{g},0}(\eta_{c}) \hskip.02cm ^{(1)}\!\phi^{(I)}_{\zeta_{g},\vec{k}_{0}}(\eta_{c}) \hskip.05cm = \hskip.05cm ^{(0)}\!\phi^{'(II)}_{\xi,0}(\eta_{c}) \hskip.02cm ^{(1)}\!\phi^{(II)}_{\xi,\vec{k}_{0}}(\eta_{c}), \label{empalmeP2}\\
&& ^{(0)}\!\phi^{'(I)}_{\zeta_{g},0}(\eta_{c}) \hskip.02cm ^{(1)}\!\phi^{'(I)}_{\zeta_{g},\vec{k}_{0}}(\eta_{c}) - a^{(I)2}(\eta_{c}) m^{2}\hskip.02cm ^{(0)}\!\phi^{(I)}_{\zeta_{g},0}(\eta_{c}) \hskip.02cm ^{(1)}\!\phi^{(I)}_{\zeta_{g},\vec{k}_{0}}(\eta_{c}) 
= \hskip.05cm ^{(0)}\!\phi^{'(II)}_{\xi,0}(\eta_{c}) \hskip.02cm ^{(1)}\!\phi^{'(II)}_{\xi,\vec{k}_{0}}(\eta_{c}) \nonumber \\
&&
- [\hskip.01cm ^{(0)}\!\phi^{'(II)}_{\xi,0}(\eta_{c}) \hskip.01cm]^{2}P_{(1)}(\eta_{c}) - a^{(II)2}(\eta_{c}) m^{2}\hskip.02cm ^{(0)}\!\phi^{(II)}_{\xi,0}(\eta_{c}) \hskip.02cm ^{(1)}\!\phi^{(II)}_{\xi,\vec{k}_{0}}(\eta_{c}). \label{empalmeP3} 
\end{eqnarray}
Making use of equation~(\ref{empalmeP2})  along with the result~(\ref{empalemresul0}), we find that:

\begin{equation}\label{empalemresul1A} 
^{(1)}\!\phi^{(I)}_{\zeta_{g},\vec{k}_{0}}(\eta_{c}) = \hskip.02cm ^{(1)}\!\phi^{(II)}_{\xi,\vec{k}_{0}}(\eta_{c}).
\end{equation}
The next step consists  in adding equations~(\ref{empalmeP1}) and (\ref{empalmeP3}). From this we obtain:

\begin{equation}
^{(0)}\!\phi^{'(I)}_{\zeta_{g},0}(\eta_{c}) \hskip.092cm ^{(1)}\!\phi^{'(I)}_{\zeta_{g},\vec{k}_{0}}(\eta_{c}) = \hskip.02cm ^{(0)}\!\phi^{'(II)}_{\xi,0}(\eta_{c}) \hskip.02cm ^{(1)}\!\phi^{'(II)}_{\xi,\vec{k}_{0}}(\eta_{c}) - [\hskip.01cm ^{(0)}\!\phi^{'(II)}_{\xi,0}(\eta_{c}) \hskip.01cm]^{2}P_{(1)}(\eta_{c}),
\end{equation}
using again the result~(\ref{empalemresul0}) on the previous equation, we find that it can be simplified to obtain:

\begin{equation}\label{diferenciamenossimple}
^{(1)}\!\phi^{'(II)}_{\xi,\vec{k}_{0}}(\eta_{c}) - \hskip.01cm ^{(1)}\!\phi^{'(I)}_{\zeta_{g},\vec{k}_{0}}(\eta_{c})   =   \hskip.01cm ^{(0)}\!\phi^{'(II)}_{\xi,0}(\eta_{c}) P_{(1)}(\eta_{c}).
\end{equation}
Now, due to continuity of the metric at the collapse hypersurface $\eta=\eta_{c}$, the Newtonian potentials on this hypersurface $\eta=\eta_{c}$ are zero. This implies that:

\begin{equation}\label{empalemresul1C}
P_{(1)}(\eta_{c}) = 0.
\end{equation}
Applying this condition on~(\ref{diferenciamenossimple}), we find that:

\begin{equation}\label{empalemresul1D}
^{(1)}\!\phi^{'(I)}_{\zeta_{g},\vec{k}_{0}}(\eta_{c}) = \hskip.02cm ^{(1)}\!\phi^{'(II)}_{\xi,\vec{k}_{0}}(\eta_{c}).
\end{equation}
Lastly, evaluating~(\ref{constricc1}) on $\eta_{c}$ and solving for $P'_{(1)}(\eta_{c})$, we arrive to:

\begin{eqnarray}\label{primerordenPetaC}
P'_{(1)}(\eta_{c}) &=& 4\pi G\hskip.02cm ^{(0)}\!\phi^{'(II)}_{\xi,0}(\eta_{c})\hskip.02cm ^{(1)}\!\phi^{(II)}_{\xi,\vec{k}_{0}}(\eta_{c}). 
\end{eqnarray}
Showing that $P'_{(1)}(\eta)$, unlike $P_{(1)}(\eta)$, is discontinuous on the collapse hypersurface $\eta=\eta_{c}$. 

To finish with this section, we consider the matching at $\eta=\eta_{c}$ at second order in $\varepsilon$. To this order, the only non-zero equations derived from the matching condition~(\ref{juntura}) are; $\langle(\delta^{(2)}\hat{T}_{\eta}\hskip.02cm^{\eta}(\eta_{c},\vec{x}))^{(I)}\rangle_{\zeta_{g}} = \langle(\delta^{(2)}\hat{T}_{\eta}\hskip.02cm^{\eta}(\eta_{c},\vec{x}))^{(II)}\rangle_{\xi}$,    $\langle(\delta^{(2)}\hat{T}_{\eta}\hskip.02cm^{l}(\eta_{c},\vec{x}))^{(I)}\rangle_{\zeta_{g}} = \langle(\delta^{(2)}\hat{T}_{\eta}\hskip.02cm^{l}(\eta_{c},\vec{x}))^{(II)}\rangle_{\xi}$, $\langle(\delta^{(2)}\hat{T}_{l}\hskip.02cm^{\eta}(\eta_{c},\vec{x}))^{(I)}\rangle_{\zeta_{g}} = \langle(\delta^{(2)}\hat{T}_{l}\hskip.02cm^{\eta}(\eta_{c},\vec{x}))^{(II)}\rangle_{\xi}$  and   $\langle(\delta^{(2)}\hat{T}_{l}\hskip.02cm^{l}(\eta_{c},\vec{x}))^{(I)}\rangle_{\zeta_{g}} = \langle(\delta^{(2)}\hat{T}_{l}\hskip.02cm^{l}(\eta_{c},\vec{x}))^{(II)}\rangle_{\xi}$, (due to the symmetries of the problem, equations $\langle(\delta^{(2)}\hat{T}_{i}\hskip.02cm^{j}(\eta_{c},\vec{x}))^{(I)}\rangle_{\zeta_{g}} = \langle(\delta^{(2)}\hat{T}_{i}\hskip.02cm^{j}(\eta_{c},\vec{x}))^{(II)}\rangle_{\xi}$ with $i\neq j$ are trivially satisfied). These equations lead to:

\begin{eqnarray}
&& ^{(0)}\!\phi^{'(I)}_{\zeta_{g},0}(\eta_{c})\hskip.05cm ^{(2)}\!\phi^{'(I)}_{\zeta_{g}}(\eta_{c},\vec{x}) + (\hskip.02cm ^{(1)}\!\phi^{'(I)}_{\zeta_{g}}(\eta_{c},\vec{x}))^{2} + \partial_{i}\hskip.02cm ^{(1)}\!\phi^{(I)}_{\zeta_{g}}(\eta_{c},\vec{x}) \hskip.01cm \partial^{i}\hskip.02cm ^{(1)}\!\phi^{(I)}_{\zeta_{g}}(\eta_{c},\vec{x})  \nonumber \\
&& + a^{2(I)}(\eta_{c}) m^{2} \hskip.01cm ^{(0)}\!\phi^{(I)}_{ \zeta_{g},0}(\eta_{c}) \hskip.01cm ^{(2)}\!\phi^{(I)}_{ \zeta_{g}}(\eta_{c},\vec{x}) + a^{2(I)}(\eta_{c}) m^{2} (\hskip.01cm ^{(1)}\!\phi^{(I)}_{ \zeta_{g}}(\eta_{c},\vec{x}))^{2} \nonumber \\
&& = \hskip.05cm ^{(0)}\!\phi^{'(II)}_{\xi,0}(\eta_{c})\hskip.05cm ^{(2)}\!\phi^{'(II)}_{\xi}(\eta_{c},\vec{x}) - 4 \varepsilon \hskip.05cm ^{(0)}\!\phi^{'(II)}_{\xi,0}(\eta_{c})\tilde{\Phi}_{(1)}(\eta_{c},\vec{x})\hskip.05cm ^{(1)}\!\phi^{'(II)}_{\xi}(\eta_{c},\vec{x}) \nonumber \\
&& + 4\varepsilon^{2} ( \hskip.02cm ^{(0)}\!\phi^{'(II)}_{\xi,0}(\eta_{c}) )^{2}(\tilde{\Phi}_{(1)}(\eta_{c},\vec{x}))^{2} - \varepsilon^{2} (\hskip.02cm ^{(0)}\!\phi^{'(II)}_{\xi,0}(\eta_{c}))^{2}\tilde{\Phi}_{(2)}(\eta_{c},\vec{x}) + (\hskip.02cm ^{(1)}\!\phi^{'(II)}_{\xi}(\eta_{c},\vec{x}))^{2}  \nonumber \\
&& + \partial_{i}\hskip.02cm ^{(1)}\!\phi^{(II)}_{\xi}(\eta_{c},\vec{x}) \hskip.01cm \partial^{i}\hskip.02cm ^{(1)}\!\phi^{(II)}_{\xi}(\eta_{c},\vec{x}) + a^{2(II)}(\eta_{c}) m^{2} \hskip.01cm ^{(0)}\!\phi^{(II)}_{ \xi,0}(\eta_{c}) \hskip.02cm ^{(2)}\phi^{(II)}_{\xi}(\eta_{c},\vec{x}) \nonumber \\
&& + a^{2(II)}(\eta_{c}) m^{2} (\hskip.01cm ^{(1)}\!\phi^{(II)}_{\xi}(\eta_{c},\vec{x}))^{2}, \\
&&
\nonumber \\
&& ^{(0)}\!\phi^{'(I)}_{\zeta_{g},0}(\eta_{c})\partial_{l}\hskip.05cm ^{(2)}\!\phi^{(I)}_{\zeta_{g}}(\eta_{c},\vec{x}) + 2\hskip.02cm ^{(1)}\!\phi^{'(I)}_{\zeta_{g}}(\eta_{c},\vec{x})\hskip.05cm  \partial_{l}\hskip.05cm ^{(1)}\!\phi^{(I)}_{\zeta_{g}}(\eta_{c},\vec{x})   
= \hskip.05cm ^{(0)}\!\phi^{'(II)}_{\xi,0}(\eta_{c})\partial_{l}\hskip.05cm ^{(2)}\!\phi^{(II)}_{\xi}(\eta_{c},\vec{x})  \nonumber \\
&&
+ 2\hskip.05cm ^{(1)}\!\phi^{'(II)}_{\xi}(\eta_{c},\vec{x})\hskip.05cm  \partial_{l}\hskip.05cm ^{(1)}\!\phi^{(II)}_{\xi}(\eta_{c},\vec{x}) + 4 \varepsilon \hskip.05cm ^{(0)}\!\phi^{'(II)}_{\xi,0}(\eta_{c})\hskip.05cm \tilde{\Psi}_{(1)}(\eta_{c}) \partial_{l}\hskip.05cm ^{(1)}\!\phi^{(II)}_{\xi}(\eta_{c},\vec{x}), \\
&& 
\nonumber \\
&& ^{(0)}\!\phi^{'(I)}_{\zeta_{g},0}(\eta_{c})\partial_{l}\hskip.05cm ^{(2)}\!\phi^{(I)}_{\zeta_{g}}(\eta_{c},\vec{x}) + 2\hskip.02cm ^{(1)}\!\phi^{'(I)}_{\zeta_{g}}(\eta_{c},\vec{x})\hskip.05cm  \partial_{l}\hskip.05cm ^{(1)}\!\phi^{(I)}_{\zeta_{g}}(\eta_{c},\vec{x}) 
 = \hskip.05cm ^{(0)}\!\phi^{'(II)}_{\xi,0}(\eta_{c})\partial_{l}\hskip.05cm ^{(2)}\!\phi^{(II)}_{\xi}(\eta_{c},\vec{x}) \nonumber \\
&& 
+ 2\hskip.02cm ^{(1)}\!\phi^{'(II)}_{\xi}(\eta_{c},\vec{x})\hskip.05cm  \partial_{l}\hskip.05cm ^{(1)}\!\phi^{(II)}_{\xi}(\eta_{c},\vec{x})  - 4 \varepsilon \hskip.05cm ^{(0)}\!\phi^{'(II)}_{\xi,0}(\eta_{c})\hskip.05cm \tilde{\Phi}_{(1)}(\eta_{c}) \partial_{l}\hskip.05cm ^{(1)}\!\phi^{(II)}_{\xi}(\eta_{c},\vec{x}), \\
\nonumber \\
&&
^{(0)}\!\phi^{'(I)}_{\zeta_{g},0}(\eta_{c})\hskip.05cm ^{(2)}\!\phi^{'(I)}_{\zeta_{g}}(\eta_{c},\vec{x})   
+ (\hskip.05cm ^{(1)}\!\phi^{'(I)}_{\zeta_{g}}(\eta_{c},\vec{x}))^{2} - \partial_{i}\hskip.05cm ^{(1)}\!\phi^{(I)}_{\zeta_{g}}(\eta_{c},\vec{x}) \hskip.01cm \partial^{i}\hskip.05cm ^{(1)}\!\phi^{(I)}_{\zeta_{g}}(\eta_{c},\vec{x})  \nonumber \\
&& - a^{2(I)}(\eta_{c}) m^{2} \hskip.01cm ^{(0)}\!\phi^{(I)}_{\zeta_{g},0}(\eta_{c})\hskip.01cm ^{(2)}\!\phi^{(I)}_{ \zeta_{g}}(\eta_{c},\vec{x}) - a^{2(I)}(\eta_{c}) m^{2} (\hskip.02cm ^{(1)}\!\phi^{(I)}_{ \zeta_{g}}(\eta_{c},\vec{x}))^{2} \nonumber \\
&&
= \hskip.05cm ^{(0)}\!\phi^{'(II)}_{\xi,0}(\eta_{c})\hskip.05cm ^{(2)}\!\phi^{'(II)}_{\xi}(\eta_{c},\vec{x}) - 4\varepsilon \hskip.05cm ^{(0)}\!\phi^{'(II)}_{\xi,0}(\eta_{c})\tilde{\Phi}_{(1)}(\eta_{c},\vec{x})\hskip.05cm ^{(1)}\!\phi^{'(II)}_{\xi}(\eta_{c},\vec{x}) \nonumber \\
&& + 4\varepsilon^{2}( \hskip.05cm ^{(0)}\!\phi^{'(II)}_{\xi,0}(\eta_{c}) )^{2}(\tilde{\Phi}_{(1)}(\eta_{c},\vec{x}))^{2}  - \varepsilon^{2} (\hskip.05cm ^{(0)}\!\phi^{'(II)}_{\xi,0}(\eta_{c}))^{2}\tilde{\Phi}_{(2)}(\eta_{c},\vec{x}) + (\hskip.05cm ^{(1)}\!\phi^{'(II)}_{\xi}(\eta_{c},\vec{x}))^{2} \nonumber \\
&& - \partial_{i}\hskip.05cm ^{(1)}\!\phi^{(II)}_{\xi}(\eta_{c},\vec{x}) \hskip.01cm \partial^{i}\hskip.05cm ^{(1)}\!\phi^{(II)}_{\xi}(\eta_{c},\vec{x}) - a^{2(II)}(\eta_{c}) m^{2} \hskip.01cm ^{(0)}\!\phi^{(II)}_{ \xi,0}(\eta_{c}) \hskip.02cm ^{(2)}\!\phi^{(II)}_{\xi}(\eta_{c},\vec{x}) \nonumber \\
&& - a^{2(II)}(\eta_{c}) m^{2} (\hskip.01cm ^{(1)}\!\phi^{(II)}_{\xi}(\eta_{c},\vec{x}))^{2}. 
\end{eqnarray}

Using the results (\ref{empalemresul1A}), (\ref{empalemresul1C}) and (\ref{empalemresul1D}), on the previous system of equations, we find:

\begin{eqnarray}
&& ^{(0)}\!\phi^{'(I)}_{\zeta_{g},0}(\eta_{c})\hskip.05cm ^{(2)}\!\phi^{'(I)}_{\zeta_{g}}(\eta_{c},\vec{x}) + a^{2(I)}(\eta_{c}) m^{2} \hskip.01cm ^{(0)}\!\phi^{(I)}_{ \zeta_{g},0}(\eta_{c}) \hskip.01cm ^{(2)}\!\phi^{(I)}_{ \zeta_{g}}(\eta_{c},\vec{x}) 
 = \hskip.05cm ^{(0)}\!\phi^{'(II)}_{\xi,0}(\eta_{c})\hskip.05cm ^{(2)}\!\phi^{'(II)}_{\xi}(\eta_{c},\vec{x}) \nonumber \\
&& - \varepsilon^{2} (\hskip.05cm ^{(0)}\!\phi^{'(II)}_{\xi,0}(\eta_{c}))^{2}\tilde{\Phi}_{(2)}(\eta_{c},\vec{x}) + \hskip.02cm a^{2(II)}(\eta_{c}) m^{2} \hskip.01cm ^{(0)}\!\phi^{(II)}_{\xi,0}(\eta_{c}) \hskip.01cm ^{(2)}\!\phi^{(II)}_{\xi}(\eta_{c},\vec{x}), \label{ebe} \\
&&
\nonumber \\
&& 
^{(0)}\!\phi^{'(I)}_{\zeta_{g},0}(\eta_{c})\partial_{l}\hskip.05cm ^{(2)}\!\phi^{(I)}_{\zeta_{g}}(\eta_{c},\vec{x})  
 = \hskip.05cm ^{(0)}\!\phi^{'(II)}_{\xi,0}(\eta_{c})\partial_{l}\hskip.05cm ^{(2)}\!\phi^{(II)}_{\xi}(\eta_{c},\vec{x}), \label{ebee} \\
&& 
\nonumber \\
&& ^{(0)}\!\phi^{'(I)}_{\zeta_{g},0}(\eta_{c})\hskip.05cm ^{(2)}\!\phi^{'(I)}_{\zeta_{g}}(\eta_{c},\vec{x}) - a^{2(I)}(\eta_{c}) m^{2} \hskip.01cm ^{(0)}\!\phi^{(I)}_{\zeta_{g},0}(\eta_{c})\hskip.01cm ^{(2)}\!\phi^{(I)}_{ \zeta_{g}}(\eta_{c},\vec{x}) = \hskip.05cm ^{(0)}\!\phi^{'(II)}_{\xi,0}(\eta_{c})\hskip.05cm ^{(2)}\!\phi^{'(II)}_{\xi}(\eta_{c},\vec{x}) \nonumber \\
&& - \varepsilon^{2} (\hskip.05cm ^{(0)}\!\phi^{'(II)}_{\xi,0}(\eta_{c}))^{2}\tilde{\Phi}_{(2)}(\eta_{c},\vec{x}) 
- a^{2(II)}(\eta_{c}) m^{2} \hskip.01cm ^{(0)}\!\phi^{(II)}_{ \xi,0}(\eta_{c}) \hskip.1cm ^{(2)}\!\phi^{(II)}_{\xi}(\eta_{c},\vec{x}).  
\label{ebeee}
\end{eqnarray}
Using~(\ref{empalemresul0}) on the equation~(\ref{ebee}) we get;

\begin{equation}\label{carcomple}
\partial_{l}\hskip.01cm ^{(2)}\!\phi^{(I)}_{\zeta_{g}}(\eta_{c},\vec{x}) = \partial_{l}\hskip.01cm ^{(2)}\!\phi^{(II)}_{\xi}(\eta_{c},\vec{x}), 
\end{equation}
while subtracting equations~(\ref{ebe}) and (\ref{ebeee}), and then using $a^{2(I)}(\eta_{c})=a^{2(II)}(\eta_{c})$ along with~(\ref{empalemresul0}) we find;  

\begin{equation}\label{2condicionjuntura}
^{(2)}\!\phi^{(I)}_{\zeta_{g}}(\eta_{c},\vec{x}) = \hskip.02cm ^{(2)}\!\phi^{(II)}_{\xi}(\eta_{c},\vec{x}). 
\end{equation}
Now we use~(\ref{expectacionhcontinu2}) and~(\ref{carcexpectacionh2}) to expand~(\ref{2condicionjuntura}), finding the equation: 

\begin{eqnarray}
&&
^{(2)}\!\phi^{(II)}_{\xi,0}(\eta_{c}) + \Big[ \hskip.02cm ^{(2)}\!\phi^{(II)}_{\xi, 2\vec{k}_{0} }(\eta_{c}) e^{ 2i\vec{k}_{0}\cdot\vec{x} } + c.c \Big] = \hskip.02cm ^{(2)}\!\phi^{(I)}_{\zeta_{g},0}(\eta_{c}) + \Big[ \hskip.02cm ^{(2)}\!\phi^{(I)}_{\zeta_{g}, 2\vec{k}_{0}}(\eta_{c}) e^{ 2i\vec{k}_{0}\cdot\vec{x} } + c.c \Big] .\label{Anuevaultima2}
\end{eqnarray}
If we take $\partial_{l}$ on both sides of equation~(\ref{Anuevaultima2}), we see this is equivalent to the condition (\ref{carcomple}). Therefore, the condition~(\ref{2condicionjuntura})  also guarantees~(\ref{carcomple}).
Besides, applying  on equation~(\ref{Anuevaultima2}) the fact that  the functions  $\{1, e^{ \pm 2i\vec{k}_{0}\cdot\vec{x} } \}$ are linearly independent, we find that:

\begin{equation}\label{edez}
^{(2)}\!\phi^{(I)}_{\zeta_{g},0}(\eta_{c}) = \hskip.02cm ^{(2)}\!\phi^{(II)}_{\xi,0}(\eta_{c}), \hskip.4cm  ^{(2)}\!\phi^{(I)}_{\zeta_{g},2\vec{k}_{0}}(\eta_{c}) = \hskip.02cm ^{(2)}\!\phi^{(II)}_{\xi,2\vec{k}_{0}}(\eta_{c}). 
\end{equation}

Now,  we sum equations~(\ref{ebe}) and (\ref{ebeee}), obtaining:

\begin{eqnarray}\label{ebeeee}
&& \varepsilon^{2}\hskip.005cm (\hskip.01cm ^{(0)}\!\phi^{'(II)}_{\xi,0}(\eta_{c}))^{2}\hskip.005cm \tilde{\Phi}_{(2)}(\eta_{c},\vec{x}) = \hskip.02cm ^{(0)}\!\phi^{'(II)}_{\xi,0}(\eta_{c})\hskip.02cm ^{(2)}\!\phi^{'(II)}_{\xi}(\eta_{c},\vec{x}) - \hskip.02cm ^{(0)}\!\phi^{'(I)}_{\zeta_{g},0}(\eta_{c})\hskip.02cm ^{(2)}\!\phi^{'(I)}_{\zeta_{g}}(\eta_{c},\vec{x}), \nonumber \\
&&   
\end{eqnarray}
using~(\ref{empalemresul0}) on the previous equation, we find;

\begin{eqnarray}\label{ebeeeee}
&& \varepsilon^{2}\hskip.02cm ^{(0)}\!\phi^{'(II)}_{\xi,0}(\eta_{c}) \hskip.02cm \tilde{\Phi}_{(2)}(\eta_{c},\vec{x}) = \hskip.02cm ^{(2)}\!\phi^{'(II)}_{\xi}(\eta_{c},\vec{x}) - \hskip.02cm ^{(2)}\!\phi^{'(I)}_{\zeta_{g}}(\eta_{c},\vec{x}). 
\end{eqnarray}
Given continuity of the metric on the hypersurface $\eta=\eta_{c}$, we obtain:

\begin{equation}\label{2potencialEnColap}
\tilde{\Phi}_{(2)}(\eta_{c},\vec{x}) = 0
\hskip0.31cm \textup{which according to (\ref{ordpertHas2Phi}) implies that:} \hskip0.51cm a_{(2)}(\eta_{c}) = P_{(2)}(\eta_{c}) = 0. 
\end{equation}
Now in the same manner we did with~(\ref{2condicionjuntura}), we will use~(\ref{expectacionhcontinu2}) and
 (\ref{carcexpectacionh2}) to expand~(\ref{ebeeeee}). After applying the property of liner independence for the functions $\{1, e^{ \pm 2i\vec{k}_{0}\cdot\vec{x} } \}$ we find:

\begin{eqnarray}
&&
^{(2)}\!\phi^{'(II)}_{\xi,0}(\eta_{c}) = \hskip.02cm ^{(2)}\!\phi^{'(I)}_{\zeta_{g},0}(\eta_{c}),
\label{edez2} 
\\
&&  
^{(2)}\!\phi^{'(II)}_{\xi, 2\vec{k}_{0} }(\eta_{c}) = \hskip.02cm ^{(2)}\!\phi^{'(I)}_{\zeta_{g},2\vec{k}_{0}}(\eta_{c}). 
\label{solnuevaultima2}
\end{eqnarray}     
Starting from the matching condition, we aim to obtain the condition $P'_{(2)}(\eta_{c})$ in terms of $^{(2)}\!\phi^{(II)}_{\xi,0}(\eta_{c})$, $^{(2)}\!\phi^{(II)}_{\xi,\vec{k}_{0}}(\eta_{c})$ and  $^{(2)}\!\phi^{(II)}_{\xi,2\vec{k}_{0}}(\eta_{c})$.  Evaluating (\ref{casodos1}) at $\eta=\eta_{c}$, taking into account the conditions $P_{(1)}(\eta_{c}) = P_{(2)}(\eta_{c}) = 0$, we obtain that:

\begin{eqnarray}\label{condicionB2}
P'_{(2)}(\eta_{c}) &=& 2\pi G \Big\{ \hskip.02cm ^{(0)}\!\phi^{'(II)}_{\xi,0}(\eta_{c}) \hskip.02cm ^{(2)}\!\phi^{(II)}_{\xi,2\vec{k}_{0}}(\eta_{c}) + \hskip.02cm ^{(1)}\!\phi^{'(II)}_{\xi,\vec{k}_{0}}(\eta_{c}) \hskip.02cm ^{(1)}\!\phi^{(II)}_{\xi,\vec{k}_{0}}(\eta_{c}) \Big\}.
\end{eqnarray}
Therefore, once the initial data $^{(0)}\!\phi^{'(II)}_{\xi,0}(\eta_{c}) =\hskip.02cm ^{(0)}\!\phi^{'(II)}_{\zeta_{g}, 0 }(\eta_{c})$, $^{(1)}\!\phi^{(II)}_{\xi,\vec{k}_{0}}(\eta_{c}) =\hskip.02cm ^{(1)}\!\phi^{(II)}_{\zeta_{g}, \vec{k}_{0} }(\eta_{c})$, $^{(1)}\!\phi^{'(II)}_{\xi,\vec{k}_{0}}(\eta_{c}) =\hskip.02cm ^{(1)}\!\phi^{'(II)}_{\zeta_{g}, \vec{k}_{0} }(\eta_{c})$ and  $^{(2)}\!\phi^{(II)}_{\xi,2\vec{k}_{0}}(\eta_{c}) =\hskip.02cm ^{(2)}\!\phi^{(II)}_{\zeta_{g}, 2\vec{k}_{0} }(\eta_{c})$ are specified, so will be the initial condition (\ref{condicionB2}).
Then, the conditions $P_{(2)}(\eta_{c})=0$ and (\ref{condicionB2}), determine a particular solution $P_{(2)}(\eta)$ of (\ref{ruidoNewDinamicaP2CasA}).
\\

Lastly, the equations of motion for the tensor modes $^{(2)}\!h_{T}(\eta)$, $^{(2)}\!H_{T}(\eta)$, $^{(2)}\!h_{L}(\eta)$ 
and  $^{(2)}\!H_{L}(\eta)$ either correspond to equations of motion without sources, that is, to equations~(\ref{ondah}) and  (\ref{ondaH}), or with sources as~(\ref{CCero}) and (\ref{3waveLongitudinalexp}). The sources being linear combinations of $P^{2}_{(1)}(\eta)$ and $(\hskip.01cm ^{(1)}\!\phi^{(II)}_{\xi,\vec{k}_{0}}(\eta))^{2}$. 

However, as previously shown, to second order in $\varepsilon$, there appear no restriction equations that might allow to impose conditions at $\eta = \eta_{c}$ on the functions $^{(2)}\!h_{T}(\eta)$, $^{(2)}\!H_{T}(\eta)$, $^{(2)}\!h_{L}(\eta)$, $^{(2)}\!H_{L}(\eta)$ and their first derivatives with respect to $\eta$. 
\\

Nevertheless, to finish the analysis at second order in perturbations, there remains to provide the initial values 
$\hskip.01cm ^{(1)}\!\phi^{(I)}_{\zeta_{g},\vec{k}_{0}}(\eta_{c}) = \hskip.01cm ^{(1)}\!\phi^{(II)}_{\xi,\vec{k}_{0}}(\eta_{c})$, $\hskip.01cm ^{(1)}\!\phi^{'(I)}_{\zeta_{g},\vec{k}_{0}}(\eta_{c})= \hskip.01cm ^{(1)}\!\phi^{'(II)}_{\xi,\vec{k}_{0}}(\eta_{c})$, $\hskip.01cm ^{(2)}\!\phi^{(I)}_{\zeta_{g},0}(\eta_{c}) = \hskip.01cm ^{(2)}\!\phi^{(II)}_{\xi,0}(\eta_{c})\hskip.1cm$, $\hskip.01cm ^{(2)}\!\phi^{'(I)}_{\zeta_{g},0}(\eta_{c}) = \hskip.01cm ^{(2)}\!\phi^{'(II)}_{\xi,0}(\eta_{c})\hskip.1cm$,
$\hskip.01cm ^{(2)}\!\phi^{(I)}_{\zeta_{g},2\vec{k}_{0}}(\eta_{c}) = \hskip.01cm ^{(2)}\!\phi^{(II)}_{\xi,2\vec{k}_{0}}(\eta_{c})\hskip.01cm$ and $\hskip.01cm ^{(2)}\!\phi^{'(I)}_{\zeta_{g},2\vec{k}_{0}}(\eta_{c})=\hskip.01cm^{(2)}\!\phi^{'(II)}_{\xi,2\vec{k}_{0}}(\eta_{c})$.
\\

For the remaining of our work, we ignore the physics that allows to describe the self-induced collapse, (the transition from the state $|\xi^{(I)}\rangle$ to $|\zeta^{(I)}_{g}\rangle$ ). We assume that the post-collapse state is randomly chosen, triggered by the quantum uncertainties of the operators  $\hat{\phi}^{(I)}_{\vec{k}}(\eta_{c})$, taken on the pre-collapse state $|\xi^{(I)}\rangle$.

We assume that the self-induced collapse is in a certain way, an imprecise measurement of the operators\footnote{ Starting from the operators $\hat{\phi}_{\vec{k}}^{(I)}(\eta,\vec{x})$, $\hat{\phi}_{\vec{k}}^{(I)\dag}(\eta,\vec{x})$ defined on (\ref{largos}), we define $\hat{\phi}_{\vec{k}}^{(I)}(\eta)$, $\hat{\phi}_{\vec{k}}^{(I)\dag}(\eta)$ as; $$\hat{\phi}_{\vec{k}}^{(I)}(\eta,\boldsymbol{x}) = \frac{ \hat{\phi}_{\vec{k}}^{(I)}(\eta)}{L^{3/2}} e^{ i\vec{k}\cdot\vec{x} }, \hskip.5cm \hat{\phi}_{\vec{k}}^{(I)\dag}(\eta,\boldsymbol{x}) = \frac{\hat{\phi}_{\vec{k}}^{(I)\dag}(\eta)}{L^{3/2}}e^{ -i\vec{k}\cdot\vec{x} }.$$ } $\hat{\phi}_{\vec{k}}^{(I)}(\eta)$. Since we are dealing with non hermitian operators, the first step is to write them down as a linear combination of hermitian operators. Therefore, starting from the definition,


\begin{equation}\label{guiaseparacionRI}
\hat{\phi}^{(I)}_{\vec{k}}(\eta) = v^{(I)}_{\vec{k}}(\eta)\hat{a}_{\vec{k}} + v^{(I)\ast}_{-\vec{k}}(\eta)\hat{a}_{-\vec{k}}^{\dag}, 
\end{equation}
naming\footnote{The non zero commutation relations between the operators $\hat{a}_{\vec{k}}^{(I) \mathcal{R}}$ and $\hat{a}_{\vec{k}}^{(I) \mathcal{I}}$, are; 

\begin{equation}
[\hat{a}_{\vec{k}}^{(I) \mathcal{R}},\hat{a}_{\vec{k}'}^{(I) \mathcal{R} \dag}] = (\delta_{\vec{k},\vec{k}'} + \delta_{\vec{k},-\vec{k}'}), 
\quad[\hat{a}_{\vec{k}}^{(I)\mathcal{I}},\hat{a}_{\vec{k}'}^{(I) \mathcal{I} \dag}]= (\delta_{\vec{k},\vec{k}'} - \delta_{\vec{k},-\vec{k}'}), 
\end{equation}
different to the standard commutation relations.}  $\hat{a}_{\vec{k}}^{(I) \mathcal{R}} \equiv \frac{1}{\sqrt{2}}(\hat{a}_{\vec{k}}^{(I)} + \hat{a}_{-\vec{k}}^{(I)})$, (corresponding to the real part of $\hat{a}_{\vec{k}}^{(I)}$ ), whereas 
$\hat{a}_{\vec{k}}^{(I) \mathcal{I}} \equiv \frac{-i}{\sqrt{2}}(\hat{a}_{\vec{k}}^{(I)} - \hat{a}_{-\vec{k}}^{(I)} )$, (imaginary part of $\hat{a}_{\vec{k}}^{(I)}$ ), along with;


\begin{equation}\label{realeimaginario}
\hat{\phi}_{\vec{k}}^{(I) \mathcal{R},\mathcal{I}}(\eta) = \frac{1}{\sqrt{2}}\Big[ v^{(I)}_{\vec{k}}(\eta)\hat{a}_{\vec{k}}^{(I) \mathcal{R},\mathcal{I}} + v^{(I)\ast}_{\vec{k}}(\eta)\hat{a}_{\vec{k}}^{\dag (I) \mathcal{R},\mathcal{I}}\Big],
\end{equation}

equation~(\ref{guiaseparacionRI}) can be written as,

\begin{equation}\label{guiaseparacionRI22}
\hat{\phi}^{(I)}_{\vec{k}}(\eta) = \hat{\phi}_{\vec{k}}^{(I)\mathcal{R}}(\eta)+i\hat{\phi}_{\vec{k}}^{(I)\mathcal{I}}(\eta). 
\end{equation}
From~(\ref{realeimaginario}), we see that the operators $\hat{\phi}_{\vec{k}}^{(I) \mathcal{R},\mathcal{I}}(\eta)$ are hermitian, since, 

$$\hat{\phi}_{\vec{k}}^{\dag (I) \mathcal{R},\mathcal{I}}(\eta) = \frac{1}{\sqrt{2}}\Big[ v^{(I)\ast}_{\vec{k}}(\eta)\hat{a}_{\vec{k}}^{\dag (I) \mathcal{R},\mathcal{I}} + v^{(I)}_{\vec{k}}(\eta)\hat{a}_{\vec{k}}^{(I) \mathcal{R},\mathcal{I}}\Big] \Rightarrow \hat{\phi}_{\vec{k}}^{(I) \mathcal{R},\mathcal{I}}(\eta) = \hat{\phi}_{\vec{k}}^{\dag (I) \mathcal{R},\mathcal{I}}(\eta).
$$

Previous to the self-induced collapse, the quantum state corresponding to the mode $\vec{k}_{0}$, we have that:
$^{(1)}\!\phi^{(I)\mathcal{R},\mathcal{I}}_{\xi,\vec{k}_{0}}(\eta) = \langle\xi^{(I)}|\hat{\phi}_{\vec{k}}^{\dag (I) \mathcal{R},\mathcal{I}}(\eta) | \xi^{(I)} \rangle = 0$, (valid for $\eta<\eta_{c}$). Now, introducing the hypothesis of self-induced collapse, which states that at the collapse hypersurface $\eta = \eta_{c}$, the state $| \xi^{(I)} \rangle$  collapses to the state $| \zeta^{(I)}_{g} \rangle$, being this the post-collapse state $| \zeta^{(I)}_{g} \rangle$ such that;

\begin{eqnarray}\label{collapse1lunes}
^{(1)}\!\phi^{(I)\mathcal{R},\mathcal{I}}_{\zeta_{g},\vec{k}_{0}}(\eta_{c}) &\equiv& \langle\zeta^{(I)}_{g}|\hat{\phi}_{\vec{k}_{0}}^{(I) \mathcal{R},\mathcal{I}}(\eta_{c}) | \zeta^{(I)}_{g} \rangle = x^{\mathcal{R},\mathcal{I}}_{\vec{k}_{0}} \sqrt{  \langle \xi^{(I)} | [\Delta\hat{\phi}^{(I)}_{\vec{k}_{0}}(\eta_{c})]^{2} |\xi^{(I)} \rangle  } \nonumber \\
&=& x^{\mathcal{R},\mathcal{I} }_{\vec{k}_{0}} \sqrt{\frac{1}{2}} |v^{(I)}_{\vec{k}_{0}}(\eta_{c})|.
\end{eqnarray}
The restriction $\xi^{(II)}_{(1),\vec{k}_{0}} = \xi^{(II)}_{(1),-\vec{k}_{0}}$ required in (\ref{Resultado2}) implies that $x^{\mathcal{I}}_{\vec{k}_{0}}=0.$   Again this  is  in  contrast   with the first order  treatment \cite{DiezTaSudarD}, where it where such overall   phases   are physically irrelevant.
  In the  treatment up to second order in perturbation theory,   one can not  ignore  certain  relative  phases such as that which  would be   involved at this point. 
  Therefore, from   the value of 
$^{(1)}\!\phi^{(I)\mathcal{R},\mathcal{I}}_{\zeta_{g},\vec{k}_{0}}(\eta_{c})$, the quantities  $^{(1)}\!\phi^{(I)}_{\zeta_{g},\vec{k}_{0}}(\eta_{c}) = \hskip.05cm ^{(1)}\!\phi^{(II)}_{\xi,\vec{k}_{0}}(\eta_{c})$ can be determined  from   (\ref{guiaseparacionRI22}), thus:
\begin{equation}\label{resultadocolapso}
^{(1)}\!\phi^{(I)}_{\zeta_{g},\vec{k}_{0}} (\eta_{c}) = \hskip.01cm ^{(1)}\!\phi_{\zeta_{g},\vec{k}_{0}}^{(I)\mathcal{R}}(\eta_{c}) + i\hskip.1cm ^{(1)}\!\phi_{\zeta_{g},\vec{k}_{0}}^{(I)\mathcal{I}}(\eta_{c}) = ( x^{\mathcal{R}}_{\vec{k}_{0}} + i x^{\mathcal{I}}_{\vec{k}_{0}})\sqrt{\frac{1}{2}} |v^{(I)}_{\vec{k}_{0}}(\eta_{c})|,
\end{equation}
whereas $^{(1)}\!\phi^{'(I)}_{\zeta_{g},\vec{k}_{0}}(\eta_{c}) = \hskip.05cm ^{(1)}\!\phi^{'(II)}_{\xi,\vec{k}_{0}}(\eta_{c})$ is determined through equation~(\ref{condicionesPyPdA}) along with~(\ref{empalemresul1C}). 
This means, solving the equation

\begin{equation}
^{(0)}\!\phi^{'(II)}_{\xi,0}(\eta_{c}) \hskip.01cm ^{(1)}\!\phi^{'(II)}_{\xi,\vec{k}_{0}}(\eta_{c})  + \Big( a^{2(II)}(\eta_{c}) m^{2}\hskip.01cm ^{(0)}\!\phi^{(II)}_{\xi,0}(\eta_{c}) + 3\mathcal{H}^{ (II) }(\eta_{c})\hskip.01cm ^{(0)}\!\phi^{'(II)}_{\xi,0}(\eta_{c})  \Big)\hskip.01cm ^{(1)}\!\phi^{(II)}_{\xi,\vec{k}_{0}}(\eta_{c}) = 0. 
\end{equation}

Considering the above, we obtain $^{(1)}\!\phi^{'(II)}_{\xi,\vec{k}_{0}}(\eta_{c})$ as function of $^{(1)}\!\phi^{(II)}_{\xi,\vec{k}_{0}}(\eta_{c})$. Having done this, we obtain  the initial data required for the construction at first order in the perturbation of the SSC-II and its matching to the SSC-I.

Regarding the second order matching in $\varepsilon$, starting from~(\ref{Aseparaexpectacionh2}) and its derivative with respect to $\eta$, that is:

\begin{eqnarray}
&&L^{3/2} \hskip.05cm\phi^{(I)}_{\zeta_{g},0}(\eta) = \Big( v^{(I)}_{0}(\eta) \zeta^{(I)}_{0 } + c.c \Big), \label{virrey} \\
&&
L^{3/2} \hskip.05cm\phi^{'(I)}_{\zeta_{g},0}(\eta) = \Big( v^{'(I)}_{0}(\eta) \zeta^{(I)}_{0 } + c.c \Big). \label{virrey2}
\end{eqnarray}
From the above, we see that the quantities $\phi^{(I)}_{\zeta_{g},0}(\eta)$ and $\phi^{'(I)}_{\zeta_{g},0}(\eta)$ only
contribute to the calculation at first order in $\varepsilon$. Hence, we conclude that  $^{(2)}\!\phi^{(I)}_{\zeta_{g},0}(\eta_{c}) = \hskip.05cm ^{(2)}\!\phi^{'(I)}_{\zeta_{g},0}(\eta_{c}) = 0$. However, using  (\ref{separaexpectacionh2subcasoB}) along with the motion equations for the variables $\delta^{(1)}v^{(II)+}_{\vec{k}}$, $\delta^{(1)}v^{(II)-}_{\vec{k}}$ and $\theta^{(II)}_{\vec{k}}$: that is, equations (\ref{Klv1masII}), (\ref{Klv1menII}) and (\ref{segperturIIC0}), we find that the motion equation for $^{(2)}\!\phi^{(I)}_{\zeta_{g},0}(\eta)$ is a non-homogenous equation. Therefore, despite the initial data $^{(2)}\!\phi^{(II)}_{\xi,0}(\eta_{c})$ and $^{(2)}\!\phi^{'(II)}_{\xi,0}(\eta_{c})$ being zero, in general, the quantity $^{(2)}\!\phi^{(II)}_{\xi,0}(\eta)$ will not be so.

Implementing these results on the equation~(\ref{restricciona}), or on~(\ref{Eee2fincero}) and taking into account the continuity of the metric at the collapse hypersurface, we find the initial data $a^{'}_{(2)}(\eta_{c})$; 

\begin{eqnarray}
&& \mathcal{H}^{(II)}(\eta_{c}) a^{'}_{(2)}(\eta_{c}) - |P^{'}_{(1)}(\eta_{c})|^{2} = -\frac{4}{3}\pi G \Big\{ |\hskip.05cm^{(1)}\!\phi^{'(II)}_{\xi,\vec{k}_{0}}(\eta_{c})|^{2}
+ \Big( a^{2(II)}(\eta_{c})m^{2} + k_{0}^{2}\Big) \hskip.05cm |\hskip.01cm ^{(1)}\!\phi^{(II)}_{\xi,\vec{k}_{0}}(\eta_{c})|^{2} \Big\}. 
\end{eqnarray}
There only remains the initial data $^{(2)}\!\phi^{(II)}_{\xi,2\vec{k}_{0}}(\eta_{c}) = \hskip.05cm ^{(2)}\!\phi^{(I)}_{\zeta_{g},2\vec{k}_{0}}(\eta_{c}) \hskip.051cm$ and  $ \hskip.051cm^{(2)}\!\phi^{'(II)}_{\xi,2\vec{k}_{0}}(\eta_{c}) = \hskip.05cm^{(2)}\!\phi^{'(I)}_{\zeta_{g},2\vec{k}_{0}}(\eta_{c})$ to be determined.

According to equation~(\ref{cultoseparaexpectacionh2}), the quantity $^{(2)}\!\phi^{(I)}_{\zeta_{g},2\vec{k}_{0}}(\eta)$ is given by ;
$$L^{3/2} \hskip.05cm ^{(2)} \!\phi^{(I)}_{\zeta_{g},2\vec{k}_{0}}(\eta) = v^{(I)}_{ 2\vec{k}_{0} }(\eta) \zeta^{(I)}_{(2), 2\vec{k}_{0} } +  v^{(I) \ast }_{ -2\vec{k}_{0} }(\eta) \zeta^{(I)\ast}_{(2), -2\vec{k}_{0} }$$
This quantity will be determined once the numbers $\zeta^{(I)}_{ (2), 2\vec{k}_{0} }$ and $\zeta^{(I)\ast}_{(2), -2\vec{k}_{0} }$ are known. These numbers appear at second  order in the expansion on  $\varepsilon$.

On the other hand, according to the results $^{(2)}\!\phi^{(I)}_{\zeta_{g},2\vec{k}_{0}}(\eta_{c}) = \hskip.02cm ^{(2)}\!\phi^{(II)}_{\xi,2\vec{k}_{0}}(\eta_{c})$ and $^{(2)}\!\phi^{'(I)}_{\zeta_{g},2\vec{k}_{0}}(\eta_{c}) =\hskip.05cm ^{(2)}\!\phi^{'(II)}_{\xi,2\vec{k}_{0}}(\eta_{c})$ derived from the matching condition at $\varepsilon^{2}$ order, we must calculate the contribution at second order of equation~(\ref{Bseparaexpectacionh2subcasoB})  evaluated at 
$\eta=\eta_{c}$. In the same manner, for its first derivative with respect to conformal time. These numbers are respectively:

\begin{eqnarray}
L^{3/2}\hskip.05cm ^{(2)}\!\phi^{(II)}_{\xi,2\vec{k}_{0}}(\eta_{c}) &=&  \delta^{(0)}v^{(II)}_{ 2\vec{k}_{0} }(\eta_{c})\xi^{(II)}_{(2), 2\vec{k}_{0} }  + \delta^{(0)}v^{(II) \ast }_{ -2\vec{k}_{0} }(\eta_{c}) \xi^{(II)\ast}_{(2), -2\vec{k}_{0} }  +  \varepsilon^{2} \Big[ \delta^{(2)}\tilde{v}^{(II)+}_{0}(\eta_{c}) \xi^{(II)}_{(0),0} \nonumber \\
&& + \hskip.01cm (\delta^{(2)}\tilde{v}^{(II)-}_{0}(\eta_{c}))^{\ast}\hskip.05cm\xi_{(0),0}^{(II)\ast} \Big], \label{necioB} \\
\nonumber \\
L^{3/2}\hskip.05cm ^{(2)}\!\phi^{'(II)}_{\xi,2\vec{k}_{0}}(\eta_{c}) &=& \delta^{(0)}v^{'(II)}_{ 2\vec{k}_{0} }(\eta_{c})\xi^{(II)}_{(2), 2\vec{k}_{0} } +  \delta^{(0)}v^{'(II) \ast }_{ -2\vec{k}_{0} }(\eta_{c}) \xi^{(II)\ast}_{ (2), -2\vec{k}_{0} } + \varepsilon^{2} \Big[ \delta^{(2)}\tilde{v}^{'(II)+}_{0}(\eta_{c}) \xi^{(II)}_{(0),0} \nonumber \\
&& + \hskip.01cm  (\delta^{(2)}\tilde{v}^{'(II)-}_{0}(\eta_{c}))^{\ast}\hskip.05cm \xi_{(0),0}^{(II)\ast} \Big]. \label{necio2B}
\end{eqnarray}

The numbers $\delta^{(2)}\tilde{v}^{'(II)+}_{0}(\eta_{c})$ and $\delta^{(2)}\tilde{v}^{'(II)-}_{0}(\eta_{c})$
are obtained by evaluating (\ref{BVanormaliza22V}) and (\ref{Vanormaliza22V}) at $\eta=\eta_{c}$. Therefore, taking into account the conditions~(\ref{empalemresul1C}) and (\ref{2potencialEnColap}), which guarantee continuity of the metric on the collapse hypersurface, we obtain:

\begin{equation}\label{lunesinicio00}
\delta^{(2)}\tilde{v}^{(II)\pm}_{\vec{k}}(\eta_{c}) = \delta^{(2)}\tilde{v}^{'(II)\pm}_{\vec{k}}(\eta_{c}) = 0. 
\end{equation}
Hence, the system of equations~(\ref{necioB}) and (\ref{necio2B}), reduces to;

\begin{eqnarray}
L^{3/2}\hskip.05cm ^{(2)}\!\phi^{(II)}_{\xi,2\vec{k}_{0}}(\eta_{c}) &=&  \delta^{(0)}v^{(II)}_{ 2\vec{k}_{0} }(\eta_{c})\xi^{(II)}_{(2), 2\vec{k}_{0} }  + \delta^{(0)}v^{(II) \ast }_{ -2\vec{k}_{0} }(\eta_{c}) \xi^{(II)\ast}_{(2), -2\vec{k}_{0} }, \label{necioBRes} \\
\nonumber \\
L^{3/2}\hskip.05cm ^{(2)}\!\phi^{'(II)}_{\xi,2\vec{k}_{0}}(\eta_{c}) &=& \delta^{(0)}v^{'(II)}_{ 2\vec{k}_{0} }(\eta_{c})\xi^{(II)}_{(2), 2\vec{k}_{0} } +  \delta^{(0)}v^{'(II) \ast }_{ -2\vec{k}_{0} }(\eta_{c}) \xi^{(II)\ast}_{ (2), -2\vec{k}_{0} }  \label{necio2BRes}
\end{eqnarray}

To finish with the construction at second order in perturbation theory,  the numbers $\zeta^{(I)}_{(2), 2\vec{k}_{0} }$ and $\zeta^{(I)\ast}_{(2),-2\vec{k}_{0}}$ ought to be specified. These numbers determine the post-collapse state $|\zeta^{(I)}_{g}\rangle$ to second order in $\varepsilon$. Therefore, in a similar manner to the way we proceeded at fist order in the perturbation expansion, we assume that:
 
\begin{equation}\label{collapse2lunes}
^{(2)}\!\phi^{(I)\mathcal{R},\mathcal{I}}_{\zeta_{g},2\vec{k}_{0}}(\eta_{c}) \equiv \hskip.05cm ^{(2)}\Big( \langle\zeta^{(I)}_{g}|\hat{\phi}_{2\vec{k}_{0}}^{(I) \mathcal{R},\mathcal{I}}(\eta) | \zeta^{(I)}_{g} \rangle \Big) = x^{\mathcal{R},\mathcal{I}}_{2\vec{k}_{0}} \sqrt{  \langle 0^{(I)} | [\Delta\hat{\phi}^{(I)}_{2\vec{k}_{0}}(\eta_{c})]^{2} |0^{(I)} \rangle  } = x^{\mathcal{R},\mathcal{I}}_{2\vec{k}_{0}} \sqrt{\frac{1}{2}} |v^{(I)}_{2\vec{k}_{0}}(\eta_{c})|
\end{equation}
The difference with respect to (\ref{collapse1lunes}) being that (\ref{collapse2lunes}) contributes from second order in the perturbation expansion onwards.  
\\

Then, once $^{(2)}\!\phi^{(I)\mathcal{R},\mathcal{I}}_{\zeta_{g},2\vec{k}_{0}}(\eta_{c})$ is known, and according to~(\ref{guiaseparacionRI22}), the number $^{(2)}\!\phi^{(I)}_{\zeta_{g},2\vec{k}_{0}}(\eta_{c}) =\hskip.05cm ^{(2)}\!\phi^{(II)}_{\xi,2\vec{k}_{0}}(\eta_{c})$ is given by;

\begin{equation}
^{(2)}\!\phi^{(I)}_{\zeta_{g},2\vec{k}_{0}}(\eta_{c}) = \hskip.05cm ^{(2)}\!\phi^{(I)\mathcal{R}}_{\zeta_{g},2\vec{k}_{0}}(\eta_{c}) + i\hskip.05cm^{(2)}\!\phi^{(I)\mathcal{I}}_{\zeta_{g},2\vec{k}_{0}}(\eta_{c})  = ( x^{\mathcal{R}}_{2\vec{k}_{0}} + i x^{\mathcal{I}}_{2\vec{k}_{0}}) \sqrt{\frac{1}{2}} |v^{(I)}_{2\vec{k}_{0}}(\eta_{c})|.
\end{equation}

Lastly, by solving the system (\ref{casodos1})-(\ref{Eee2finmas}), evaluated at $\eta=\eta_{c}$, we can find 
$^{(2)}\!\phi^{'(II)}_{\xi,2\vec{k}_{0}}(\eta_{c})$ as function of $^{(2)}\!\phi^{(II)}_{\xi,2\vec{k}_{0}}(\eta_{c})$. With the aid of the junction condition~(\ref{edez}) and (\ref{edez2}) can be understood as  $^{(2)}\!\phi^{'(I)}_{\zeta_{g},2\vec{k}_{0}}(\eta_{c})$, a function of $^{(2)}\!\phi^{(I)}_{\zeta_{g},2\vec{k}_{0}}(\eta_{c})$,  that is;

\begin{eqnarray}
&& \Big\{ - \Big( 3\mathcal{H}^{(I)} \hskip.05cm ^{(0)}\!\phi^{'(I)}_{\zeta_{g},0}  + a^{ 2(I) } m^{ 2 } \hskip.05cm^{(0)}\!\phi^{(I)}_{\zeta_{g},0} \Big)\hskip.05cm ^{(2)}\!\phi^{(I)}_{\zeta_{g},2\vec{k}_{0}}(\eta) - \hskip.03cm ^{(0)}\!\phi^{'(I)}_{\zeta_{g},0} \hskip.03cm ^{(2)}\!\phi^{'(I)}_{\zeta_{g},2\vec{k}_{0}}(\eta) \Big\} \Big|_{\eta=\eta_{c}} = \Big\{ \Big( \hskip.02cm ^{(1)}\!\phi^{'(I)}_{\zeta_{g},\vec{k}_{0}}(\eta)\Big)^{2} 
\nonumber \\
&&  
+ (a^{ 2(I) } m^{ 2 }- k_{0}^{2}) \Big(\hskip.05cm^{(1)}\!\phi^{(I)}_{\zeta_{g},\vec{k}_{0}}(\eta)\Big)^{2}  + 3\mathcal{H}^{(I)} \hskip.05cm ^{(1)}\!\phi^{'(I)}_{\zeta_{g},\vec{k}_{0}}(\eta) \hskip.1cm ^{(1)}\!\phi^{(I)}_{\zeta_{g},\vec{k}_{0}}(\eta) \Big\} \Big|_{\eta=\eta_{c}}. 
\end{eqnarray}

Therefore, having given the data $^{(0)}\!\phi^{(II)}_{\xi,0}(\eta_{c})$, $^{(0)}\!\phi^{'(II)}_{\xi,0} (\eta_{c})$, $^{(1)}\!\phi^{(II)}_{\xi,\vec{k}_{0}}(\eta_{c})$, $^{(1)}\!\phi^{'(II)}_{\xi,\vec{k}_{0}} (\eta_{c})$, $^{(2)}\!\phi^{(II)}_{\xi,0}(\eta_{c})$, $^{(2)}\!\phi^{'(II)}_{\xi,0}(\eta_{c})$, $^{(2)}\!\phi^{(II)}_{\xi,2\vec{k}_{0}}(\eta_{c})$, and $^{(2)}\!\phi^{'(II)}_{\xi,2\vec{k}_{0}}(\eta_{c})$, along with the conditions~(\ref{empalemresul0}), (\ref{empalemresul1A}), (\ref{empalemresul1D}), (\ref{edez}),  (\ref{edez2}) and (\ref{solnuevaultima2}), we completely know the initial data required for the construction to second order in $\varepsilon$ of the SSC-II and its junction or matching to the SSC-II.

\section{Conclusions}
\label{sec:concl}

In this  work  we have  extended   to  second order in perturbation theory the application of the SSC   formalism   for the incorporation of   spontaneous 
collapse of quantum  states as   a  source  of the emergence of the seeds of structure in inflationary cosmology.  This     contributes to   
demonstrate  the robustness of  the   formalism, at   least  in its application to situations that  represent    small  deviations from homogeneity and   isotropy.

In the process, we  uncover  the need to incorporate the    tensor perturbations  in  these higher-order treatments, a feature that distinguishes it from the  first-order ones  in which  the   scalar perturbations can be considered  in  isolation. On the   other hand,  the inclusion of tensor perturbations is  not  a sufficient condition for the construction of SSC-II up to second order in perturbations, and  in fact, a  specific matching  between  the excitation of   the   different  field   modes  is  found  to  be necesary. In fact  the construction of the SSC-II is possible only if , in addition to including tensor perturbations in the metric, one also considers the contributions of new excited modes: $\pm 2\vec{k}_{0}$, which can be disregarded up to first order in perturbations, (or in other words, up to first order in the perturbation, the state in the SSC-II just needs the modes $\vec{0}$ and $\pm\vec{k}_{0}$ to  be excited, whereas at second order in the perturbation, the  excited modes  must include   $\vec{0}$, $\pm\vec{k}_{0}$ and $\pm2\vec{k}_{0}$ ). 
 The  excitation of these modes  must be   phase correlated as discussed  in  the context of equation~(\ref{problematica1}). In the same manner that Eq.~(\ref{problematica1}) arises at orders higher than first, it is likely  that new  type  of constraints  on the excitation of   different modes   might  arise   from  considerations at   higher orders. 
 From the physical point of view,   the  first result   is  not really unexpected as  it   is  well known that  scalar perturbations act as sources  of   tensor perturbations, and in fact, as it is shown in \cite{Leon_Majhi_Okon_Sud1},\cite{Leon_Majhi_Okon_Sud2}, in the context of semiclassical  treatments involving  spontaneous  collapse,  these  second-order effects  might be the only sources of tensor modes, in contrast with   the  standard treatments where tensor modes are generated in  close analogy   with  scalar  perturbations.  A    fact that leads  in standard  treatments to  expectations  of   realtivelly  high  amplitudes   for the  B polarization  modes in  the Cosmic Microwave Background, which  have not been observed to this point. The present   work, which   contributes  to   establish the feasibility and reliability  of  extending  the  SSC formalism to higher  orders,   can therefore,    be  considered  as   offering further support  to the   analysis carried out in \cite{Leon_Majhi_Okon_Sud1} \cite{Leon_Majhi_Okon_Sud2}, which  depends on  a  second  order  treatment.

On the other hand, the  sheer complexity of the   resulting treatment   indicates that   attempts to go  beyond  second order,   would probably  represent  monumental tasks and, the question of convergence of the perturbative treatment  will   remain   for the foreseeable   future  an  open one.  Although   at fist sight,  we  have found   nothing  here  that    can be taken as   indicative  that  there  will be   some  obstacle to  such    convergence.   

We  conclude,  by  confirming extendability to higher  order in perturbation theory, that  the  SSC formalism   introduced  in \cite{DiezTaSudarD}   possesses  a   high degree of  robustness  and  that   it  offers  a  solid  ground to explore   certain  issues  at the interface of quantum  theory and gravitation.   Of course,   this  should  be restricted  to  situations   where a semi-classical  treatment  is not  ruled out by indications that  one  is dealing  with Planck  scale physics,  or that   the  quantum aspects are not  so   large as  to render the notion of  classical space-time irremediably lost.

\section*{Acknowledgements}

DS acknowledges partial financial support from DGAPA-UNAM project IG100316 and by CONACyT project 101712, as well as the sabbatical fellowship from CO-MEX-US (Fullbright--Garcia Robles) and from DGAPA-UNAM (Paspa). ER is grateful to FAPEMIG for supporting her visit in 2016 to the Federal University of Juiz de Fora, MG, Brazil, where part of this work was done.



\end{document}